\documentclass[aps,prb,twocolumn,superscriptaddress,amssymb,amsfonts,longbibliography,10pt]{revtex4-2}

\usepackage{amsmath}
\usepackage{amssymb}
\usepackage{graphicx}
\usepackage{bm}
\usepackage{braket}
\usepackage{array}
\usepackage{bbold}
\usepackage[normalem]{ulem}
\usepackage[dvipsnames]{xcolor}
\usepackage[
    colorlinks,
    linkcolor={blue!80!black},
    citecolor={blue!80!black},
    urlcolor={blue!80!black}
]{hyperref}

\newcommand{\bP}{\hat P}
\def\vec{\mathbf}
\newcommand{\bH}{\hat H}

\def\vec{\mathbf}

\begin{document}
\title{
Drude weight of an interacting flat-band metal
}
\author{Ohad Antebi}
\thanks{These authors contributed equally.}
\affiliation{Department of Condensed Matter Physics, Weizmann Institute of Science, Rehovot 76100, Israel}
\author{Johannes Mitscherling}
\thanks{These authors contributed equally.}
\affiliation{Department of Physics, University of California,  Berkeley, California 94720, USA}
\author{Tobias Holder}
\thanks{tobiasholder@tauex.tau.ac.il}
\affiliation{School of Physics and Astronomy, Tel Aviv University, Tel Aviv 69978, Israel}
\date{\today}

\begin{abstract}
    Flatband systems form a new class of materials that challenge the conventional wisdom of transport. 
    The intrinsically strong electronic correlations combined with the vanishing kinetic energy scale suggest a sensitive dependence of transport properties on the flat band states and make interacting flat bands promising candidates for exotic quantum transport. 
    Utilizing the Drude weight, we investigate the low-frequency spectral properties of the electrical conductivity within a controlled analytic treatment of the many-body response at temperatures above the bandwidth and the interaction strength and below the bandgap. 
    Focusing on this new transport regime, we demonstrate the potential of a quantum geometric approach for interacting systems and intermediate temperatures. The derived spectral weight yields unexplored four-point geometric contributions unrelated to the quantum metric, which questions the previously proposed projection methods. For long-ranged interactions, we show that the low-frequency spectral weight reduces to the variance of the Berry curvature.
\end{abstract}

\maketitle

\emph{Introduction.---}
Recent progress in the fabrication and characterization of quantum materials and heterostructures~\cite{Liu2016, Giustino2021} has allowed the systematic exploration of engineered electronic states that reside predominantly in isolated bands with strongly quenched kinetic energy~\cite{Regnault2022}. These \emph{approximate} flat bands exhibit a rich experimental phenomenology, ranging from correlated insulating states~\cite{Cao2018a, Zhou2021a}, fractional quantum Hall physics~\cite{Zeng2023, Cai2023, Kang2024}, flavor cascades~\cite{Zhou2021, Zondiner2020} and unconventional superconductivity~\cite{Cao2018, Tian2023} to Planckian transport and strange metallicity~\cite{Polshyn2019, Cao2020, Checkelsky2024, Ye2024}. These experimental results challenge the theoretical understanding of electron transport as a mostly semiclassical phenomenon since quantum fluctuations are pronouncedly enhanced. Indeed, perfectly flat bands with attractive interaction counter-intuitively show stable superconductivity with a large superfluid stiffness~\cite{Peotta2015, Liang2017, Hofmann2020, Torma2022, Herzogarbeitman2022, Huhtinen2022, Mao2023, Hofmann2023}, which originates from the nontrivial momentum dependence of the Bloch states. Namely, it was first shown within mean-field theory~\cite{Peotta2015, Liang2017} that the superfluid weight in a superconducting flat band is proportional to the interaction and the quantum metric~\cite{Provost1980, Marzari2012, Roy2014} of the flat band. It is a natural question whether a similar effect can lead to conduction in the nonsuperconducting normal state, where the large density of states suggests a metallic state whereas the strongly suppressed velocity points towards an insulating behavior. However, the analysis of this question is usually complicated by the presence of strong correlation effects, making definite answers difficult. 
Previous approaches, therefore, employed phenomenological assumptions on the relaxation rate~\cite{Mitscherling2022, Huhtinen2023} or focused on fine-tuned disordered flatband models~\cite{Bouzerar2022} providing evidence for an important role of the quantum metric,
but it is unclear to what extent these models can capture the effects of a diverging density of states when approaching the flatband limit.

\begin{figure}[b!]
    \centering
    \includegraphics[width=0.237\textwidth]{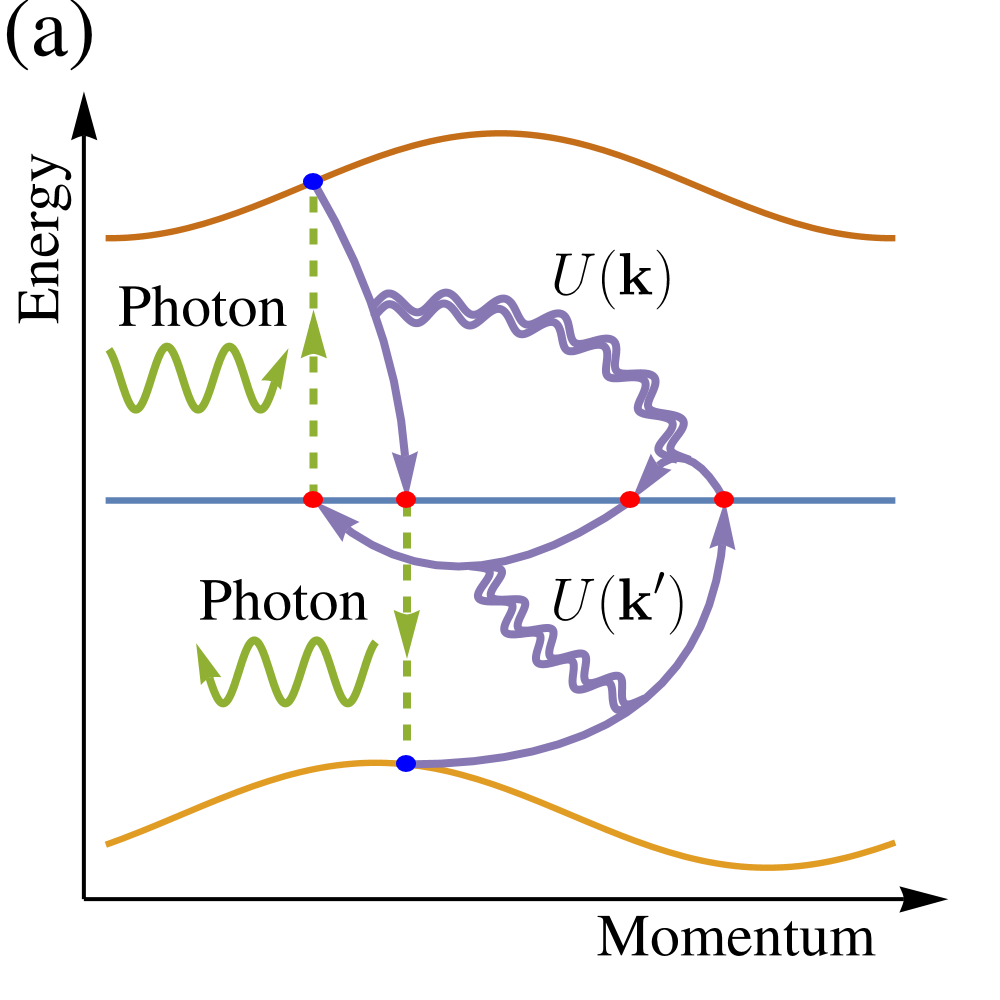}
    \includegraphics[width=0.237\textwidth]{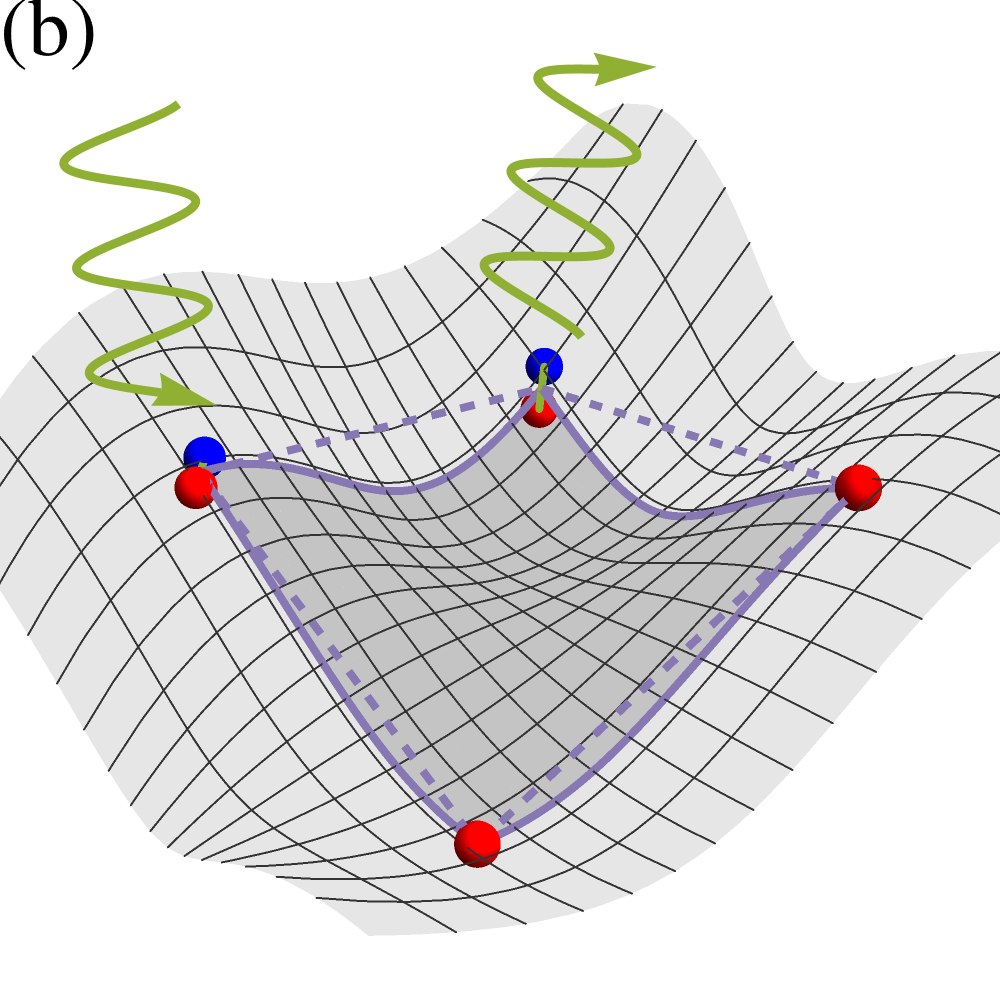}
    \caption{
    (a) 
    A typical scattering process contributing to the Drude weight of an isolated flat band involves two virtual excitations to remote bands mediated by photons and interaction.
    (b) The same process but 
    depicted in the manifold of flat band states, where distances and phases are described by quantum state geometry.
    The quantum metric and Berry curvature emerge only if all momenta in this four-point geometric object are close to each other.
    }
    \label{fig:sketch}
\end{figure}

In this Letter, we take a complementary approach by assuming a clean multiband system without any prior relaxation process and aim to calculate the part of the spectral weight that contributes to transport in the presence of a repulsive density-density interaction. We focus on systems hosting an isolated flat band near the chemical potential. It is commonly believed that the intrinsically strong interaction scale, which naturally exceeds the flatband bandwidth, opposes a perturbative approach. However, we identify a new approach using the temperature as a regularizing energy scale that allows a perturbative diagrammatic treatment in a controlled analytic manner for temperatures above the bandwidth and the interaction strength and below the bandgap to remote bands.
 
In combination with a generalized quantum geometric formalism, we employ this approach to explore the intriguing interplay between the charge response of a partially filled flat band with nontrivial flatband states and interaction in this rather unconventional \emph{semiquantum} transport regime. In particular, we identify the leading-order many-body scattering processes contributing to the low-frequency spectral weight of the electrical conductivity in flat bands (Fig.~\ref{fig:sketch}). Notably, these interband transitions lead to wavefunctions overlaps at distinct momenta corresponding to distances and phases in complex projective space. Drawing inspiration from newly developed geometric concepts to approach the complex quantum state structure of multiorbital systems~\cite{Holder2020, Kaplan2024, Ahn2020, Ahn2022, Mera2022, Chen2022, Avdoshkin2023, Bouhon2023, Jankowski2023, Hirschmann2024, Verma2024, Komissarov2024, Avdoshkin2024}, we express our results via gauge-invariant band projectors, facilitating both the interpretation and enabling a straightforward numerical evaluation. Interestingly, we find that the variance of the Berry curvature across the Brillouin zone is a measure of the spectral weight if the interaction decays rapidly in momentum space.

\emph{Spectral weight and Drude weight.---}
We aim to calculate the low-frequency spectral weight of a clean, interacting system in the normal state. To this end, a direct way would be to invoke the many-body Kubo formula for the current-current response. However, this approach is difficult as one needs to faithfully capture the many-body spectrum and states with their respective Boltzmann weights, and identify all leading order contributions to the low-frequency regime. Even to first-order perturbation theory, such a calculation can be daunting as the flatband states are macroscopically degenerate. Therefore, we employ a different path, which utilizes the fact that all the flatband states have the same energy. In such a case, the flatband's entire spectral weight resides on the dc frequency. Thus, the Drude weight and the spectral weight are equivalent. We stress that we do not argue in favor of an infinite dc conductivity, as the interaction can smear the spectral weight without changing its value.

While the concept of a Drude weight as the charge stiffness is deceptively simple, its derivation is still rather delicate~\cite{Scalapino1993, Resta2002, Resta2018} and requires careful treatment. Within linear response, one calculates the response of the electric current $\braket{\hat{j}_{q}^{a}}$ to an electromagnetic gauge field $A_{q}^{b}$, where $q=(iq_0,\vec{q})$, and $iq_0$ is a bosonic Matsubara frequency. We denote Cartesian coordinates via $a,b,c,d$. The current-current correlation function is 
\begin{equation}\label{eq:corr_func}
    \Pi^{ab}_{q}=-\frac{1}{\beta V}\left.\braket{\hat{j}^{a}_{q}\,\hat{j}^{b}_{-q}}\right|_{A=0} \, ,
\end{equation}
where $\beta=1/T$ is inverse temperature, $V$ is the system volume, and $\langle ...\rangle$ is the expectation value. As a consequence of the electromagnetic gauge invariance, the response for a constant and uniform electromagnetic gauge field ($A_{q=0}$) is zero. We keep this in mind as we perturbatively expand $\Pi^{ab}_{q}-\Pi^{ab}_{0}$ while preserving gauge invariance at every perturbation order.

For the charge response of the normal state, one must always take the limit of $\vec{q}\rightarrow 0$ before taking the frequency limit $\omega\rightarrow 0$. Note that the reverse order of limits yields the superfluid stiffness \cite{Scalapino1993}. We hence focus on $\vec{q}=0$, and denote the response function as $\Pi^{ab}(iq_0)=\Pi^{ab}_{\vec{q}=0,iq_0}$. 
The zero frequency limit of the current-current correlation function will be of key importance. We define the limiting value after analytic continuation by
\begin{equation}\label{eq:drude_weight}
    D^{ab}= \pi\lim_{\omega+i\eta\rightarrow0}\left.\left[\Pi^{ab}(iq_0)-\Pi^{ab}(0)\right]\right|_{iq_0\rightarrow \omega+i\eta},
\end{equation}
with small but positive $\eta$. 
This value is the Drude weight, which manifests in the conductivity tensor as a $\delta$-function frequency response of the real part, and a  $1/\omega$ decay in the imaginary part~\cite{Resta2018}. It arises from terms in the current-current correlation function $\Pi^{ab}(0)$ that are not smoothly connected to $\Pi^{ab}(iq_0)$ via the analytic continuation. 

A straightforward way to observe this noncommutativity of limits is in terms of the Matsubara summations over products of propagators in Eq.~\eqref{eq:corr_func}, which have strictly real energies in a clean system. The summation is done by computing residues at the various poles. For simple poles, this leads to expressions involving the Fermi-Dirac distribution function $n_{F}(\epsilon)=(e^{\beta\epsilon}+1)^{-1}$. For higher-order poles, the residue instead introduces derivatives of this function. The peculiarity of a nonzero Drude weight is a result of two poles that are separated along the imaginary axis by the nonzero external bosonic Matsubara frequency $iq_0\neq 0$, which merge into a high-order pole on setting $iq_0=0$. Since the Fermi-Dirac distribution is periodic in bosonic Matsubara frequencies, i.e., $n_{F}(\epsilon+iq_0)=n_{F}(\epsilon)$, the analytic continuation followed by the limit, $iq_0\rightarrow \omega+i\eta \rightarrow 0$, is not the same as setting $iq_0=0$, and only the latter will produce the derivatives of the Fermi-Dirac distribution. 

\emph{Multiorbital system with density-density interaction.---}
We consider a translation-invariant system on a Bravais lattice with periodic boundary conditions and a total of $N_{c}$ units cells. The Hamiltonian is given by $\hat{H}=\hat{H}_{0}+\hat{H}_{\text{int}}$ with
\begin{align}\label{eq:hamiltonian_kin}
    \hat{H}^{}_{0} &= \sum_{n}\sum_\vec{p}\epsilon^{}_{n,\vec{p}}\,\hat c_{n,\vec{p}}^{\dagger}\hat c^{}_{n,\vec{p}} \, ,
    \\\label{eq:hamiltonian_int}
    \hat{H}^{}_{\text{int}} &=\frac{1}{2N_{c}}\sum_{\vec{k}}U(\vec{k}):\!\hat \rho^{}_{\vec{k}}\hat \rho^{}_{-\vec{k}}\!: \, ,
    \\\label{eq:hamiltonian_density}
    \hat\rho^{}_{\vec{k}}
    &=\sum_{n,m}\sum_{\vec{p}}\hat c_{n,\vec{p}}^{\dagger}\braket{u^{}_{n,\vec{p}}|u^{}_{m,\vec{p}+\vec{k}}}\hat c^{}_{m,\vec{p}+\vec{k}} \, ,
\end{align}
where $\epsilon_{n,\vec{p}}=E_{n,\vec{p}}-\mu$ are the single-particle energies with respect to the chemical potential $\mu$. The colons represent normal ordering, and $U(\vec{k})=U(-\vec{k})$ is a real and symmetric interaction. The annihilation and creation operators $\hat c_{n,\vec{p}}$ and $\hat c^\dagger_{n,\vec{p}}$ are for a Bloch states of band $n$ with quasimomentum $\vec{p}$, whose unit-cell periodic part is given by $\ket{u_{n,\vec{p}}}$. The overlap between these periodic Bloch states at different bands and quasimomenta is at the heart of the quantum state geometry of translation invariant systems. For clarity, we assume that all real-space orbitals are centered at the same position in the unit cell. 

Any Hamiltonian term that depends solely on spatially local density operators is left invariant when coupled to a uniform (but time-dependent) electromagnetic gauge field. Thus, by choosing the density-density interaction in Eq.~\eqref{eq:hamiltonian_int}, it is guaranteed that the current operator of the noninteracting system is preserved. Other forms of the interaction lead to additional contributions to the Drude weight due to interaction-induced hopping. 

\emph{Semiquantum limit.---}
We set the chemical potential $\mu$ to be closest to the band of interest~\footnote{We assume the chemical potential is substantially closer to one band than to all others, making the selection of the active band obvious even after introducing electron-electron interactions.}, denoted by band index $f$. The bandwidth of this band is defined by $\left.W=\max_{\vec{p}}\epsilon_{f,\vec{p}}-\min_{\vec{p}}\epsilon_{f,\vec{p}}\right.$. For a perfectly dispersionless band, we have $W=0$, which is equivalent to having $\epsilon_{f,\vec{p}}=\epsilon_{f}$ independent of quasimomentum $\vec{p}$. In this case, the Fermi-Dirac distribution function in the flat band is also quasimomentum independent and given by the filling factor $\nu$ of the flat band, i.e., $n_{F}(\epsilon_f)=\nu$. Its derivative is $n_{F}'(\epsilon_{f})=-\beta\nu(1-\nu)$. The flatness of the band naturally leads us to consider the regime where $W\ll T$, and in fact, the above discussion of the Fermi-Dirac distribution holds, up to corrections of the order of $W/T$, for narrow bands with weak but nonzero dispersion. By choosing the interaction energy scale to be smaller than temperature, $U\ll T$, we find that $U/T$ gives us a small parameter needed for a controlled expansion, even for $W\ll U$, which is naturally the case for flat and narrow bands. To complete the picture, we define the energy gap $\Delta=\min_{n\neq f,\vec{p}',\vec{p}}|\epsilon_{n,\vec{p}'}-\epsilon_{f,\vec{p}}|$ as the closest energy to the flat/narrow band. We do not require remote bands to be dispersive. The parameter regime of interest is known as the semiquantum limit~\cite{Spivak2006}, which we summarize as
\begin{align}\label{eq:semiquantum_limit}
    W \ll U\ll T \ll \Delta \, .
\end{align}
In this parameter regime, the width of the Fermi-Dirac distribution $T$ is much larger than the renormalized bandwidth $U$. Simultaneously, remote bands below (above) the flat bands are completely occupied (unoccupied), which is markedly distinct from any classical electron gas~\cite{Tai2023}. 

\begin{figure}
    \centering
    \includegraphics[width=0.13\textwidth]{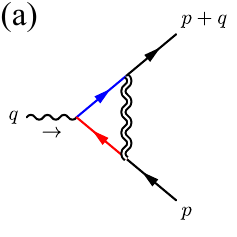}\hspace{1cm}
    \includegraphics[width=0.13\textwidth]{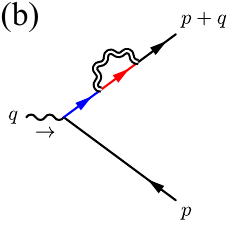}\\[2mm]
    \includegraphics[width=0.11\textwidth]{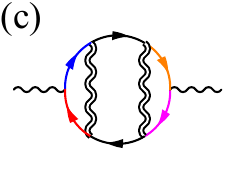}
    \includegraphics[width=0.11\textwidth]{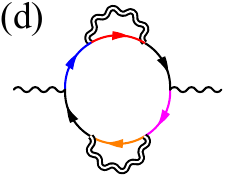}
    \includegraphics[width=0.11\textwidth]{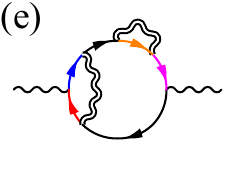}
    \includegraphics[width=0.11\textwidth]{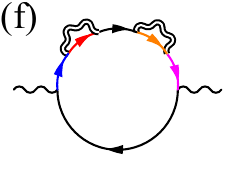}
    \caption{
    The on-shell electron-photon scattering processes [(a) and (b)] lead to the diagrams contributing to the low-frequency spectral weight in narrow bands [(c)-(f)]. 
    Wiggly lines denote the electromagnetic gauge field and double wiggly lines denote the density-density interaction.
    Colored fermion lines denote propagators which are not restricted to the flat band, while black fermion lines are always in the flat band.}
    \label{fig:narrow_band_processes}
\end{figure}

\begin{figure}[b!]
    \centering
    \includegraphics[width=0.13\textwidth]{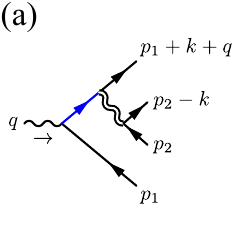}\\
    \includegraphics[width=0.11\textwidth]{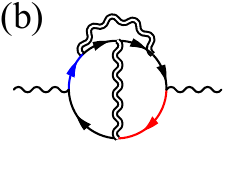}
    \includegraphics[width=0.11\textwidth]{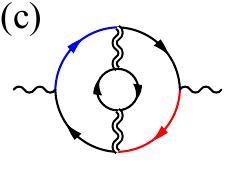}
    \includegraphics[width=0.11\textwidth]{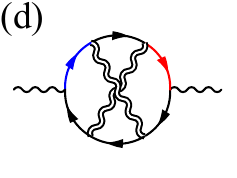}\\
    \includegraphics[width=0.11\textwidth]{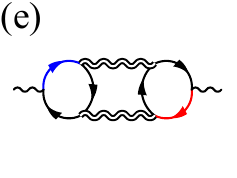}
    \includegraphics[width=0.11\textwidth]{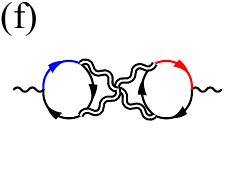}
    \includegraphics[width=0.11\textwidth]{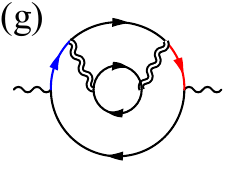}
    \includegraphics[width=0.11\textwidth]{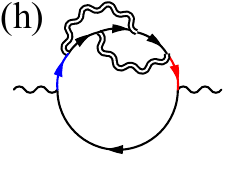}
    \caption{The scattering process (a) and corresponding diagrams [(b)-(h)] that contribute for exactly flat bands in the same notation as in Fig.~\ref{fig:narrow_band_processes}. All other diagrams are subleading.}
    \label{fig:flat_band_processes}
\end{figure}

\emph{Power-counting scheme.---}
We have now set the stage for calculating the low-frequency spectral weight via the Drude weight for a flat/narrow band within the semiquantum regime. To evaluate Eq.~\eqref{eq:drude_weight}, we first calculate the current-current correlation function given in Eq.~\eqref{eq:corr_func}. Although this is usually a nontrivial feat in the presence of an interaction term $U\gg W$, we show via an energy-scale power-counting scheme that the correlation function can here be calculated within the usual diagrammatic expansion in interaction vertices~\cite{SupplMat}.

Consider a diagram with two current vertices and $N_U$ interaction vertices. Each interaction vertex introduces a factor of $UT$ along with two propagators. The two current vertices introduce each a factor of $\Delta$ from the interband current vertex weight. The intraband term is typically negligible for the narrow band as it is $W/\Delta$ times smaller than the interband terms. We include it for completeness and denote by $N_W$ the number of intraband current vertices. The Drude weight is thus given by $\left.D\sim T \Delta^2 (W/\Delta)^{N_W}(UT)^{N_U}\times M\right.$. Here, $M$ accounts for the Matsubara loops, which contain a total of $2N_U+2$ propagators, of which $N_{R}$ are for remote bands and the rest are for flat bands. The parts of the Matsubara sum relevant to the Drude weight are estimated as follows: Each remote band contributes a $1/\Delta$ to the sum, while each flat band contributes a $1/T$, from the derivative of the Fermi-Dirac distribution. Therefore, we expect that $\left.M\propto (1/\Delta)^{N_{R}}(1/T)^{(2N_U+2-N_{R})}\right.$. Overall, this implies the scaling of the Drude weight to be
\begin{align}\label{eq:drude_estimate}
    D\sim (T/\Delta)^{N_{R}-2}\,(W/\Delta)^{N_W}\,(U/T)^{N_U}\,T\,.    
\end{align}
This estimate suggests that if all current vertices are interband ($N_W=0$) and only two propagators are remote ($N_{R}=2$), the leading order terms in the Drude weight remain finite even in the extremal limit of Eq.~\eqref{eq:semiquantum_limit} where $W\rightarrow0$ and $\Delta\rightarrow\infty$.
\\

\emph{Drude weight for flat and narrow bands.---}
For the exactly flat band ($N_W=0$), we find that at least two remote propagators ($N_{R}\ge2$) and at least two vertices ($N_U\ge2$) are required to obtain a nonzero Drude weight. For a narrow band, terms with fewer remote propagators and fewer interactions vertices may also lead to a nonzero Drude weight, e.g., $D\sim W^{2}/T$, but will be subleading. Following our power-counting scheme in Eq.~\eqref{eq:drude_estimate}, $N_{R}=N_U=2$ is expected to be the leading-order contribution to the Drude weight in the semiquantum limit, with $D\sim U^2/T$. Explicitly calculating the contributions of all shown diagrams in Fig.~\ref{fig:narrow_band_processes} and \ref{fig:flat_band_processes}, the Drude weight of isolated flat bands can therefore be stated to leading order as~\cite{SupplMat},
\begin{align}
\label{eq:drude_result}
    &D^{ab}(T)
    =\,
    \frac{\pi e^{2}}{T V_c}\int_{\mathbf{k},\mathbf{k}'}
    \!\!U(\vec{k})\,U(\vec{k}')
    \,\bigg[\,\nu(1-\nu)\,\,\mathfrak{g}_{1}^{ab}(\vec{k},\vec{k}';\nu)
    \nonumber\\%[1mm]
&\,\,\,\,\,+\,\delta^{}_{W,0}\,\big[\nu(1-\nu)\big]^{2}\Big[
    \mathfrak{g}_{2}^{ab}(\vec{k})\,\tilde\delta(\vec{k},\vec{k}')-\mathfrak{g}_{3}^{ab}(\vec{k},\vec{k}')
    \Big]
    \!\bigg] \, ,
\end{align}
where $e$ is the electron charge, $V_{c}=V/N_{c}$ is the unit cell volume, and we define the dimensionless momentum integral $\int_\mathbf{k}\equiv V_c\int_\text{BZ}d^d\mathbf{k}/(2\pi)^d$ over the $d$-dimensional Brillouin zone and $\tilde\delta(\vec{k},\vec{k}')=\delta(\vec{k}-\vec{k}')(2\pi)^d/V_c$. 

The various tensors $\mathfrak{g}_i$ on the right-hand side of Eq.~\eqref{eq:drude_result} are all real and symmetric tensors of quantum state geometric origin, i.e., they encode differential geometric properties of the complex projective manifold formed by the band eigenstates. 
In topological band theory, the quantum geometric tensor, which includes the quantum metric and Berry curvature, is typically sufficient to encode such features.
However, the $\mathfrak{g}_i$ constitute strictly more general objects which go beyond the quantum geometric tensor. Together with Eq.~\eqref{eq:drude_result}, they are the main results of this Letter. They are stated most conveniently using a projector notation.
Denoting $ \bP_{n,\vec{p}}=\ket{u_{n,\vec{p}}}\bra{u_{n,\vec{p}}}$ as the projector onto band $n$, the projector onto the (nondegenerate) flat band can be abbreviated as $ \bP_{\vec{p}}= \bP_{f,\vec{p}}$, while the projection weighted by the occupation function is $ \bP_{\text{occ},\vec{p}}^{[\nu]}=\sum_{n<f} \bP_{n,\mathbf{p}}+\nu  \bP_\mathbf{p}$. Then the tensor $\mathfrak{g}_1$ takes the form
\begin{align}\label{eq:g1_tensor}
    &\mathfrak{g}_{1}^{{a}{b}}(\vec{k},\vec{k}';\nu)
    =\!\frac{1}{2}\text{Re}\bigg[\!\int_\mathbf{p}\!\!    
    \text{tr}\Big[
     \bP^{}_{\vec{p}}\,
    \partial^{}_{{a}}\big( \bP^{}_{\vec{p}}\, \bP^{[\nu]}_{\text{occ},\vec{p}+\vec{k}}\,  \bP^{}_{\vec{p}}\big)\,
     \nonumber \\
    &\hspace{20mm}\times
    \bP^{}_{\vec{p}}\partial^{}_{{b}}\big( \bP^{}_{\vec{p}}\, \bP^{[\nu]}_{\text{occ},\vec{p}+\vec{k}'}\, \bP^{}_{\vec{p}}\big)\!\Big]
    \!\bigg]+({a}\leftrightarrow{b})\,
\end{align}
with momentum derivative $\partial_a\bP^{}_{\vec{p}}\equiv \partial_{p_a}\bP^{}_{\vec{p}}$ and trace tr. This tensor corresponds to the diagrams in Fig.~\ref{fig:narrow_band_processes}, amounting to the dressed current vertex and dressed propagator. Intuitively, these diagrams represent quasimomentum-preserving ($\vec{q}=0$) on-shell electron-photon scattering processes in the narrow band. Consequently, $\mathfrak{g}_{1}$ is the only term that is retained at exactly zero frequency
for nonzero $W$, which is reminiscent of the low-temperature mean-field result in normal metals~\cite{Resta2018}. The other tensors correspond to the diagrams in Fig.~\ref{fig:flat_band_processes}, whose contributions to the Drude weight are unique to the exact flat band ($W=0$). They read
\begin{align}\label{eq:g2_tensor}
    &\mathfrak{g}_{2}^{{a}{b}}(\vec{k})
    = \text{Re}\bigg[\!\int_{\mathbf{p}}\!\!
    \text{tr}\Big[ \bP^{}_{\vec{p}}\partial^{}_{{a}}\big( \bP^{}_{\vec{p}} \bP^{}_{\vec{p}+\vec{k}}\big) \bP^{}_{\vec{p}+\vec{k}}\Big]\, 
    \nonumber \\ & \hspace{30mm}\times
    \!\!\!\int_{\mathbf{p}'}\!\!\!
    \text{tr}\Big[ \bP^{}_{\vec{p}'}\,\partial^{}_{{b}}\big( \bP^{}_{\vec{p}'}\, \bP^{}_{\vec{p}'+\vec{k}}\big)\, \bP^{}_{\vec{p}'+\vec{k}}\Big]\!
    \nonumber \\ & \hspace{0mm}+\!\!\!\int_{\mathbf{p}}\!\!
    \text{tr}\Big[ \bP^{}_{\vec{p}}\partial_{{a}}\big( \bP^{}_{\vec{p}}\bP^{}_{\vec{p}+\vec{k}}\big)\bP^{}_{\vec{p}+\vec{k}}\partial^{}_{{b}}\big(\bP^{}_{\vec{p}+\vec{k}}\bP^{}_{\vec{p}}\big)\!\Big]\!\int_{\mathbf{p}'}\!\!\!\!
    \text{tr}\Big[\bP^{}_{\vec{p}'+\vec{k}}\bP^{}_{\vec{p}'}\Big]\!\bigg]
\end{align}
and
\begin{align}\label{eq:g3_tensor}
    &\mathfrak{g}_{3}^{{a}{b}}(\vec{k},\vec{k}')=
    \text{Re}\bigg[\!\int_\mathbf{p}\!\!
    \text{tr}\Big[ \bP^{}_{\vec{p}}\,\partial^{}_{{a}}\big( \bP^{}_{\vec{p}}\, \bP^{}_{\vec{p}+\vec{k}}\big)\, \bP^{}_{\vec{p}+\vec{k}}\nonumber\\&\hspace{8mm}\times  \bP^{}_{\vec{p}+\vec{k}+\vec{k}'}\, \bP^{}_{\vec{p}+\vec{k}'}\,\partial^{}_{{b}}\big( \bP^{}_{\vec{p}+\vec{k}'}\, \bP^{}_{\vec{p}}\big)\!\Big]\!\bigg]+({a}\leftrightarrow{b}) \,.
\end{align}
They are enabled by a simple yet profound property of the flat band that intraband transitions between different quasimomenta have zero energy, i.e., $\forall\vec{p},\vec{p}':\,\epsilon_{f,\vec{p}}=\epsilon_{f,\vec{p}'}$. This feature allows for new on-shell electron-photon scattering processes that include a quasimomentum shift. 
Equations~\eqref{eq:g2_tensor} and \eqref{eq:g3_tensor} represent the spectral weight that is normally spread at least across the bandwidth $W$ but collapses to exactly zero frequency for a flat band~\cite{Mitscherling2018, Mitscherling2020, Ahn2021}.

\emph{Berry curvature variance.---} %Quantum state geometry beyond quantum metric
According to Eq.~\eqref{eq:drude_result}, the Drude weight of a flat band involves a complicated collection of Bloch band projectors onto multiple momenta, very different from the quantum metric $g_{\vec{p}}^{ab}$ and Berry curvature $\Omega_{\vec{p}}^{ab}$, which are much simpler objects involving exactly one momentum, defined via the trace expression $\text{tr}\big[ \bP_{\vec{p}}(\partial_{a} \bP_{\vec{p}})(\partial_{b} \bP_{\vec{p}})\big]=g_{\vec{p}}^{ab}-i\,\Omega_{\vec{p}}^{ab}/2$. However, one can expect that such momentum-local geometric objects will reemerge within a momentum expansion of the $\mathfrak{g}_i^{ab}$ \cite{Avdoshkin2023}. This expansion is justified for an interaction $U(\vec{k})$ which is peaked close to zero momentum transfer. For a nondegenerate flatband, we expand in small $\vec{k},\vec{k}'$, and find at leading order 
\footnote{
Using the identities $ \bP_{n,\vec{p}}(\partial_{a} \bP_{n,\vec{p}}) \bP_{n,\vec{p}}=0$ and $(1- \bP_{n,\vec{p}})(\partial_{a} \bP_{n,\vec{p}})(1- \bP_{n,\vec{p}})=0$, one can show that all the tensors in Eqs.~\eqref{eq:drude_result}
vanish when at least one momentum argument is zero, i.e., $\vec{k}=0$ or $\vec{k}'=0$. Since the interaction is even in momentum, the lowest order that survives is second-order in momentum $\vec{k}$. The only contributing term is $\mathfrak{g}_{2}^{{a}{b}}(\vec{k})$.}
\begin{align}\label{eq:D_simplified}
    D^{ab}(T) &\approx 
    \,\delta_{W,0}\,[\nu(1-\nu)]^{2}\frac{\pi e^{2}}{TV_{c}} \!\sum_{c,d}\!
    \bigg(\!
    \int_{\vec{k}}
    [U(\vec{k})]^{2}\,k_{c}\,k_{d}\!
    \bigg)
    \nonumber\\
    &\hspace{7mm}\times\!
    \!
    \int_{\vec{p}}\!\bigg[\!
    \Big(\Omega^{ac}_{\vec{p}}-\int_{\vec{p}'}\!\!\Omega^{ac}_{\vec{p}'}\Big)\!\!\,
    \Big(\Omega^{bd}_{\vec{p}}-\int_{\vec{p}'}\!\!\Omega^{bd}_{\vec{p}'}\Big)\!
    \bigg] \, .
\end{align}
Even though the Drude weight is substantially simplified, it contains geometric properties, i.e. the variance of the Berry curvature, that differ from previous estimates for both the Drude weight and superfluid weights and cannot be captured using a low-frequency projection of the interaction to the flat band. In particular, note that for the broadly used spatially local Hubbard interaction such a momentum expansion is never justified and the complete expression in Eq.~\eqref{eq:drude_result} has to be employed.  

\emph{Low-frequency spectral weight of flat bands.---}
Equipped with Eq.~\eqref{eq:drude_result}, we are now in the position to re-express the flatband Drude weight in the form of the many-body Kubo formula. To this end, we rewrite Eq.~\eqref{eq:drude_result} as a sum over scattering processes
\begin{align}\label{eq:drude_result_manybody}
    &D^{ab}(T)= \frac{\pi}{TV_c}\bigg[\nu(1-\nu)\!\!\int_{\vec{p}}\!\!\text{Re}\Big[\mathcal{M}_{\gamma_{a}\rightarrow\overline{\vec{p}}\vec{p}}\,\overline{\mathcal{M}_{\gamma_{b}\rightarrow\overline{\vec{p}}\vec{p})}}
    \,\Big]
    \nonumber\\
    &+\delta_{W,0}[\nu(1-\nu)]^{2}\!\!\int_{\{\vec{p}_i\}}\!\!\!\!\!\!\!
    \text{Re}\Big[\mathcal{M}_{\gamma_{a}\rightarrow\overline{\vec{p}_1}\overline{\vec{p}_2}\vec{p}_3\vec{p}_4}\,\overline{\mathcal{M}_{\gamma_{b}\rightarrow\overline{\vec{p}_1}\overline{\vec{p}_2}\vec{p}_3\vec{p}_4}}
    \Big]\!\bigg],
\end{align}
where $\mathcal{M}_{\gamma_{a}\rightarrow\psi}$ represents a scattering amplitude for a photon of polarity $a$ to an electron-hole state $\psi$, as shown in Figs.~\ref{fig:narrow_band_processes} and \ref{fig:flat_band_processes}. We note that in the form above, the positive semidefiniteness of the Drude weight is evident. The scattering amplitudes are given by
\begin{align}
    \label{eq:scattering1}
    &\mathcal{M}_{\gamma_{a}\rightarrow\overline{\vec{p}}\vec{p}}
    =e \!\int_{\vec{k}} U(\vec{k})\bra{u_{\vec{p}}}\partial^{}_{{a}}( \bP^{}_{\vec{p}} \bP^{[\nu]}_{\text{occ},\vec{p}+\vec{k}} \bP^{}_{\vec{p}})\ket{u_{\vec{p}}} ,\\
% \end{align}
%and
% \begin{align}
    \label{eq:scattering2}
    &\mathcal{M}_{\gamma_{a}\rightarrow\overline{\vec{p}_1}\overline{\vec{p}_2}\vec{p}_3\vec{p}_4}
    \nonumber\\
    &\hspace{4mm}=\frac{1}{2}e 
    \Big[\!
    \big[\,
    U(\vec{p}_2-\vec{p}_3)\bra{u_{\vec{p}_1}}\partial_{a}( \bP_{\vec{p}_1} \bP_{\vec{p}_4})\ket{u_{\vec{p}_4}} \braket{u_{\vec{p}_2}|u_{\vec{p}_3}}
    \nonumber\\
    &\hspace{12mm}-(\vec{p}_1\leftrightarrow\vec{p}_2)\big]
    -(\vec{p}_3\leftrightarrow\vec{p}_4)\Big]
\tilde{\delta}_{\vec{p}_1+\vec{p}_2,\vec{p}_3+\vec{p}_4}\,,
\end{align}
where $\ket{u_{\vec{p}}}$ is shorthand notation for the flat band noninteracting Bloch eigenstate $\ket{u_{f,\vec{p}}}$. We compare Eq.~\eqref{eq:drude_result_manybody} with the real part of the general many-body Kubo formula for the longitudinal conductivity, which reads
\begin{align}
    \mathrm{Re}\big[\sigma^{aa}(\omega)\big]=\pi\frac{1-e^{-\beta\omega}}{\omega}\!\!\sum_{\Psi,\Psi'}\!\frac{e^{-\beta E_\Psi}}{Z}|j^{a}_{\Psi\Psi'}|^{2}\delta(\omega-E_{\Psi'\Psi})\, .
    \label{eq:manybody}
\end{align}
Here, $\Psi$ and $\Psi'$ denote many-body eigenstates, $E_{\Psi\Psi'}\equiv E_{\Psi}-E_{\Psi'}$ is the difference between the many-body energies, $Z$ is the partition function, and $j^{a}_{\Psi\Psi'} = \braket{\Psi|\hat{j}^{a}|\Psi'}$ is the matrix element of the many-body current operator. For $|\omega|\ll T$, we note that the current matrix element connecting two flatband many-body states is given, to leading order in interaction, by the scattering amplitudes in Eqs.~\eqref{eq:scattering1} and \eqref{eq:scattering2}. The occupancy number $\nu$ of any flatband electron determines the fraction of many-body states permitting such scattering processes. Thus, without evaluating the many-body energies, we conclude that the low-frequency ac-conductivity spectral weight is given by the Drude weight in Eq.~\eqref{eq:drude_result} evaluated with $\delta_{W,0}=1$ via the sum rule $D^{aa}=\int_{-\Lambda}^{\Lambda}\mathrm{d}\omega\,\mathrm{Re}[\sigma^{aa}(\omega)]$ for an adequately chosen energy window $U\ll\Lambda\ll T$. A nonzero bandwidth $W>0$ or a band broadened by interaction would modify the many-body energies $E_{\Psi}$, but would not affect our conclusion, as long as the integration energy window exceeds this bandwidth $W\ll U\ll \Lambda$. Any corrections to the low-frequency spectral weight due to intraband scattering processes are subleading of order $W/U$.

\emph{Conclusions and outlook.---}
We identified all contributions of leading order $U^2/T$ to the low-energy spectral weight of an isolated interacting flatband within the semiquantum temperature regime $W\ll U\ll T\ll \Delta$, including vertex corrections not captured by the ladder approximation. Within this new distinct transport regime for flatband systems, both two- and four-particle scattering processes are enabled, which we characterize by four-point geometric contributions to the spectral weight.
These contributions exemplify the physical relevance of novel independent geometric contributions that have recently been revealed in quantum state geometric characterizations of the Bloch state manifold beyond the standard quantum geometric tensor \cite{Avdoshkin2023, Avdoshkin2024, Avdoshkin2024a}.
A reduction to single-momentum geometric objects is possible for long-range interaction, where we identify the variance of the Berry curvature as the key geometric quantity, see Eq.~\eqref{eq:D_simplified}.

These structures are visible neither in previous results on the longitudinal conductivity for noninteracting systems with finite quasiparticle lifetimes~\cite{Mitscherling2022, Kruchkov2022, Huhtinen2023} nor in mean-field estimates of the superfluid weight \cite{Peotta2015, Torma2022}. 
Indeed, it has been noticed before that the conjectured dependence on the quantum metric can vary based on the procedure by which the groundstate projection is performed~\cite{Hofmann2022, Thumin2024}. Our analysis shows the importance of vertex corrections for response functions of flatband systems, which challenge the application of commonly used infinite re-summations of diagrams potentially.

The presented formalism not only sheds light on the underlying analytical structure of the quantum states in interacting multiorbital lattice systems but is also amenable to direct numerical evaluation due to its transparent gauge-invariant projector formulation.
Our results show that the projection procedure to the flat band is more delicate than expected. Namely, a projection of the interaction itself is insufficient for response functions, which contain two or more interband processes. 
This finding exposes the unusually strong quantum fluctuations in flatbands system, which remain relevant even if their energy exceeds the temperature, thereby adding further evidence that the conventional wisdom of scale separation is not straightforward in gapless quantum many-body systems~\cite{Mandal2015, Ye2022, Li2024}.

\begin{acknowledgments}
We thank Alexander Avdoshkin, 
Erez Berg,
Dan S. Borgnia, 
Johannes S. Hofmann,
Iliya Esin,
Walter Metzner, 
Joel E. Moore,
Raquel Queiroz,
Gil Refael,
J\"org Schmalian, 
Ady Stern,
Hengxin Tan, and
Evyatar Tulipman for helpful discussions and their valuable feedback.
O.A.\ acknowledges support by the European Research Council (ERC) under grant HQMAT (Grant No. 817799), by the ERC under the European Union’s Horizon 2020 research and innovation programme (Grant No. LEGOTOP 788715), the DFG (CRC/Transregio 183, EI 519/71), and by the ISF Quantum Science and Technology (2074/19). 
J.M.\ acknowledges support by the German National Academy of Sciences Leopoldina through Grant No. LPDS 2022-06.
T.H.\ acknowledges financial support by the 
European Research Council (ERC) under grant QuantumCUSP
(Grant No. 101077020). 
\end{acknowledgments}

% \bibliography{references}
\bibliography{main_arxiv}

%------------------------------------------------------------------------------

\clearpage

\appendix

\begin{widetext}

\section*{Supplemental Material for \\[2mm] ``Drude weight of an interacting flat-band metal''}

In the Supplemental Material (SM), we provide comprehensive details on the derivation of the Drude weight of a flatband metal in the semiquantum limit. We establish (i) the basic definitions for the current-current response function and (ii) the multiorbital system with density-density interaction. We continue by (iii) deriving the necessary higher-order Matsubara frequency summations and conclude by (iv) performing the diagrammatic expansion of the current-current response function for narrow and flat-band systems to second order in interaction and leading order in the semiquantum limit.

\section{The current-current response function}

We define the electrical current $J_{q}^{a}[\vec{A}]$ in spatial direction $a$ as a function of $q=(iq_0,\vec{q})$ with bosonic Matsubara frequency $iq_0$ and lattice momentum $\vec{q}$ and the electromagnetic gauge potential $\vec{A}$ via the functional derivative of the grand canonical potential
\begin{align}
    J_{q}^{a}[\vec{A}]&\equiv-\frac{1}{V}\frac{\delta \Omega(\vec{A})}{\delta A_{-q}^{a}} \, .
\end{align}
The grand canonical potential $\Omega(\vec{A})$ is defined via the partition function
\begin{align}
    Z[\vec{A}] &= \int\!\!D[\bar{\psi}\psi]\,e^{-S[\bar{\psi},\psi,\vec{A}]}
    \equiv e^{-\beta\Omega(\vec{A})} \, ,
\end{align}
using the Feynman path integral formalism over Grassmann fields $\bar \psi$ and $\psi$ with euclidean action $S$. The inverse temperature is denoted by $\beta\equiv 1/T$. We are interested in expanding $J_{q}^{a}[\vec{A}]$ to linear order in $\vec{A}$. We see that  
\begin{align}
    \frac{\delta J^{a}_q[\vec{A}]}{\delta A_{q}^{b}}\Bigg|_{\vec{A}=0}
    &=
    \frac{1}{\beta V}\frac{\delta^{2} \ln Z[\vec{A}]}{\delta A_{-q}^{a}\delta A_{q}^{b}}\Bigg|_{\vec{A}=0}
    =\frac{1}{\beta V} 
    \Bigl \langle
    \frac{\delta S}{\delta A^{a}_{-q}}\frac{\delta S}{\delta A^{b}_{q}}
    -\frac{\delta^{2} S}{\delta A^{b}_{q}\delta A^{a}_{-q}}
    \Bigr \rangle
    \Bigg|_{\vec{A}=0}
    =  \frac{1}{\beta V} 
    \Bigl \langle
    \frac{\delta S}{\delta A^{a}_{-q}}\frac{\delta S}{\delta A^{b}_{q}}
    \Bigr \rangle
    \Bigg|_{\vec{A}=0}
    -
    \big(q=0\big)\, ,
\end{align}
under variation with respect to the vector potential with the system volume $V$. We identify the subtraction of the $q=0$ contribution of the first term. The current-current correlation function is defined as
\begin{align}
    \Pi^{ab}_{q} \equiv -\frac{1}{\beta V} 
    \Bigl \langle
    \frac{\delta S}{\delta A^{a}_{-q}}\frac{\delta S}{\delta A^{b}_{q}}
    \Bigr \rangle
    \Bigg|_{\vec{A}=0} \, ,
\end{align}
which is given in Eq.~\eqref{eq:corr_func} in the main text by identifying the current operator $\hat{j}^{a}_{q}\equiv\delta S/\delta A^{a}_{-q}$. The expression for the expansion of the electrical current thus reads 
\begin{align}
    J^{a}_q[\vec{A}] = J^{a}_q[0]-\sum_b(\Pi^{ab}_{q}-\Pi^{ab}_{0})\,A_{q}^{b}+\mathcal{O}(A^{2}) \, .
\end{align}
The equation above reveals the connection between the current-current response function and the Drude weight, as given by Eq.~\eqref{eq:drude_weight} in the main text.

\section{Multiorbital system with density-density interaction}

The model is described by the action $S$ at $\vec{A}=0$ and its functional derivative - the current operator $\hat j^a_q$. The electromagnetic gauge field couples only to the kinetic part of the action due to the considered density-density interaction, see main text. Thus, we can write 
\begin{align}
    S[\bar{\psi},\psi,\vec{A}] &= S_{0}[\bar{\psi},\psi,\vec{A}]+S_{\text{int}}[\bar{\psi},\psi] \, .
\end{align}
Considering the non-interacting Hamiltonian $\hat  \bH_0$ in Eq.~\eqref{eq:hamiltonian_kin} in the main text, the corresponding non-interacting part $S_0$ of the action is given by
\begin{align}
    S_{0}[\bar{\psi},\psi,\vec{A}=0]= -\sum_{p}\sum_{m}\bar{\psi}_{m,p}\,G^{-1}_{m,p}\,\psi_{m,p}
    &= \sum_{p}\sum_{m}\bar{\psi}_{m,p}(-ip_0+\epsilon_{m,\vec{p}})\psi_{m,p} \, ,
\end{align}
with sum over fermionic Matsubara frequencies and lattice momentum $p=(ip_0,\vec{p})$ as well as band indices $m$. We include the chemical potential $\mu$ into the band dispersions $\epsilon_{m,\vec{p}}=E_{m,\vec{p}}-\mu$. The (band) Green's functions read 
\begin{equation}
    G_{m,p} = -\braket{\psi_{p,m}\bar{\psi}_{p,m}}=\frac{1}{ip_0-\epsilon_{m,\vec{p}}} \, .
\end{equation}
The Green's function is diagonal in band and momentum/frequency index. Considering the density-density interaction $\hat  \bH_\text{int}$ defined in Eq.~\eqref{eq:hamiltonian_int} in the main text, the interacting part of the action is given by
\begin{align}
    S_{\text{int}}[\bar{\psi},\psi]&=\frac{1}{2N_c\beta}\sum_{p_{i},m_{i}}[U_{p_1,p_2,p_3,p_4}]_{m_1,m_2}^{m_3,m_4}\,\bar{\psi}_{p_1,m_1}\bar{\psi}_{p_2,m_2}\psi_{p_3,m_3}\psi_{p_4,m_4}\,
    \delta_{p_1+p_2,p_3+p_4}\, ,
\end{align}
with the number of unit cells $N_c$ and interaction vertex 
\begin{align}
    \label{eq:interaction_vertex}
    [U_{p_1,p_2,p_3,p_4}]_{m_1,m_2}^{m_3,m_4}&=
    U(\vec{p}_3-\vec{p}_2)\braket{u_{m_2,\vec{p}_2}|u_{m_3,\vec{p}_3}}\braket{u_{m_1,\vec{p}_1}|u_{m_4,\vec{p}_4}} \, .
\end{align}
The Kronecker delta assures frequency and momentum conservation. The interaction is frequency-independent. We assume it to be real and symmetric in momentum $U(\vec{k})=U(-\vec{k})=[U(\vec{k})]^{*}$. The interaction vertex includes the overlap between Bloch states $|u_{m_i,\vec{p}_i}\rangle$ of different bands and momenta, which arise from the change from orbital basis to band basis in the particle density given in Eq.~\eqref{eq:hamiltonian_density} in the main text. The current operator at $\vec{q}=0$ is given by
\begin{align}
    \label{eq:current_vertex}
    \hat{j}_{q}^{a}\left.[\bar{\psi},\psi,\vec{A}=0]\right|_{\vec{q}=0} &= \frac{\delta S}{\delta A_{-q}^{a}}\Bigg|_{\vec{A}=0,\vec{q}=0}
    =e\sum_{ip_0,\,\vec{p}}\sum_{m,m'}
    \bra{u_{m,\vec{p}}}(\partial_{a} \bH_{\vec{p}})
    \ket{u_{m',\vec{p}}}\,
    \bar{\psi}_{m,\vec{p},ip_0}
    \psi_{m',\vec{p},ip_0+iq_0}^{\phantom{\dagger}}
\end{align}
with $q=(iq_0,\vec{q})$ and electric charge $e$. The current operator involves the momentum derivative $\partial_a\equiv \partial_{p_a}$ of the Bloch Hamiltonian $ \bH_\vec{p}$ in orbital basis, which in general involves both the momentum derivative of the band dispersion as diagonal components as well as interband transitions related to the interband Berry connection.

\section{Higher-order Matsubara frequency summations} 
\label{sec:matsubara_summations}

In the following, we are deriving the relevant Matsubara summations required to evaluate the second-order diagrammatic expansion of the current-current correlation function. We keep the notation very general and specify it when needed. 

\subsection{Summations involving a single Matsubara frequency}

Consider the Matsubara summation over a single fermionic Matsubara frequency $ip_0$ with $N$ poles
\begin{align}
    \label{eq:definition_SFN}
    S_{F}^{(N)}(\xi_1,\xi_2,\dots,\xi_N) \equiv 
    \frac{1}{\beta}\sum_{ip_0}
    \prod_{j=1}^{N}
    \frac{1-\delta_{ip_0,\xi_j}}{ip_0-\xi_j} \, ,
\end{align}
and, similarly, the Matsubara summation over a single bosonic Matsubara frequency $ik_0$
\begin{align}
    \label{eq:bosonic_matsubara}
    S_{B}^{(N)}(\xi_1,\xi_2,\dots,\xi_N) \equiv 
    \frac{1}{\beta}\sum_{ik_0}
    \prod_{j=1}^{N}
    \frac{1-\delta_{ik_0,\xi_j}}{ik_0-\xi_j} \, ,
\end{align}
where $\xi_{j}\in\mathbb{C}$ are complex numbers excluding the respective fermionic or bosonic Matsubara frequencies as a potential pole. To keep track of these and other exclusions in the following, we explicitly denote this exclusion via $1-\delta_{a,b}$ using the Kronecker delta function. For a consistent definition, we understand this identity such that the corresponding term is set to zero regardless of the value of the part with which it is multiplied. Defining the complex function
\begin{align}
    g(z)=\prod_{j=1}^{N}
    \frac{1}{z-\xi_j}\,,
\end{align}
the Matsubara sum is evaluated by contour deformation using a properly chosen weighting function $h(z)$. These contour deformations produce a sum over residues at the poles of $g(z)$
\begin{align}
    S_{F/B}^{N}(\xi_1,\xi_2,\dots,\xi_N) = -\frac{1}{\beta}\sum_{z_0\in\text{poles of $g(z)$}}\underset{z=z_0}{\text{Res}}\lbrace g(z)h(z)\rbrace \, .
\end{align}
Typically, we will choose $h(z)=-\beta \,n_{F}(z)$ for fermionic Matsubara frequency summations and $h(z)=\beta \,n_{B}(z)$  for bosonic ones, where $n_{F}(z)=(e^{\beta z}+1)^{-1}$ is the Fermi-Dirac distribution function, and $n_{B}(z)=(e^{\beta z}-1)^{-1}$ is the Bose-Einstein distribution function. However, if the Matsubara sum does not converge on the left-half plane $(\text{Re}\,z<0)$, as is typical for summations over a single Green's function, then one uses $h(z)=-\beta (n_{F}(z)-1)$ for fermionic Matsubara frequency summations and $h(z)=\beta (n_{B}(z)-1)$ for bosonic ones. Performing the summation for strictly distinct poles without convergence issues ($N\ge 2$), we obtain
\begin{align}
    \forall i\neq j: |\xi_i-\xi_j| > 0 \implies 
    \begin{cases}
        S_{F}^{(N)}(\xi_1,\xi_2,\dots,\xi_N) 
        = \sum_{j=1}^{N}
        \Big(n_{F}(\xi_j)\prod_{l=1,l\neq j}^{N}\frac{1}{\xi_j-\xi_l}\Big)
        \, ,
        \\[2mm]
        S_{B}^{(N)}(\xi_1,\xi_2,\dots,\xi_N) 
        = -\sum_{j=1}^{N}
        \Big(n_{B}(\xi_j)\prod_{l=1,l\neq j}^{N}\frac{1}{\xi_j-\xi_l}\Big)
        \, .
    \end{cases}
    \label{eq:GeneralMatsubaraSum}
\end{align}
The assumption of simple and distinct poles implies well-defined denominators in the expressions above. 
For coinciding (high-order) poles, the same expressions generally hold if we interpret them as a limit, i.e. $S_{F/B}^{(N)}(\xi_1,\xi_1,\xi_3,\dots,\xi_N)=\lim_{\xi_2\rightarrow \xi_1}S_{F/B}^{(N)}(\xi_1,\xi_2,\xi_3,\dots,\xi_N)$ for coinciding $\xi_1=\xi_2$.

\subsection{Summation identities}\label{sec:sum_identities}

The following identities will be useful to reduce the number of poles within typical Matsubara summations conveniently. By definition, the Matsubara summation is totally symmetric under permutations of the poles, that is,
\begin{align}
    S_{F/B}^{(N)}(\xi_1,\xi_2,\dots,\xi_{N})=S_{F/B}^{(N)}(\xi_{\sigma(1)},\xi_{\sigma(2)},\dots,\xi_{\sigma(N)})\,,
\end{align}
for any permutation $\sigma$. Shifting all poles by a bosonic frequency $ik_0$ does not alter a Matsubara summation, that is,
\begin{align}
    S_{F/B}^{(N)}(\xi_1,\xi_2,\dots,\xi_{N})
    &=
    S_{F/B}^{(N)}(\xi_1+ik_0,\xi_2+ik_0,\dots,\xi_{N}+ik_0) \, ,
\end{align}
which follows from the fact that different fermionic and bosonic Matsubara frequencies are related by a bosonic Matsubara frequency. By definition, performing a sign flip on each pole produces a sign for an odd number of poles, that is,
\begin{align}
    S_{F/B}^{(N)}(\xi_1,\xi_2,\dots,\xi_{N})
    =
    (-1)^{N}S_{F/B}^{(N)}(-\xi_1,-\xi_2,\dots,-\xi_{N})\, .
\end{align}
It is possible to split and merge Matsubara summations via an additional bosonic Matsubara summation, that is,
\begin{align}
    &\frac{1}{\beta}\sum_{ik_0}S_{F/B}^{(N)}(\xi_1,\xi_2,\dots,\xi_{M},\xi_{M+1}-ik_0,\xi_{M+2}-ik_0,\dots,\xi_{N}-ik_0) \nonumber\\
    &= S_{F/B}^{(M)}(\xi_1,\xi_2,\dots,\xi_{M})\,
    S_{F/B}^{(N-M)}(\xi_{M+1},\xi_{M+2},\dots,\xi_{N})\, ,
    \label{eq:matsubara_splitting}
\end{align}
which follows by relabeling the additional bosonic Matsubara summation to a separate second fermionic Matsubara summation. This identity allows us to reduce higher-order Matsubara summations involving a large number of poles into products of Matsubara summations with a small number of poles only. Thus, only a minimal number of Matsubara summations have to be explicitly performed. 

\subsection{Analytic continuation and dc limit}
\label{sec:limits}

There are a few recurring forms of Matsubara summations that we will encounter in the evaluation of the current-current correlation function, which we will derive in the following. Let $\epsilon_{j}\in\mathbb{R}$ be strictly real for all $j$, corresponding to the band dispersions in the upcoming application. Let $iq_0$ be a bosonic Matsubara frequency. Using Eq.~\eqref{eq:GeneralMatsubaraSum} for two poles $N=2$, we obtain
\begin{alignat}{2}
    \label{eq:S2Result}
    S_{F}^{(2)}(\epsilon_{1}-iq_0,\epsilon_{2}) &= 
    \begin{cases}
        \frac{n_{F}(\epsilon_{1})-n_{F}(\epsilon_{2})}{-iq_0+\epsilon_{1}-\epsilon_{2}} &\text{for} \hspace{0.5cm}\epsilon_{1}\neq\epsilon_{2}\,,
        \\[2mm]
        n_{F}'(\epsilon_{1})\,\delta_{iq_0,0} &\text{for} \hspace{0.5cm}\epsilon_{1}=\epsilon_{2}\,.
    \end{cases}
\end{alignat}
The case $\epsilon_1=\epsilon_2$ is derived via the limit $\epsilon_2\rightarrow \epsilon_1$. For this, we see that the numerator vanishes whereas the denominator remains finite for $iq_0\neq 0$. In the case of $iq_0=0$, we explicitly recover the definition of the derivative. Both cases are compactly summarized by using a Kronecker delta $\delta_{iq_0,0}$. Analytically continuing the bosonic frequency $iq_0\rightarrow \omega+i\eta$ and taking the dc limit, we find
\begin{align}\label{eq:s2f_limit}
    \lim_{\omega+i\eta\,\rightarrow\,0}\,S_{F}^{(2)}(\epsilon_{1}-iq_0,\epsilon_{2})\Big|_{iq_0\rightarrow\omega+i\eta} &=
    S_{F}^{(2)}(\epsilon_{1},\epsilon_{2})
    -n_{F}'(\epsilon_{1})\,\delta_{\epsilon_{1},\epsilon_{2}}\,.
\end{align}
The equation above is the basis for the standard Drude weight for a single dispersive band, where the quasiparticle velocities are weighted by the Fermi function derivative. It arises from the non-analyticity of $S^{(2)}_F$ when its arguments coincide. This yields the mathematical reason for a finite Drude weight as described around Eq.~\eqref{eq:drude_weight} in the main text, which we will harness throughout the derivation. For $N=3$ poles we have
\begin{align}
    S_{F}^{(3)}(\epsilon_{1}-iq_0,\epsilon_{2},\epsilon_{3}) &= 
    \begin{cases}
    \frac{n_{F}(\epsilon_{1})}{(-iq_0+\epsilon_{1}-\epsilon_{2})(-iq_0+\epsilon_{1}-\epsilon_{3})}
    +\frac{n_{F}(\epsilon_{2})}{-(-iq_0+\epsilon_{1}-\epsilon_{2})(\epsilon_{2}-\epsilon_{3})}
    +\frac{n_{F}(\epsilon_{3})}{(-iq_0+\epsilon_{1}-\epsilon_{3})(\epsilon_{2}-\epsilon_{3})} &\text{for} \hspace{0.3cm} \epsilon_{1}\neq\epsilon_{2}\neq\epsilon_{3}\neq\epsilon_{1}\,,
    \\[2mm]
    \frac{n_{F}(\epsilon_{1})-n_{F}(\epsilon_{2})}{(-iq_0+\epsilon_{1}-\epsilon_{2})^{2}}
    -\frac{n_{F}'(\epsilon_{2})}{(-iq_0+\epsilon_{1}-\epsilon_{2})} &\text{for} \hspace{0.3cm} \epsilon_{1}\neq\epsilon_{2}=\epsilon_{3}\neq \epsilon_1\,,
    \\[2mm]
    \frac{n_{F}(\epsilon_{3})-n_{F}(\epsilon_{1})}{(-iq_0+\epsilon_{1}-\epsilon_{3})(\epsilon_{1}-\epsilon_{3})}
    +\frac{n_{F}'(\epsilon_{1})}{\epsilon_{1}-\epsilon_{3}}\,\delta_{iq_0,0} &\text{for} \hspace{0.3cm} \epsilon_{1}=\epsilon_{2}\neq\epsilon_{3}\neq \epsilon_1\,,
    \\[2mm]
    \frac{n_{F}(\epsilon_{2})-n_{F}(\epsilon_{1})}{(-iq_0+\epsilon_{1}-\epsilon_{2})(\epsilon_{1}-\epsilon_{2})}
    +\frac{n_{F}'(\epsilon_{1})}{\epsilon_{1}-\epsilon_{2}}\,\delta_{iq_0,0}
    &
    \text{for} \hspace{0.3cm} \epsilon_{1}\neq\epsilon_{2}\neq\epsilon_{3}= \epsilon_1\,,
    \\[2mm]
    \frac{n_{F}'(\epsilon_{1})}{iq_0}(1-\delta_{iq_0,0})+\frac{1}{2}\,n_{F}''(\epsilon_{1})\,\delta_{iq_0,0} &\text{for} \hspace{0.3cm} \epsilon_{1}=\epsilon_{2}=\epsilon_{3}= \epsilon_1\,.
    \end{cases}
\end{align}
We find after analytic continuation and taking the dc limit
\begin{align}
    \lim_{\omega+i\eta\rightarrow0}\left[S_{F}^{(3)}(\epsilon_{1}-iq_0,\epsilon_{2},\epsilon_{3})\right]_{iq_0\rightarrow\omega+i\eta} &=
    S_{F}^{(3)}(\epsilon_{1},\epsilon_{2},\epsilon_{3})
    -n_{F}'(\epsilon_{1})
    \left[
    \frac{1}{\epsilon_{1}-\epsilon_{3}}
    \delta_{\epsilon_{1},\epsilon_{2}}
    +
    \frac{1}{\epsilon_{1}-\epsilon_{2}}
    \delta_{\epsilon_{1},\epsilon_{3}}
    \right]
    (1-\delta_{\epsilon_{2},\epsilon_{3}}) \nonumber
    \\
    &+\left[
    -\frac{1}{2}n_{F}''(\epsilon_{1})
    +
    n_{F}'(\epsilon_{1})
    \lim_{\omega+i\eta\rightarrow0}\frac{1}{\omega+i\eta}
    \right]
    \delta_{\epsilon_{1},\epsilon_{2}}\delta_{\epsilon_{2},\epsilon_{3}} \, ,
\end{align}
where we see that the dc limit is only well-defined for $\epsilon_2\neq\epsilon_3$. Extending this procedure to higher-order Matsubara summations $N\geq 3$, we find that merging high-order poles leads to this and similar divergences. Most importantly for our goal to identify a finite Drude weight, we see that the dc limit after analytic continuation of $iq_0$ yields a finite difference from explicitly setting $iq_0=0$ when two simple poles are separated {\it exactly} by the bosonic Matsubara frequency $iq_0$. As a third example, we give the expression for $N=4$ restricting ourselves to $(\epsilon_1\neq\epsilon_2) \cap (\epsilon_3\neq\epsilon_4)$, which shows no divergences, to avoid the quite lengthy general expression that is not necessary for the upcoming derivations. Under this assumption we find 
\begin{align}
    (\epsilon_1\neq\epsilon_2) \cap (\epsilon_3\neq\epsilon_4)
    \implies
    \lim_{\omega+i\eta\rightarrow0}
    &
    \bigg[S_{F}^{(4)}(\epsilon_{1}-iq_0,\epsilon_{2},\epsilon_{3}-iq_0,\epsilon_{4})\bigg]_{iq_0\rightarrow\omega+i\eta} 
    \nonumber\\
    &=
    S_{F}^{(4)}(\epsilon_{1},\epsilon_{2},\epsilon_{3},\epsilon_{4})
    -
    \frac{n_{F}'(\epsilon_{1})}{(\epsilon_{1}-\epsilon_{3})(\epsilon_{1}-\epsilon_{2})}
    \delta_{\epsilon_{1},\epsilon_{4}}
    -
    \frac{n_{F}'(\epsilon_{3})}{(\epsilon_{3}-\epsilon_{1})(\epsilon_{3}-\epsilon_{4})}
    \delta_{\epsilon_{2},\epsilon_{3}} \,.
\end{align}
Again, we explicitly see that the non-analyticity arises when two simple poles are exactly separated by the bosonic Matsubara frequency, denoted by the respective Kronecker delta.

\subsection{Matsubara summations involving multiple frequencies}

We will repeatedly encounter fermionic Matsubara sums with three independent Matsubara frequencies and overall six poles. Using the identities in Sec.~\ref{sec:sum_identities}, these typically decouple into the following two products, which we label as
\begin{align}\label{eq:l_1}
    L_{1}(iq_0;\epsilon_1,\epsilon_2;\epsilon_3,\epsilon_4;\epsilon_5,\epsilon_6) &\equiv S_{F}^{(2)}(\epsilon_1-iq_0,\epsilon_2)\,
    S_{F}^{(2)}(\epsilon_3-iq_0,\epsilon_4)\,
    S_{F}^{(2)}(\epsilon_5-iq_0,\epsilon_6)\, ,
    \\\label{eq:l_2}
    L_{2}(iq_0;\epsilon_1,\epsilon_2,\epsilon_3;\epsilon_4,\epsilon_5,\epsilon_6) &\equiv \frac{1}{\beta}\sum_{ik_0}
    S_{F}^{(3)}(\epsilon_1-iq_0,\epsilon_2,\epsilon_3-ik_0)\,
    S_{F}^{(3)}(\epsilon_4-iq_0,\epsilon_5,\epsilon_6-ik_0) \, .
\end{align}
As obvious from the previous discussions, a general derivation is possible but very tedious. In light of a transparent derivation, we focus on those aspects that are required for performing the analytic continuation followed by taking the dc limit for each product above. We keep the expressions as general as possible but invoke a constraint such as $(\epsilon_1\neq\epsilon_2) \cap (\epsilon_3\neq\epsilon_4)$ if needed. These restrictions correspond to taking only the interband transitions of the current vertex into account, which is sufficient for the leading-order contribution to the Drude weight of narrow and flat bands in the semiquantum limit defined in Eq.~\eqref{eq:semiquantum_limit} in the main text.   

A slightly different part is required for the contributions involving a frequency-dependent self-energy, which will occur for the second-order terms in interaction. Here, we define 
\begin{align}
    L_{3}(ip_0;\epsilon_1;\epsilon_2;\epsilon_3) &=
    \frac{1}{\beta^{2}}\sum_{ip_0',ip_0''}\frac{1}{(ip_0''-\epsilon_{1})(ip_0'+ip_0''-ip_0-\epsilon_{2})(ip_0'-\epsilon_{3})}\,.
\end{align}
In contrast to the other typical Matsubara sums, the external frequency is fermionic, as this will correspond to the poles of the self-energy of the fermion propagator.

\subsubsection{Product of three \texorpdfstring{$N=2$}{N=2} Matsubara summations}

The first expression, Eq.~\eqref{eq:l_1}, is straightforwardly evaluated by using our general results derived in Sec.~\ref{sec:limits}. Using Eq.~\eqref{eq:s2f_limit} we find seven terms subtracted from or added to $L_1$ evaluated at zero,
\begin{align}
    &\lim_{\omega+i\eta\,\rightarrow\,0}\,L_{1}(iq_0)\Big|_{iq_0\rightarrow\omega+i\eta} 
    =
    \Big[S_{F}^{(2)}(\epsilon_{1},\epsilon_{2})
    -n_{F}'(\epsilon_{1})\delta_{\epsilon_{1},\epsilon_{2}}
    \Big]
    \Big[S_{F}^{(2)}(\epsilon_{3},\epsilon_{4})
    -n_{F}'(\epsilon_{3})\delta_{\epsilon_{3},\epsilon_{4}}
    \Big]
    \Big[S_{F}^{(2)}(\epsilon_{5},\epsilon_{6})
    -n_{F}'(\epsilon_{5})\delta_{\epsilon_{5},\epsilon_{6}}
    \Big]
    \nonumber\\
    &=L_{1}(0)
    \nonumber\\
    &-n_{F}'(\epsilon_{1})\,S_{F}^{(2)}(\epsilon_{3},\epsilon_{4})\,S_{F}^{(2)}(\epsilon_{5},\epsilon_{6})\,\delta_{\epsilon_{1},\epsilon_{2}}
    \,\,\,-n_{F}'(\epsilon_{3})\,S_{F}^{(2)}(\epsilon_{1},\epsilon_{2})\,S_{F}^{(2)}(\epsilon_{5},\epsilon_{6})\,\delta_{\epsilon_{3},\epsilon_{4}}
    \,\,\,-n_{F}'(\epsilon_{5})\,S_{F}^{(2)}(\epsilon_{1},\epsilon_{2})\,S_{F}^{(2)}(\epsilon_{3},\epsilon_{4})\,\delta_{\epsilon_{5},\epsilon_{6}}
    \nonumber\\[1mm]
    &+n_{F}'(\epsilon_{1})\,n_{F}'(\epsilon_{3})\,S_{F}^{(2)}(\epsilon_{5},\epsilon_{6})\,\delta_{\epsilon_{1},\epsilon_{2}}\,\delta_{\epsilon_{3},\epsilon_{4}}
    +n_{F}'(\epsilon_{1})\,n_{F}'(\epsilon_{5})\,S_{F}^{(2)}(\epsilon_{3},\epsilon_{4})\,\delta_{\epsilon_{1},\epsilon_{2}}\,\delta_{\epsilon_{5},\epsilon_{6}}
    +n_{F}'(\epsilon_{3})\,n_{F}'(\epsilon_{5})\,S_{F}^{(2)}(\epsilon_{1},\epsilon_{2})\,\delta_{\epsilon_{3},\epsilon_{4}}\,\delta_{\epsilon_{5},\epsilon_{6}}
    \nonumber\\[2mm]
    &-n_{F}'(\epsilon_{1})\,n_{F}'(\epsilon_{3})\,n_{F}'(\epsilon_{5})\,\delta_{\epsilon_{1},\epsilon_{2}}\,\delta_{\epsilon_{3},\epsilon_{4}}\,\delta_{\epsilon_{5},\epsilon_{6}} \, ,
\end{align}
involving the product of multiple Fermi function derivatives $n'_F$ and $S^{(2)}_F$ when the arguments of $L_1$ coincide. For $(\epsilon_1\neq\epsilon_2) \cap (\epsilon_3\neq\epsilon_4)$, this expression reduces to the compact result 
\begin{align}
    (\epsilon_1\neq\epsilon_2) \cap (\epsilon_3\neq\epsilon_4)\implies \lim_{\omega+i\eta\,\rightarrow\,0}\,L_{1}(iq_0)\Big|_{iq_0\rightarrow\omega+i\eta} 
    &=L_{1}(0)
    -n_{F}'(\epsilon_{5})\,S_{F}^{(2)}(\epsilon_{1},\epsilon_{2})\,S_{F}^{(2)}(\epsilon_{3},\epsilon_{4})\,\delta_{\epsilon_{5},\epsilon_{6}}\, .
    \label{eq:L1_analyzed}
\end{align}
The explicit expression for $S^{(2)}_F$ with distinct arguments was given in Eq.~\eqref{eq:S2Result}. 

\subsubsection{Convoluted product of two \texorpdfstring{$N=3$}{N=3} Matsubara summations}

The second sum, Eq.~\eqref{eq:l_2}, is more complicated. It is given by a bosonic-frequency summation over a product of two single-fermionic-frequency sums. As both factors of the product have the same form, we explicitly write down only one of them and have
\renewcommand{\arraystretch}{2} % Adjusts row spacing
\begin{align}
    S_{F}^{(3)}&\big(\epsilon_1-iq_0,\epsilon_2,\epsilon_3-ik_0\big) =
    \nonumber\\
    &
    \begin{array}{lcl}
    \phantom{=}
    \frac{(1-\delta_{\epsilon_1,\epsilon_3}\delta_{ik_0,iq_0})}{ik_0-(\epsilon_3-\epsilon_1+iq_0)}
    &\times&
    \Big[
    \frac{(1-\delta_{\epsilon_1,\epsilon_2}\delta_{iq_0,0})}{(-iq_0+\epsilon_1-\epsilon_2)}
    n_{F}(\epsilon_1)
    +\frac{1}{2}n_{F}'(\epsilon_1)\delta_{\epsilon_1,\epsilon_2}\delta_{iq_0,0}
    \Big]
    \\
    +\,
    \frac{(1-\delta_{\epsilon_2,\epsilon_3}\delta_{ik_0,0})}{ik_0-(\epsilon_3-\epsilon_2)}
    &\times&
    \Big[
    \frac{(1-\delta_{\epsilon_1,\epsilon_2}\delta_{iq_0,0})}{(\epsilon_2-\epsilon_1+iq_0)}
    n_{F}(\epsilon_2)
    +\frac{1}{2}n_{F}'(\epsilon_2)\delta_{\epsilon_1,\epsilon_2}\delta_{iq_0,0}
    \Big]
    \\
    +\,
    \frac{(1-\delta_{\epsilon_1,\epsilon_3}\delta_{ik_0,iq_0})
    (1-\delta_{\epsilon_2,\epsilon_3}\delta_{ik_0,0})}{(ik_0-(\epsilon_3-\epsilon_1+iq_0))(ik_0-(\epsilon_3-\epsilon_2))}
    &\times&
    \Big[
    n_{F}(\epsilon_3)
    -n_{F}(\epsilon_1)
    \delta_{\epsilon_1,\epsilon_2}\delta_{iq_0,0}
    \Big]
    \\
    +\,
    \delta_{\epsilon_2,\epsilon_3}
    \delta_{ik_0,0}
    &\times&
    \Big[\frac{-n_{F}(\epsilon_2)}{(\epsilon_2-\epsilon_1+iq_0)^{2}}
    +\frac{n_{F}'(\epsilon_2)}{\epsilon_2-\epsilon_1+iq_0}\Big]
    (1-\delta_{\epsilon_1,\epsilon_2}\delta_{iq_0,0})
    \\
    +\,
    \delta_{\epsilon_1,\epsilon_3}
    \delta_{ik_0,iq_0}
    &\times&
    \Big(\Big[\frac{-n_{F}(\epsilon_1)}{(-iq_0+\epsilon_1-\epsilon_2)^{2}}
    +
    \frac{n_{F}'(\epsilon_1)}{-iq_0+\epsilon_1-\epsilon_2}\Big]
    (1-\delta_{\epsilon_1,\epsilon_2}\delta_{iq_0,0})
    +
    \frac{1}{2}n_{F}''(\epsilon_1)
    \delta_{\epsilon_1,\epsilon_2}
    \Big) \, .
    \\
    \stackrel{\quad\underbrace{\phantom{\quad\quad\quad\quad\quad\quad\quad\quad\quad\quad\quad\quad}}}{\equiv\vec{y}(ik_0;iq_0;\epsilon_1,\epsilon_2,\epsilon_3)}
    &
    &
    \stackrel{\underbrace{\phantom{\quad\quad\quad\quad\quad\quad\quad\quad\quad\quad\quad\quad\quad\quad\quad\quad\quad\quad\quad\quad\quad\quad\quad\quad\quad\quad\quad}}}{\equiv\vec{x}(iq_0;\epsilon_1,\epsilon_2,\epsilon_3)}
    \end{array}
\end{align}
This expression can be written as an inner product of two vectors $\vec{x}$ and $\vec{y}$ with five entries, where the latter depends on $ik_0$ but the former does not. The Matsubara sum $L_{2}(iq_0)$ contains, in general, $5\times5=25$ nonzero elements within this decomposition. We write them in a tensor form 
\begin{align}
    L_{2}(iq_0) &= 
    \frac{1}{\beta}\sum_{ik_0}S_{F}^{(3)}(\epsilon_1-iq_0,\epsilon_2,\epsilon_3-ik_0)\,S_{F}^{(3)}(\epsilon_4-iq_0,\epsilon_5,\epsilon_6-ik_0)
    \nonumber\\&= 
    \frac{1}{\beta}\sum_{ik_0}\sum_{i,j=1}^{5}
    x_{i}(iq_0;\epsilon_1,\epsilon_2,\epsilon_3)\,
    y_{i}(ik_0;iq_0;\epsilon_1,\epsilon_2,\epsilon_3)
    \,
    y_{j}(ik_0;iq_0;\epsilon_4,\epsilon_5,\epsilon_6)
    \,
    x_{j}(iq_0;\epsilon_4,\epsilon_5,\epsilon_6)
    \nonumber\\
    &\equiv\sum_{i,j=1}^{5}x_{i}(iq_0;\epsilon_1,\epsilon_2,\epsilon_3)\,A_{ij}(iq_0;\epsilon_1,\epsilon_2,\epsilon_3,\epsilon_4,\epsilon_5,\epsilon_6)\,x_{j}(iq_0;\epsilon_4,\epsilon_5,\epsilon_6) \, ,
    \label{eq:L2_matrixform}
\end{align}
where we introduce the $5\times 5$ matrix $A_{ij}$ including the Matsubara sum over $ik_0$. This summation couples the real poles $\lbrace\epsilon_1,\epsilon_2,\epsilon_3\rbrace$ and $\lbrace\epsilon_4,\epsilon_5,\epsilon_6\rbrace$. The matrix $A_{ij}$ can be decomposed into 4 blocks
\begin{align}\label{eq:A_ij_block_mat}
    &A_{ij}(iq_0) \equiv 
    \begin{pmatrix}
         A^{(1,1)}_{3\times 3}(iq_0) & A^{(1,2)}_{3\times 2}(iq_0)
         \\
         A^{(2,1)}_{2\times 3}(iq_0) & A^{(2,2)}_{2\times 2}(iq_0)
    \end{pmatrix}_{\!\!\!ij} \, ,
\end{align}
which read
{\allowdisplaybreaks
\begin{align}
    &A^{(1,1)}_{3\times 3}(iq_0)
    \equiv
    \begin{pmatrix}
        S_{B}^{(2)}(\omega^{}_{31},\omega^{}_{64}) & S_{B}^{(2)}(\omega^{}_{31}+iq_0,\omega^{}_{65})  & S_{B}^{(3)}(\omega^{}_{31},\omega^{}_{64},\omega^{}_{65}-iq_0)   
        \\
        S_{B}^{(2)}(\omega^{}_{32},\omega^{}_{64}+iq_0) & S_{B}^{(2)}(\omega^{}_{32},\omega^{}_{65})& S_{B}^{(3)}(\omega^{}_{32},\omega^{}_{64}+iq_0,\omega^{}_{65})
        \\
        S_{B}^{(3)}(\omega^{}_{31},\omega^{}_{32}-iq_0,\omega^{}_{64}) & S_{B}^{(3)}(\omega^{}_{31}+iq_0,\omega^{}_{32},\omega^{}_{65})& 
        S_{B}^{(4)}(\omega^{}_{31}+iq_0,\omega^{}_{32},\omega^{}_{64}+iq_0,\omega^{}_{65})&
        \end{pmatrix} \, ,\\[2mm]
    &A^{(1,2)}_{3\times 2}(iq_0)
    \equiv
    \begin{pmatrix}
         -\frac{(1-\delta_{\epsilon_1,\epsilon_3}\,\delta_{iq_0,0})\,\delta_{\epsilon_5,\epsilon_6}}{\beta\,(\epsilon_3-\epsilon_1+iq_0)}
         & -\frac{(1-\delta_{\epsilon_1,\epsilon_3})\,\delta_{\epsilon_4,\epsilon_6} }{\beta\,(\epsilon_3-\epsilon_1)} 
         \\[2mm]
           -\frac{(1-\delta_{\epsilon_2,\epsilon_3})\,\delta_{\epsilon_5,\epsilon_6} }{\beta\,(\epsilon_3-\epsilon_2)}
         & -\frac{(1-\delta_{\epsilon_2,\epsilon_3}\,\delta_{iq_0,0})\,\delta_{\epsilon_4,\epsilon_6} }{\beta\,(-iq_0+\omega_{3,2})} 
         \\[2mm]
          \frac{(1-\delta_{\epsilon_1,\epsilon_3}\,\delta_{iq_0,0})\,
        (1-\delta_{\epsilon_2,\epsilon_3})\,\delta_{\epsilon_5,\epsilon_6} }{\beta\,(\epsilon_3-\epsilon_1+iq_0)\,(\epsilon_3-\epsilon_2)}
        & \frac{(1-\delta_{\epsilon_1,\epsilon_3})\,
        (1-\delta_{\epsilon_2,\epsilon_3}\,\delta_{iq_0,0})\,\delta_{\epsilon_4,\epsilon_6} }{\beta\,(\epsilon_3-\epsilon_1)\,(-iq_0+\epsilon_3-\epsilon_2)} 
        \end{pmatrix}\, ,\\[2mm]
    &A^{(2,1)}_{2\times 3}(iq_0)
    \equiv
    \begin{pmatrix}
        \frac{-(1-\delta_{\epsilon_4,\epsilon_6}\,\delta_{iq_0,0})\,\delta_{\epsilon_2,\epsilon_3}}{\beta\,(\epsilon_6-\epsilon_4+iq_0)}
        & \frac{-(1-\delta_{\epsilon_5,\epsilon_6})\,\delta_{\epsilon_2,\epsilon_3} }{\beta\,(\epsilon_6-\epsilon_5)} 
        & \frac{(1-\delta_{\epsilon_4,\epsilon_6}\,\delta_{iq_0,0})\,
        (1-\delta_{\epsilon_5,\epsilon_6})\,\delta_{\epsilon_2,\epsilon_3} }{\beta\,(\epsilon_6-\epsilon_4+iq_0)\,(\epsilon_6-\epsilon_5)} 
        \\[2mm]
        \frac{-(1-\delta_{\epsilon_4,\epsilon_6})\,\delta_{\epsilon_1,\epsilon_3}}{\beta\,(\epsilon_6-\epsilon_4)} 
        & \frac{-(1-\delta_{\epsilon_5,\epsilon_6}\,\delta_{iq_0,0})\,\delta_{\epsilon_1,\epsilon_3}}{\beta\,(-iq_0+\epsilon_6-\epsilon_5)} 
        & \frac{(1-\delta_{\epsilon_4,\epsilon_6})\,
        (1-\delta_{\epsilon_5,\epsilon_6}\,\delta_{iq_0,0})\,\delta_{\epsilon_1,\epsilon_3}}{\beta\,(\epsilon_6-\epsilon_4)\,(-iq_0+\epsilon_6-\epsilon_5)} 
        \end{pmatrix} \, ,\\[2mm]
    &A^{(2,2)}_{2\times 2}(iq_0)
    \equiv
    \begin{pmatrix}
        \frac{1}{\beta}\,\delta_{\epsilon_2,\epsilon_3}\,\delta_{\epsilon_5,\epsilon_6} 
        & \frac{1}{\beta}\,\delta_{\epsilon_2,\epsilon_3}\,\delta_{\epsilon_4,\epsilon_6}\,\delta_{iq_0,0}  
        \\
        \frac{1}{\beta}\,\delta_{\epsilon_1,\epsilon_3}\,\delta_{\epsilon_5,\epsilon_6}\,\delta_{iq_0,0}
        & \frac{1}{\beta}\,\delta_{\epsilon_1,\epsilon_3}\,\delta_{\epsilon_4,\epsilon_6} 
        \end{pmatrix} \, .
\end{align}}
\!\!We introduced the short notation $\omega_{ij}\equiv \epsilon_i-\epsilon_j$ for the differences between the real-valued parts of the poles. It is obvious that the evaluation of $L_{2}(iq_0)-L_{2}(0)$ is very challenging without further constraints. In anticipation of the diagrammatic expansion for flat and narrow bands in the semiquantum limit, we are only interested in cases where $(\epsilon_1\neq\epsilon_2)\cap(\epsilon_4\neq\epsilon_5)$. This corresponds to the scenario of excluding intraband transitions at the current vertices, which are sub-leading for the considered physics. Under this assumption, we find that the analytic continuation of $\vec{x}(iq_0;\epsilon_1,\epsilon_2,\epsilon_3)$ is smoothly connected to $iq_0=0$ upon taking the limit leading to
\begin{align}\label{eq:x_analytic}
    \epsilon_1\neq\epsilon_2 \implies \lim_{\omega+i\eta\rightarrow0}\vec{x}(iq_0;\epsilon_1,\epsilon_2,\epsilon_3)|_{iq_0\rightarrow \omega+i\eta} = \vec{x}(iq_0=0;\epsilon_1,\epsilon_2,\epsilon_3)\,,
\end{align}
with the five components 
\begin{align}
    x_{i}(iq_0=0;\epsilon_1,\epsilon_2,\epsilon_3)
    =
    \begin{pmatrix}
    \frac{n_{F}(\epsilon_1)}{\epsilon_1-\epsilon_2}\,\,,
    &
    \frac{n_{F}(\epsilon_2)}{\epsilon_2-\epsilon_1}\,\,,
    &
    n_{F}(\epsilon_3)\,\,,
    &
    -\frac{n_{F}(\epsilon_2)}{(\epsilon_2-\epsilon_1)^{2}}
    +\frac{n_{F}'(\epsilon_2)}{\epsilon_2-\epsilon_1}\,\,,
    &
    -\frac{n_{F}(\epsilon_1)}{(\epsilon_1-\epsilon_2)^{2}}
    +
    \frac{n_{F}'(\epsilon_1)}{\epsilon_1-\epsilon_2}
    \end{pmatrix}_{\!\!i} \, .
\end{align}
The property \eqref{eq:x_analytic} hugely simplifies the analysis since it implies that the only source of non-analyticity in $L_{2}(iq_0)$ can be found within the matrix $A_{ij}(iq_0)$. Carefully examining the blocks of the matrix $A_{ij}(iq_0)$ in Eq.~\eqref{eq:A_ij_block_mat}, we find that either $\omega_{31}=\omega_{65}$ or $\omega_{32}=\omega_{64}$ or $\delta_{\epsilon_1,\epsilon_3}=\delta_{\epsilon_5,\epsilon_6}=1$ or $\delta_{\epsilon_2,\epsilon_3}=\delta_{\epsilon_4,\epsilon_6}=1$ are required to hold for the occurrence of the desired non-analyticity. We will only consider cases where $\omega_{31}=\omega_{65}=0$ or $\omega_{32}=\omega_{64}=0$ since these correspond to the only cases that are not sub-extensive after taking the thermodynamic limit of the final result obtained in the diagrammatic expansion. Therefore, the relevant possibilities are reduced  to $(\epsilon_3=\epsilon_1)\cap(\epsilon_6=\epsilon_5)$ or $(\epsilon_3=\epsilon_2)\cap(\epsilon_6=\epsilon_4)$. These two cases cannot be satisfied at the same time because they contradict $\epsilon_1\neq\epsilon_2$ and $\epsilon_4\neq\epsilon_5$. The two possibilities are related to each other in $L_{2}(iq_0;\epsilon_1,\epsilon_2,\epsilon_3;\epsilon_4,\epsilon_5,\epsilon_6)$ by the replacements $\epsilon_1\leftrightarrow\epsilon_2$, $\epsilon_4\leftrightarrow\epsilon_5$ and $iq_0\leftrightarrow-iq_0$ so that we will focus only on the first case and determine the second by this relation. We find 
\renewcommand{\arraystretch}{0.5}
\begin{align}
    &\epsilon_3=\epsilon_1\neq\epsilon_2 \cap 
    \epsilon_6=\epsilon_5\neq\epsilon_4 \implies
    \nonumber\\[4mm]
    &
    A_{ij}(iq_0)-A_{ij}(0) 
    \nonumber\\
    &\hspace{5mm}
    =\begin{pmatrix}
         0 & 
         \mbox{\normalsize $
         S_{B}^{(2)}(iq_0,0)-(iq_0=0)
         $}
         & 
         \mbox{\normalsize $
         S_{B}^{(3)}(-iq_0,0,\omega^{}_{64})-(iq_0=0)
         $}
         & -\frac{1}{\beta}\frac{1}{iq_0} 
         & 0
         \\[3mm]
         0 & 0 & 0  &  -\frac{1}{\beta}\frac{1}{\epsilon_1-\epsilon_2} 
         & 0
         \\[2mm]
         0 & 
         \mbox{\normalsize $
         \hspace{3mm}
         S_{B}^{(3)}(iq_0,0,\omega^{}_{12})-(iq_0=0)
         \hspace{3mm}
         $}
         & 
         \mbox{\normalsize $
         \hspace{2mm}
         S_{B}^{(4)}(iq_0,0,\omega^{}_{12},\omega^{}_{64}+iq_0)-(iq_0=0)
         \hspace{2mm}
         $}
           & 
           \mbox{\normalsize $
           \hspace{5mm}
           \frac{1}{\beta}\frac{1}{iq_0(\epsilon_1-\epsilon_2)} \hspace{5mm}
           $}
        & 0
        \\[3mm]
        0
        & 0
        & 0
        & 0
        & 0 \\[2mm]
        -\frac{1}{\beta}\frac{1}{\epsilon_6-\epsilon_4}  
        & \frac{1}{\beta}\frac{1}{iq_0}
        & -\frac{1}{\beta}\frac{1}{iq_0(\epsilon_6-\epsilon_4)}
        & -\frac{1}{\beta} 
        & 0
    \end{pmatrix}\, ,
\end{align}
where we again use the short notation $\omega_{ij}=\epsilon_i-\epsilon_j$ and $(iq_0=0)$ when the same quantity evaluated at $iq_0=0$ is subtracted. The remaining bosonic Matsubara summations have to be evaluated carefully due to the exclusion of one or two poles within the infinite sum, see definition in Eq.~\eqref{eq:bosonic_matsubara}. Via the residue theorem and after expanding in powers of $iq_0$ we find
\begin{alignat}{3}
    &S_{B}^{(2)}(iq_0,0) - (iq_0=0)
    &&= \frac{2}{\beta(iq_0)^{2}}+\frac{\beta}{12} \, , \\
    &S_{B}^{(3)}(iq_0,0,\omega_{jk}) - (iq_0=0)
    &&=-\frac{2}{\beta\omega_{jk}(iq_0)^2}-\frac{1}{\beta(iq_0)\omega_{jk}^2}-\frac{\beta}{12 \omega_{jk}} + \mathcal{O}(iq_0) \, , \\
    &S_{B}^{(4)}(iq_0,0,\omega_{jk}, \omega_{lm}+iq_0) - (iq_0=0)
    &&= \frac{2}{\beta \omega_{jk} \omega_{lm} (iq_0)^2}
    -\frac{\omega_{jk}- \omega_{lm}}{\beta \omega_{jk}^2  \omega_{lm}^2 (iq_0)} -\frac{1}{\beta\omega_{jk}^2  \omega_{lm}^2}+\frac{\beta}{12 \omega_{jk} \omega_{lm}} + \mathcal{O}(iq_0) \, .
\end{alignat}
We expect that only the contributions that are independent of $iq_0$ contribute to the final expression after analytic continuation and the dc limit. In particular, the higher-order contributions will vanish within this limit. Inserting the results into Eq.~\eqref{eq:L2_matrixform} we find 
{\allowdisplaybreaks
\begin{align}
    &\epsilon_3=\epsilon_1\neq\epsilon_2 \cap 
    \epsilon_6=\epsilon_5\neq\epsilon_4 \implies 
    \nonumber\\[4mm]
    &L_{2}(iq_0)-L_{2}(0) 
    \nonumber\\
    &\hspace{5mm}=\frac{n_{F}(\epsilon_1)}{\epsilon_1-\epsilon_2}
    \Bigg[\frac{n_{F}(\epsilon_6)}{\epsilon_6-\epsilon_4}
    \Bigg(\!\!
    \bigg(\frac{2}{\beta(iq_0)^{2}}+\frac{\beta}{12}\bigg)
    +\bigg(\!\!-\frac{2}{\beta(iq_0)^2}+\frac{1}{\beta(iq_0)(\epsilon_6-\epsilon_4)}-\frac{\beta}{12}\bigg)\!\!\Bigg)
    -\frac{1}{\beta}\frac{1}{iq_0}
    \Bigg(\!\!-\frac{n_{F}(\epsilon_6)}{(\epsilon_6-\epsilon_4)^{2}}
    +
    \frac{n_{F}'(\epsilon_6)}{\epsilon_6-\epsilon_4}\Bigg)
    \Bigg]
    \nonumber\\
    &\hspace{5mm}+\frac{1}{\beta}\frac{n_{F}(\epsilon_2)}{(\epsilon_2-\epsilon_1)^{2}}
    \Bigg[-\frac{n_{F}(\epsilon_6)}{(\epsilon_6-\epsilon_4)^{2}}
    +
    \frac{n_{F}'(\epsilon_6)}{\epsilon_6-\epsilon_4}\Bigg]
    \nonumber\\
    &\hspace{5mm}+n_{F}(\epsilon_1)
    \Bigg[
    \Bigg(\!\!-\frac{2}{\beta(\epsilon_1-\epsilon_2)(iq_0)^2}-\frac{1}{\beta(iq_0)(\epsilon_1-\epsilon_2)^2}-\frac{\beta}{12 (\epsilon_1-\epsilon_2)}\Bigg)\frac{n_{F}(\epsilon_6)}{\epsilon_6-\epsilon_4}
    \nonumber\\
    &\hspace{5mm}\phantom{+n_{F}(\epsilon_1)
    \Bigg[}+\Bigg(\frac{2}{\beta (\epsilon_1-\epsilon_2) (\epsilon_6-\epsilon_4) (iq_0)^2}
        -\frac{(\epsilon_1-\epsilon_2)-(\epsilon_6-\epsilon_4)}{\beta (\epsilon_1-\epsilon_2)^2 (\epsilon_6-\epsilon_4)^2 (iq_0)}
        \Bigg)n_{F}(\epsilon_6)
        \nonumber\\
    &\hspace{5mm}\phantom{+n_{F}(\epsilon_1)
    \Bigg[}
        +\Bigg(\!\!
        -\frac{1}{\beta(\epsilon_1-\epsilon_2)^2 (\epsilon_6-\epsilon_4)^2}+\frac{\beta}{12(\epsilon_1-\epsilon_2)(\epsilon_6-\epsilon_4)}\Bigg)n_{F}(\epsilon_6)
    \nonumber\\
    &\hspace{5mm}\phantom{+n_{F}(\epsilon_1)
    \Bigg[}+
    \frac{1}{\beta}\frac{1}{iq_0(\epsilon_1-\epsilon_2)}\Bigg(\!\!-\frac{n_{F}(\epsilon_6)}{(\epsilon_6-\epsilon_4)^{2}}
    +
    \frac{n_{F}'(\epsilon_6)}{\epsilon_6-\epsilon_4}\Bigg)
    \Bigg]
    \nonumber\\
    &\hspace{5mm}+
    \Bigg[\!\!
    -\frac{n_{F}(\epsilon_1)}{(\epsilon_1-\epsilon_2)^{2}}
    +
    \frac{n_{F}'(\epsilon_1)}{\epsilon_1-\epsilon_2}
    \Bigg]
    \Bigg[
    \frac{1}{\beta}\frac{n_{F}(\epsilon_4)}{(\epsilon_6-\epsilon_4)^{2}}
    -\frac{1}{\beta}\Bigg(\!\!-\frac{n_{F}(\epsilon_6)}{(\epsilon_6-\epsilon_4)^{2}}
    +
    \frac{n_{F}'(\epsilon_6)}{\epsilon_6-\epsilon_4}\Bigg)
    \Bigg]
    \nonumber\\[3mm]
    &\hspace{5mm}+\mathcal{O}\big(iq_0\big)\\
    &=\frac{1}{\beta}\frac{1}{(\epsilon_1-\epsilon_2)(\epsilon_6-\epsilon_4)}\Bigg[-\frac{n_F(\epsilon_1)\big(n_F(\epsilon_4)+n_F(\epsilon_6)\big)}{(\epsilon_1-\epsilon_2)(\epsilon_6-\epsilon_4)}-\frac{\big(n_F(\epsilon_1)+n_F(\epsilon_2)\big)n_F(\epsilon_6)}{(\epsilon_1-\epsilon_2)(\epsilon_6-\epsilon_4)}\nonumber\\
    &\hspace{37mm}+\frac{n'_F(\epsilon_1)\big(n_F(\epsilon_4)+n_F(\epsilon_6)\big)}{\epsilon_6-\epsilon_4}+\frac{\big(n_F(\epsilon_1)+n_F(\epsilon_2)\big)n'_F(\epsilon_6)}{\epsilon_1-\epsilon_2}-n'_F(\epsilon_1) n'_F(\epsilon_6)\Bigg]+\mathcal{O}\big(iq_0\big) \, .
\end{align}
}
\!\!All the divergences in $1/(iq_0)^2$ and $1/(iq_0)$ cancel, as expected. Note that $\epsilon_2$ and $\epsilon_4$ will correspond to remote bands whereas the remaining bands will correspond to the flat band $f$. Using that $n'_F(\epsilon_f)$ scales like $\beta$ and denoting the gap to the remote band as $\Delta$ we obtain the leading term
\begin{align}
    \label{eq:L2v1}
    \epsilon_3=\epsilon_1\neq\epsilon_2 \cap 
    \epsilon_6=\epsilon_5\neq\epsilon_4 \implies L_{2}(iq_0)-L_{2}(0) &=
    -\frac{1}{\beta}\frac{n_{F}'(\epsilon_3)}{\epsilon_3-\epsilon_2}\frac{n_{F}'(\epsilon_6)}{\epsilon_6-\epsilon_4}
    \Bigg[1+\mathcal{O}\bigg(\frac{1}{\beta\Delta}\bigg)\Bigg]+\mathcal{O}\big(iq_0\big) \, ,
\end{align}
up to corrections of the order $\mathcal{O}(1/\beta\Delta)$. The second case obtained by the symmetry transformation reads
\begin{align}
    \label{eq:L2v2}
    \epsilon_3=\epsilon_2\neq\epsilon_1 \cap 
    \epsilon_6=\epsilon_4\neq\epsilon_5 \implies L_{2}(iq_0)-L_{2}(0) &=
    -\frac{1}{\beta}\frac{n_{F}'(\epsilon_3)}{\epsilon_3-\epsilon_1}\frac{n_{F}'(\epsilon_6)}{\epsilon_6-\epsilon_5}
    \Bigg[1+\mathcal{O}\bigg(\frac{1}{\beta\Delta}\bigg)\Bigg]+\mathcal{O}\big(iq_0\big) \, .
\end{align}
These results conclude the general discussion of higher-order Matsubara summations and allow us to proceed to the diagrammatic expansion of the current-current correlation function.

\subsubsection{Mixed \texorpdfstring{$N=3$}{N=3} Matsubara summation}\label{sec:L3}

The final expression we wish to evaluate is a product of three poles with two fermionic frequencies. This expression admits a relatively straightforward evaluation using the identities of the previous section. We find
\begin{align}
    L_{3}(ip_0;\epsilon_1;\epsilon_2;\epsilon_3) &=
    \frac{1}{\beta^{2}}\sum_{ip_0',ip_0''}\frac{1}{(ip_0''-\epsilon_{1})(ip_0'+ip_0''-ip_0-\epsilon_{2})(ip_0'-\epsilon_{3})}
    \\
    &=
    \frac{1}{\beta}\sum_{ip_0'}\frac{S_{F}^{(2)}(\epsilon_{1},\epsilon_{2}+ip_0-ip_0')}{ip_0'-\epsilon_{3}}
    \\
    &=
    \begin{cases}
        \frac{1}{\beta}\sum_{ip_0'}\frac{n_{F}(\epsilon_1)-n_{F}(\epsilon_2)}{(ip_0'-\epsilon_{3})(\epsilon_1-(\epsilon_2+ip_0-ip_0'))} &\hspace{10mm}\epsilon_1\neq\epsilon_2
        \\[2mm]
        \frac{1}{\beta}\sum_{ip_0'}\frac{n_{F}'(\epsilon_1)\delta_{ip_0',ip_0}}{ip_0'-\epsilon_{3}}
        &\hspace{10mm}\epsilon_1=\epsilon_2
    \end{cases}
    \\
    &=
    \begin{cases}
        \frac{[n_{F}(\epsilon_1)-n_{F}(\epsilon_2)][n_{F}(\epsilon_2-\epsilon_1+ip_0)-n_{F}(\epsilon_3)]}{ip_0-(\epsilon_3+\epsilon_1-\epsilon_2)} & \hspace{3.5mm}\epsilon_1\neq\epsilon_2
        \\[2mm]
        \frac{1}{\beta}\frac{n_{F}'(\epsilon_1)}{ip_0-\epsilon_{3}}
        &\hspace{3.5mm}\epsilon_1=\epsilon_2
    \end{cases}
    \\
    &=
    \begin{cases}
        \frac{[n_{F}(\epsilon_1)-n_{F}(\epsilon_2)][-n_{B}(\epsilon_2-\epsilon_1)-n_{F}(\epsilon_3)]}{ip_0-(\epsilon_3+\epsilon_1-\epsilon_2)} &\hspace{7mm}\epsilon_1\neq\epsilon_2
        \\[2mm]
        \frac{-n_{F}(\epsilon_1)(1-n_{F}(\epsilon_1))}{ip_0-\epsilon_{3}}
        &\hspace{7mm}\epsilon_1=\epsilon_2 \,,
    \end{cases}
\end{align}
where in the last equation we used the fact that fermion and bosonic distribution functions are related by a shift of a fermionic frequency, i.e. $n_{F}(\epsilon+ip_0)=-n_{B}(\epsilon)$. Using the identity $[n_{F}(\epsilon_2)-n_{F}(\epsilon_1)]n_{B}(\epsilon_2-\epsilon_1)=-n_{F}(\epsilon_2)(1-n_{F}(\epsilon_1))$ we find
\begin{align}
    L_{3}(ip_0;\epsilon_1;\epsilon_2;\epsilon_3) 
    &=
    \begin{cases}
        \frac{-n_{F}(\epsilon_2)(1-n_{F}(\epsilon_1))
        -[n_{F}(\epsilon_1)-n_{F}(\epsilon_2)]n_{F}(\epsilon_3)}{ip_0-(\epsilon_3+\epsilon_1-\epsilon_2)} &\hspace{15mm}\epsilon_1\neq\epsilon_2
        \\[2mm]
        \frac{-n_{F}(\epsilon_1)(1-n_{F}(\epsilon_1))}{ip_0-\epsilon_{3}}
        &\hspace{15mm}\epsilon_1=\epsilon_2
    \end{cases}
    \\
    &=
    \frac{-n_{F}(\epsilon_2)\Big(1-[n_{F}(\epsilon_1)+n_{F}(\epsilon_3)]\Big)-n_{F}(\epsilon_1)n_{F}(\epsilon_3)}{ip_0-(\epsilon_3+\epsilon_1-\epsilon_2)}\,.
\end{align}
Interestingly, one can show, that this pole corresponds to a low energy propagator (i.e. $|\epsilon_3+\epsilon_1-\epsilon_2|<\Delta$) with a weight of order unity, i.e. $(T/\Delta)^{0}\sim 1$, if and only if all energies $\epsilon_1,\epsilon_2,\epsilon_3$ are in the flat (or narrow) band.

\newpage

\section{Diagrammatic expansion for narrow and flat band systems}

The Drude weight is evaluated by calculating the current-current correlation function, see Eq.~\eqref{eq:drude_weight} in the main text. One can easily show that the perfectly flat band has no Drude weight to zeroth and first order in interaction. For a narrow band, the Drude weight originating from zeroth and first order in interaction will be subleading in our parameter regime, see Eq.~\eqref{eq:semiquantum_limit} in the main text, when compared to contributions to second order in interaction. Therefore, we turn our attention to the second order in interaction. To structure the following derivation, we separate this calculation into contributions involving self-energies and those that do not. In Fig.~\ref{fig:second_order_vertex_corrections}, we show all diagrams that do not involve a self-energy diagram. In Fig.~\ref{fig:second_order_self_energy_corrections}, we show the remaining diagrams with self-energy contributions.

\begin{figure}[b!]
    \centering
    \includegraphics[width=0.28\textwidth]{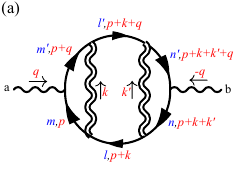} \hspace{20mm}
    \includegraphics[width=0.28\textwidth]{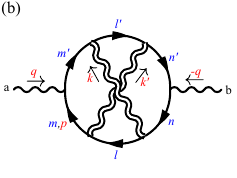}\\[2mm]
    \includegraphics[width=0.28\textwidth]{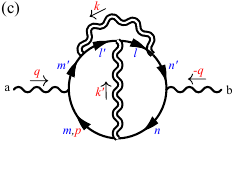}\hspace{20mm}
    \includegraphics[width=0.28\textwidth]{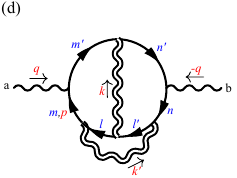}\\[2mm]
    \includegraphics[width=0.28\textwidth]{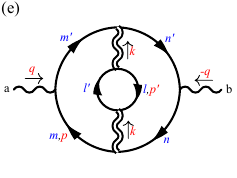}\hspace{5mm}
    \includegraphics[width=0.28\textwidth]{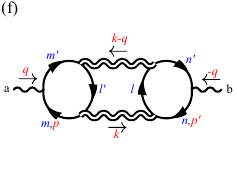}\hspace{5mm}
    \includegraphics[width=0.28\textwidth]{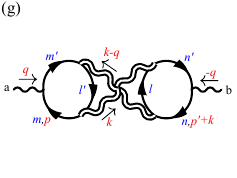}\\[2mm]
    \includegraphics[width=0.28\textwidth]{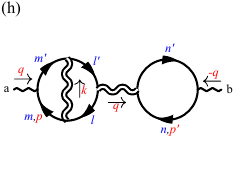}\hspace{5mm}
    \includegraphics[width=0.28\textwidth]{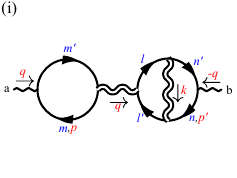}\hspace{5mm}
    \includegraphics[width=0.28\textwidth]{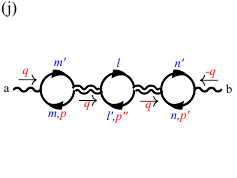}
    \caption{All second-order diagrams that may contribute to the current-current correlation function excluding self-energy contributions. Solid lines represent the band Green's functions. Single-wiggle and double-wiggle lines denote the electromagnetic gauge potentials and electron density-density interactions, respectively. The external indices $a,b$ are the spatial current and electric field directions. We indicate the band indices (blue) and the combined Matsubara frequencies and momenta indices (red). The arrow indicates the sign of the frequency/momentum to determine their conservation at each vertex. We provide the full labels in (a) and focus on the minimum labels that determine all other labels in the other diagrams. Note that $q=(iq_0,0)$ is only a bosonic Matsubara frequency and the interaction vertices do not depend on frequencies.}
    \label{fig:second_order_vertex_corrections}
\end{figure}

\newpage

\begin{figure}[t!]
    \centering
    \includegraphics[width=0.28\textwidth]{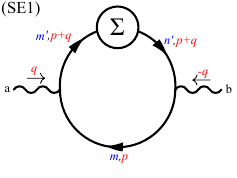} \hspace{20mm}
    \includegraphics[width=0.28\textwidth]{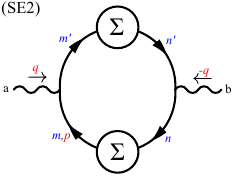}\\[2mm]
    \includegraphics[width=0.28\textwidth]{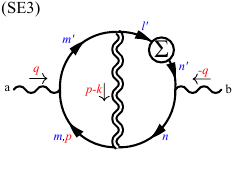}\hspace{20mm}
    \includegraphics[width=0.28\textwidth]{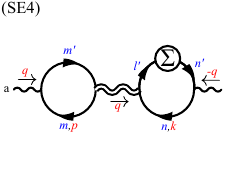}\hspace{20mm}
    \includegraphics[width=0.28\textwidth]{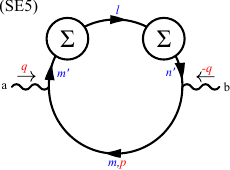}
    \caption{All (topologically distinct) second-order diagrams that may contribute to the current-current correlation functions and involve a self-energy $\hat\Sigma$. The notations are the same as in Fig.~\ref{fig:second_order_vertex_corrections}. Note that the self-energy conserves frequency and momentum.}
    \label{fig:second_order_self_energy_corrections}
\end{figure}

\subsection{Outline of the derivation}

We begin with the evaluation of the diagrams to determine their contribution to the current-current correlation function $\Pi^{ab}_q$ defined in Eq.~\eqref{eq:corr_func} in the main text. First, we focus on the diagrams that are second-order in interaction but do not involve self-energy diagrams. All possible diagrams are depicted in Fig.~\ref{fig:second_order_vertex_corrections}. The evaluation of each diagram follows the same procedure:
\begin{enumerate}
    \item Due to our choice of the interaction with the interaction vertex given in Eq.~\eqref{eq:interaction_vertex}, each diagram can be written as single or multiple traces over products of band projectors $ \bP_{m,\vec{p}}\equiv |u_{m,\vec{p}}\rangle\langle u_{m,\vec{p}}|$ with Bloch states $|u_{m,\vec{p}}\rangle$ of the respective band with dispersion $\epsilon_{m,\vec{p}}$, accompanied by derivatives of the Bloch Hamiltonian $\partial_a  \bH_\vec{p}$ at the current vertices, see Eq.~\eqref{eq:current_vertex}. The fermionic propagators give Green's functions with poles at the corresponding band energies (measured to the chemical potential), i.e. $G_{m,p}= \bP_{m,\vec{p}}/(ip_0-\epsilon_{m,\vec{p}})$. These two rules allow us to directly read off the formula by going through the diagram clockwise starting with the $(m,\vec{p})$-line in diagram $(a)$ to $(d)$, and at those solid lines, where the momentum is explicitly given, in the remaining diagrams.
    \item Besides traces over the orbital space, each diagram involves the summations over the internal momenta and Matsubara frequencies. To provide a consistent notation required for a transparent derivation, we include the summations over fermionic Matsubara frequencies $ip_0$, $ip_0'$ and $ip_0''$ within the $S_{F}^{(N)}$ defined in Eq.~\eqref{eq:definition_SFN}. In contrast, we write summations over bosonic Matsubara frequencies $ik_0$ and $ik_0'$, explicitly. We leave the fermionic signs of each of the ten diagrams, labeled $C_{(a)}$ to $C_{(j)}$, as a guide to the reader to help associate each term with its corresponding diagram in Fig.~\ref{fig:second_order_vertex_corrections}. Each diagram has a combinatorial factor of $8$ that cancels against the factor arising from the $1/2$ per each interaction term and the $1/2$ due to second-order perturbation theory. Thus, it only acquires a $(-1)$ for each fermionic loop and an additional $(-1)$ in case of an overall odd number of Green's functions. We give the mathematical formulas in Sec.~\ref{sec:mathematical_formulation} and perform the Matsubara summations in Sec.~\ref{sec:matsubara_summation} using the results from Sec.~\ref{sec:matsubara_summations}.
    \item After evaluating the Matsubara summations we use several identities for band projectors to simplify the result. In particular, we can explicitly perform various summations over the band indices, leading to geometric contributions that only involve the projector onto the flat (or narrow) band and the occupied states. We show this derivation in Sec.~\ref{sec:recombination}.
\end{enumerate}
In the second step, we continue with the evaluation of the diagrams involving self-energy contributions, where we follow the same strategy and keep the self-energy general if possible. All possible diagrams are depicted in Fig.~\ref{fig:second_order_self_energy_corrections}. Finally, we combine all contributions in a convenient compact form, which we give in the main text in Eqs.~\eqref{eq:drude_result}, \eqref{eq:g1_tensor}, \eqref{eq:g2_tensor}, and \eqref{eq:g3_tensor}.
We point out that this finite-order perturbation expansion by construction preserves particle number and Ward identities.

\subsection{Second-order diagrams in interaction excluding those involving a self-energy diagram}

As can be verified easily from the complex pole structure, the first order diagrams all vanish.
This is because for ideally flat bands where the intraband current vertex on the flatband is zero, both current vertices must be interband. In that case, all diagrams with a single interaction line cannot be decomposed into two (flatband) on-shell scattering amplitudes (cf. Figs. \ref{fig:narrow_band_processes} and \ref{fig:flat_band_processes} of the main text), and thus the Drude weight vanishes.

\subsubsection{Mathematical formulation of the diagrams}
\label{sec:mathematical_formulation}

We translate the diagrams shown in Fig.~\ref{fig:second_order_vertex_corrections} into their mathematical form in the following. We note that the diagrams can be split into four groups with different dependencies on the two interaction vertices. Going through the diagrams clockwise and using the indicated band, momentum, and frequency indices, we get 
\begin{align}
    \label{eq:diagrams}
    \Pi^{ab}_{\text{no-SE}}(iq_0) = 
    \sum_{m,m'}\sum_{n,n'}\sum_{l,l'}\Big(\Pi^{ab,mm'nn'll'}_{\text{(a,b,c,d)}}(iq_0) +\Pi^{ab,mm'nn'll'}_{\text{(e,f,g)}}(iq_0) +\Pi^{ab,mm'nn'll'}_{\text{(h,i)}}(iq_0) +\Pi^{ab,mm'nn'll'}_{\text{(j)}}(iq_0) \Big)
\end{align}
with the contribution involving the diagrams $(a)$ to $(d)$
\begin{align}
    &\Pi^{ab,mm'nn'll'}_{\text{(a,b,c,d)}}(iq_0)=-\frac{e^2}{N_c^{3}V^{}_c}\frac{1}{\beta^{2}}\sum_{ik_0,ik_0'}\sum_{\vec{p},\vec{k},\vec{k}'}
    U(\vec{k})\,U(\vec{k}')
    \nonumber \\[2mm]
    &\hspace{3mm}\times
    \Big(
    C_{(a)}\,
    \text{tr}\big[ \bP_{m,\vec{p}}(\partial_{a} \bH_{\vec{p}}) \bP_{m',\vec{p}} \bP_{l',\vec{p}+\vec{k}} \bP_{n',\vec{p}+\vec{k}+\vec{k}'}(\partial_{b} \bH_{\vec{p}+\vec{k}+\vec{k}'}) \bP_{n,\vec{p}+\vec{k}+\vec{k}'} \bP_{l,\vec{p}+\vec{k}}\big]
    \nonumber \\
    &\hspace{10mm}\times
    S_{F}^{(6)}\big(\epsilon_{m,\vec{p}},\epsilon_{m',\vec{p}}-iq_0,\epsilon_{l',\vec{p}+\vec{k}}-ik_0-iq_0,\epsilon_{n',\vec{p}+\vec{k}+\vec{k}'}-ik_0-ik_0'-iq_0,\epsilon_{n,\vec{p}+\vec{k}+\vec{k}'}-ik_0-ik_0',\epsilon_{l,\vec{p}+\vec{k}}-ik_0\big)
    \nonumber \\[4mm]
    &\hspace{6mm}
    +
    C_{(b)}\,
    \text{tr}\big[ \bP_{m,\vec{p}}(\partial_{a} \bH_{\vec{p}}) \bP_{m',\vec{p}} \bP_{l',\vec{p}+\vec{k}} \bP_{n',\vec{p}+\vec{k}+\vec{k}'}(\partial_{b} \bH_{\vec{p}+\vec{k}+\vec{k}'}) \bP_{n,\vec{p}+\vec{k}+\vec{k}'} \bP_{l,\vec{p}+\vec{k}'}\big]
    \nonumber \\
    &\hspace{10mm}\times
    S_{F}^{(6)}\big(\epsilon_{m,\vec{p}},\epsilon_{m',\vec{p}}-iq_0,\epsilon_{l',\vec{p}+\vec{k}}-ik_0-iq_0,\epsilon_{n',\vec{p}+\vec{k}+\vec{k}'}-ik_0-ik_0'-iq_0,\epsilon_{n,\vec{p}+\vec{k}+\vec{k}'}-ik_0-ik_0',\epsilon_{l,\vec{p}+\vec{k}'}-ik_0'\big)
    \nonumber \\[4mm]
    &\hspace{6mm}
    +
    C_{(c)}\,
    \text{tr}\big[ \bP_{m,\vec{p}}(\partial_{a} \bH_{\vec{p}}) \bP_{m',\vec{p}} \bP_{l',\vec{p}+\vec{k}} \bP_{l,\vec{p}+\vec{k}+\vec{k}'} \bP_{n',\vec{p}+\vec{k}'}(\partial_{b} \bH_{\vec{p}+\vec{k}'}) \bP_{n,\vec{p}+\vec{k}'}\big]
    \nonumber \\
    &\hspace{10mm}\times
    S_{F}^{(6)}\big(\epsilon_{m,\vec{p}},\epsilon_{m',\vec{p}}-iq_0,\epsilon_{l',\vec{p}+\vec{k}}-ik_0-iq_0,\epsilon_{l,\vec{p}+\vec{k}+\vec{k}'}-ik_0-ik_0'-iq_0,\epsilon_{n',\vec{p}+\vec{k}'}-ik_0'-iq_0,\epsilon_{n,\vec{p}+\vec{k}'}-ik_0'\big)
    \nonumber \\[4mm]
    &\hspace{6mm}
    +
    C_{(d)}\,
    \text{tr}\big[ \bP_{m,\vec{p}}(\partial_{a} \bH_{\vec{p}}) \bP_{m',\vec{p}} \bP_{n',\vec{p}+\vec{k}}(\partial_{b} \bH_{\vec{p}+\vec{k}}) \bP_{n,\vec{p}+\vec{k}} \bP_{l',\vec{p}+\vec{k}+\vec{k}'} \bP_{l,\vec{p}+\vec{k'}}\big]
    \nonumber \\
    &\hspace{10mm}\times
    S_{F}^{(6)}\big(\epsilon_{m,\vec{p}},\epsilon_{m',\vec{p}}-iq_0,\epsilon_{n',\vec{p}+\vec{k}}-ik_0-iq_0,\epsilon_{n,\vec{p}+\vec{k}}-ik_0,\epsilon_{l',\vec{p}+\vec{k}+\vec{k}'}-ik_0-ik_0',\epsilon_{l,\vec{p}+\vec{k}'}-ik_0'\big)
    \Big) \, ,
\end{align}
the contribution involving diagrams $(e)$ to $(g)$
\begin{align}
    &\Pi^{ab,mm'nn'll'}_{\text{(e,f,g)}}(iq_0)=-\frac{e^2}{N_c^{3}V^{}_c}\frac{1}{\beta}\sum_{ik_0}\sum_{\vec{p},\vec{p}',\vec{k}}
    U(\vec{k})\,U(\vec{k})\nonumber\\[2mm]
    &\hspace{3mm}\times\Big(
    C_{(e)}\,
    \text{tr}\big[ \bP_{m,\vec{p}}(\partial_{a} \bH_{\vec{p}}) \bP_{m',\vec{p}} \bP_{n',\vec{p}+\vec{k}}(\partial_{b} \bH_{\vec{p}+\vec{k}}) \bP_{n,\vec{p}+\vec{k}}\big]\,
    \text{tr}\big[ \bP_{l',\vec{p}'+\vec{k}} \bP_{l,\vec{p}'}\big]
    \nonumber \\
    &\hspace{28mm}\times
    S_{F}^{(4)}\big(\epsilon_{m,\vec{p}},\epsilon_{m',\vec{p}}-iq_0,\epsilon_{n',\vec{p}+\vec{k}}-ik_0-iq_0,\epsilon_{n,\vec{p}+\vec{k}}-ik_0\big)\,
    S_{F}^{(2)}\big(\epsilon_{l',\vec{p}'+\vec{k}}-ik_0,\epsilon_{l,\vec{p}'}\big)
    \nonumber \\[4mm]
    &\hspace{6mm}+
    C_{(f)}\,
    \text{tr}\big[ \bP_{m,\vec{p}}(\partial_{a} \bH_{\vec{p}}) \bP_{m',\vec{p}} \bP_{l',\vec{p}+\vec{k}}\big]\,
    \text{tr}\big[ \bP_{n',\vec{p}'}(\partial_{b} \bH_{\vec{p}'}) \bP_{n,\vec{p}'} \bP_{l,\vec{p}'+\vec{k}}\big]
    \nonumber \\
    &\hspace{28mm}\times
    S_{F}^{(3)}\big(\epsilon_{m,\vec{p}},\epsilon_{m',\vec{p}}-iq_0,\epsilon_{l',\vec{p}+\vec{k}}-ik_0\big)\,
    S_{F}^{(3)}\big(\epsilon_{n',\vec{p}'}-iq_0,\epsilon_{n,\vec{p}'},\epsilon_{l,\vec{p}'+\vec{k}}-ik_0\big)
    \nonumber \\[4mm]
    &\hspace{6mm}+
    C_{(g)}\,
    \text{tr}\big[ \bP_{m,\vec{p}}(\partial_{a} \bH_{\vec{p}}) \bP_{m',\vec{p}} \bP_{l',\vec{p}+\vec{k}}\big]\,
    \text{tr}\big[ \bP_{n',\vec{p}'+\vec{k}}(\partial_{b} \bH_{\vec{p}'+\vec{k}}) \bP_{n,\vec{p}'+\vec{k}} \bP_{l,\vec{p}'}\big]
    \nonumber \\&\hspace{28mm}\times
    S_{F}^{(3)}\big(\epsilon_{m,\vec{p}},\epsilon_{m',\vec{p}}-iq_0,\epsilon_{l',\vec{p}+\vec{k}}-ik_0\big)\,
    S_{F}^{(3)}\big(\epsilon_{n',\vec{p}'+\vec{k}}-iq_0-ik_0,\epsilon_{n,\vec{p}'+\vec{k}}-ik_0,\epsilon_{l,\vec{p}'}-iq_0\big)
    \Big) \, ,
\end{align}
the contribution involving diagrams $(h)$ and $(i)$
{\allowdisplaybreaks
\begin{align}
    &\Pi^{ab,mm'nn'll'}_{\text{(h,i)}}(iq_0)=-\frac{e^2}{N_c^{3}V^{}_c}\frac{1}{\beta}\sum_{ik_0}\sum_{\vec{p},\vec{p}',\vec{k}}
    U(\vec{k})\,U(\vec{0})\nonumber\\[2mm]
    &\hspace{3mm}\times \Big(C_{(h)}\,
    \text{tr}\big[ \bP_{m,\vec{p}}(\partial_{a} \bH_{\vec{p}}) \bP_{m',\vec{p}} \bP_{l',\vec{p}+\vec{k}} \bP_{l,\vec{p}+\vec{k}}\big]\,
    \text{tr}\big[ \bP_{n',\vec{p}'}(\partial_{b} \bH_{\vec{p}'}) \bP_{n,\vec{p}'}\big]
    \nonumber \\
    &\hspace{38mm}\times
    S_{F}^{(4)}\big(\epsilon_{m,\vec{p}},\epsilon_{m',\vec{p}}-iq_0,\epsilon_{l',\vec{p}+\vec{k}}-ik_0-iq_0,\epsilon_{l,\vec{p}+\vec{k}}-ik_0\big)\,
    S_{F}^{(2)}\big(\epsilon_{n',\vec{p}'}-iq_0,\epsilon_{n,\vec{p}'}\big)
    \nonumber \\[4mm]
    &\hspace{6mm}+
    C_{(i)}\,
    \text{tr}\big[ \bP_{m,\vec{p}}(\partial_{a} \bH_{\vec{p}}) \bP_{m',\vec{p}}\big]\,
    \text{tr}\big[ \bP_{n',\vec{p}'}(\partial_{b} \bH_{\vec{p}'}) \bP_{n,\vec{p}'} \bP_{l',\vec{p}'+\vec{k}} \bP_{l,\vec{p}'+\vec{k}}\big]
    \nonumber \\
    &\hspace{38mm}\times
    S_{F}^{(2)}\big(\epsilon_{m,\vec{p}},\epsilon_{m',\vec{p}}-iq_0\big)\,
    S_{F}^{(4)}\big(\epsilon_{n',\vec{p}'}-iq_0,\epsilon_{n,\vec{p}'},\epsilon_{l',\vec{p}'+\vec{k}}-ik_0,\epsilon_{l,\vec{p}'+\vec{k}}-ik_0-iq_0\big)\Big) \, ,
\end{align}
}
\!\!and the contribution involving diagram $(j)$
\begin{align}
    &\Pi^{ab,mm'nn'll'}_{\text{(j)}}(iq_0)=-\frac{e^2}{N_c^{3}V^{}_c}\sum_{\vec{p},\vec{p}',\vec{p}''}
    U(\vec{0})\,U(\vec{0})\nonumber\\[2mm]
    &\hspace{3mm}\times
    C_{(j)}\,
    \text{tr}\big[ \bP_{m,\vec{p}}(\partial_{a} \bH_{\vec{p}}) \bP_{m',\vec{p}}\big]\,
    \text{tr}\big[ \bP_{n',\vec{p}'}(\partial_{b} \bH_{\vec{p}'}) \bP_{n,\vec{p}'}\big]\,
    \text{tr}\big[ \bP_{l',\vec{p}''} \bP_{l,\vec{p}''}\big]
    \nonumber \\
    &\hspace{58mm}\times
    S_{F}^{(2)}\big(\epsilon_{m,\vec{p}},\epsilon_{m',\vec{p}}-iq_0\big)\,
    S_{F}^{(2)}\big(\epsilon_{n,\vec{p}'},\epsilon_{n',\vec{p}'}-iq_0\big)\,
    S_{F}^{(2)}\big(\epsilon_{l,\vec{p}''},\epsilon_{l',\vec{p}''}-iq_0\big)\, .
\end{align}
Note that the interaction does not carry any frequency dependence so diagrams $(f)$ and $(g)$ are part of the same class as $(e)$ for the spatially uniform electric field that we consider. For the same reason the diagrams $(h)$ to $(j)$ yield the interaction evaluated at zero momentum.

\subsubsection{Evaluation of the Matsubara summations}
\label{sec:matsubara_summation}

Calculating the Drude weight requires calculating $\Pi^{ab}(iq_0)-\Pi^{ab}(iq_0=0)$, see Eq.~\eqref{eq:drude_weight} in the main text. The only dependence on the bosonic Matsubara frequency $iq_0$ is in the Matsubara sum expressions above, see the second line of each diagram contribution $C_{(a)}$ to $C_{(j)}$. The strategies developed within Sec.~\ref{sec:matsubara_summations} allow us to systematically analyze the different contributions analytically although the complete expression for $\Pi^{ab}(iq_0)$ is extremely lengthy. To determine the Drude weight contribution, we start by evaluating first the Matsubara summations of each diagram separately. We use the identities derived in Sec.~\ref{sec:matsubara_summations}. In the step denoted by $\rightarrow$ we keep only the difference between the sum evaluated for $iq_0\neq 0$ and $iq_0=0$ to leading order for the semiquantum regime, specified in Eq.~\eqref{eq:semiquantum_limit} and \eqref{eq:drude_estimate} in the main text. Thus, we allow only interband transitions at each current vertex, i.e., $m\neq m'$ and $n\neq n'$ omitting sub-leading terms in the bandwidth, which is our smallest energy scale. We denote the flat band as $f$. The restriction of the band indices is indicated by the Kronecker delta $\delta_{n,m}$ for complete and transparent bookkeeping. Furthermore, we approximate the Fermi function and its derivative via 
\begin{align}
    n_F(\epsilon_{f,\vec{p}})\approx \nu\, ,\hspace{1cm}n'_F(\epsilon_{f,\vec{p}}) \approx -\beta\,\nu(1-\nu) \, ,
\end{align}
capturing the flatband filling in the regime $W\ll T$, part of the semiquantum limit specified in Eq.~\eqref{eq:semiquantum_limit} in the main text. The diagram $(a)$ involves a fermionic Matsubara summation with six poles. The particular structure involves two external bosonic Matsubara summations such that we can use the decomposition identity in Eq.~\eqref{eq:matsubara_splitting} and obtain $L_{1}(iq_0)$ analyzed in Eq.~\eqref{eq:L1_analyzed}. We get
\begin{align}
    &\frac{C_{(a)}}{\beta^{2}}\sum_{ik_0,ik_0'}
    S_{F}^{(6)}\big(\epsilon_{m,\vec{p}},\epsilon_{m',\vec{p}}-iq_0,\epsilon_{l',\vec{p}+\vec{k}}-ik_0-iq_0,\epsilon_{n',\vec{p}+\vec{k}+\vec{k}'}-ik_0-ik_0'-iq_0,\epsilon_{n,\vec{p}+\vec{k}+\vec{k}'}-ik_0-ik_0',\epsilon_{l,\vec{p}+\vec{k}}-ik_0\big)
    \nonumber\\[2mm]
    &=
    C_{(a)}\,
    S_{F}^{(2)}\big(\epsilon_{m,\vec{p}},\epsilon_{m',\vec{p}}-iq_0\big)\,
    S_{F}^{(2)}\big(\epsilon_{l',\vec{p}+\vec{k}}-iq_0,\epsilon_{l,\vec{p}+\vec{k}}\big)\,
    S_{F}^{(2)}\big(\epsilon_{n',\vec{p}+\vec{k}+\vec{k}'}-iq_0,\epsilon_{n,\vec{p}+\vec{k}+\vec{k}'}\big)
    \nonumber\\[3mm]
    &\rightarrow
    \big(iq_0=0\big)
    -
    C_{(a)}\,
    n_{F}'(\epsilon_{f,\vec{p}+\vec{k}})\,\delta^{}_{l,f}\,\delta^{}_{l',f}\,
    (1-\delta_{m,m'})\,
    (1-\delta_{n,n'})\,
    S_{F}^{(2)}\big(\epsilon_{m,\vec{p}},\epsilon_{m',\vec{p}}\big)\,
    S_{F}^{(2)}\big(\epsilon_{n',\vec{p}+\vec{k}+\vec{k}'},\epsilon_{n,\vec{p}+\vec{k}+\vec{k}'}\big)
    \nonumber\\[3mm]
    &\approx
    \big(iq_0=0\big)
    +
    C_{(a)}\,
    \beta\nu(1-\nu)\,\delta^{}_{l,f}\,\delta^{}_{l',f}\,
    (1-\delta_{m,m'})\,
    (1-\delta_{n,n'})\,
    \frac{n_{F}(\epsilon_{m,\vec{p}})-n_{F}(\epsilon_{m',\vec{p}})}{\epsilon_{m,\vec{p}}-\epsilon_{m',\vec{p}}}\,
    \frac{n_{F}(\epsilon_{n,\vec{p}+\vec{k}+\vec{k}'})-n_{F}(\epsilon_{n',\vec{p}+\vec{k}+\vec{k}'})}{\epsilon_{n,\vec{p}+\vec{k}+\vec{k}'}-\epsilon_{n',\vec{p}+\vec{k}+\vec{k}'}}\,.
\end{align}
At this stage, no further simplifications are required, in particular, to simplify the remaining Fermi functions. We will return to this result in Sec.~\ref{sec:recombination}. The Matsubara summation for diagram $(b)$ only slightly differs from the previous term by involving a $k'_0$ instead of a $k_0$ in the last pole. After several basic manipulations discussed in Sec.~\ref{sec:matsubara_summations} we identify the structure of $L_{2}(iq_0)$. We obtain
{\allowdisplaybreaks
\begin{align}
    &
    \frac{C_{(b)}}{\beta^{2}}\sum_{ik_0,ik_0'}
    S_{F}^{(6)}\big(\epsilon_{m,\vec{p}},\epsilon_{m',\vec{p}}-iq_0,\epsilon_{l',\vec{p}+\vec{k}}-ik_0-iq_0,\epsilon_{n',\vec{p}+\vec{k}+\vec{k}'}-ik_0-ik_0'-iq_0,\epsilon_{n,\vec{p}+\vec{k}+\vec{k}'}-ik_0-ik_0',\epsilon_{l,\vec{p}+\vec{k}'}-ik_0'\big)
    \nonumber\\
    &
    =\frac{C_{(b)}}{\beta}\sum_{ik_0}
    S_{F}^{(3)}\big(\epsilon_{m,\vec{p}},\epsilon_{m',\vec{p}}-iq_0,\epsilon_{l',\vec{p}+\vec{k}}-ik_0-iq_0\big)\,
    S_{F}^{(3)}\big(\epsilon_{n',\vec{p}+\vec{k}+\vec{k}'}-ik_0-iq_0,\epsilon_{n,\vec{p}+\vec{k}+\vec{k}'}-ik_0,\epsilon_{l,\vec{p}+\vec{k}'}\big)
    \nonumber\\
    &
    =-\frac{C_{(b)}}{\beta}\sum_{ik_0}
    S_{F}^{(3)}\big(\epsilon_{m,\vec{p}},\epsilon_{m',\vec{p}}-iq_0,\epsilon_{l',\vec{p}+\vec{k}}-ik_0\big)\,
    S_{F}^{(3)}\big(\!-\epsilon_{n',\vec{p}+\vec{k}+\vec{k}'},-\epsilon_{n,\vec{p}+\vec{k}+\vec{k}'}-iq_0,-\epsilon_{l,\vec{p}+\vec{k}'}-ik_0\big)
    \nonumber\\
    &
    \rightarrow
    \big(iq_0=0\big)
    +C_{(b)}\,\delta^{}_{W,0}\,\frac{n_{F}'(\epsilon_{f})\,n_{F}'(\epsilon_{f})}{\beta}\,\delta^{}_{l,f}\,\delta^{}_{l',f}\,
    \frac{\delta^{}_{m,f}\,\big(1-\delta^{}_{m',f}\big)\,\delta^{}_{n,f}\,\big(1-\delta^{}_{n',f}\big)
    +\delta^{}_{m',f}\,\big(1-\delta^{}_{m,f}\big)\,\delta^{}_{n',f}\,\big(1-\delta^{}_{n,f}\big)
    }{\big(\epsilon_{m,\vec{p}}-\epsilon_{m',\vec{p}}\big)\big(-\epsilon_{n,\vec{p}+\vec{k}+\vec{k}'}+\epsilon_{n',\vec{p}+\vec{k}+\vec{k}'}\big)}
    \nonumber\\[2mm]
    &
    \approx
    \big(iq_0=0\big)
    -C_{(b)}\,\delta^{}_{W,0}\,\beta\big[\nu(1-\nu)\big]^{2}\,\delta^{}_{l,f}\,\delta^{}_{l',f}\,
    \frac{\delta^{}_{m,f}\,\big(1-\delta^{}_{m',f}\big)\,\big(1-\delta^{}_{n',f}\big)\,\delta^{}_{n,f}
    +\big(1-\delta^{}_{m,f}\big)\,\delta^{}_{m',f}\,\delta^{}_{n',f}\,\big(1-\delta^{}_{n,f}\big)
    }{\big(\epsilon_{m,\vec{p}}-\epsilon_{m',\vec{p}}\big)\big(\epsilon_{n,\vec{p}+\vec{k}+\vec{k}'}-\epsilon_{n',\vec{p}+\vec{k}+\vec{k}'}\big)}
\end{align}
}
\!\!\!Note one essential difference to diagram $(a)$, which is relevant for all diagrams involving $L_2(iq_0)$. The general analysis of $L_2(iq_0)-(iq_0=0)$ summarized in Eqs.~\eqref{eq:L2v1} and \eqref{eq:L2v2} revealed that certain dispersions must be identical for all momenta. Thus, only perfectly flat bands with $W=0$ contribute, which we indicate by $\delta_{W,0}$. To be specific, the contribution to the Drude weight in diagram $(b)$ arises when either $m=n=f$ or $m'=n'=f$ are perfectly flat bands. The derivation for the diagrams $(c)$ to $(g)$ follows by the same line of arguments with slightly modified dependencies on the band indices and momenta. We obtain
{\allowdisplaybreaks
\begin{align}
    &
    \frac{C_{(c)}}{\beta^{2}}\sum_{ik_0,ik_0'}
    S_{F}^{(6)}\big(\epsilon_{m,\vec{p}},\epsilon_{m',\vec{p}}-iq_0,\epsilon_{l',\vec{p}+\vec{k}}-ik_0-iq_0,\epsilon_{l,\vec{p}+\vec{k}+\vec{k}'}-ik_0-ik_0'-iq_0,\epsilon_{n',\vec{p}+\vec{k}'}-ik_0'-iq_0,\epsilon_{n,\vec{p}+\vec{k}'}-ik_0'\big)
    \nonumber\\
    &=
    \frac{C_{(c)}}{\beta}\sum_{ik_0}
    S_{F}^{(3)}\big(\epsilon_{m',\vec{p}}-iq_0,\epsilon_{m,\vec{p}},\epsilon_{l',\vec{p}+\vec{k}}-ik_0\big)\,
    S_{F}^{(3)}\big(\epsilon_{n',\vec{p}+\vec{k}'}-iq_0,\epsilon_{n,\vec{p}+\vec{k}'},\epsilon_{l,\vec{p}+\vec{k}+\vec{k}'}-ik_0\big)
    \nonumber\\
    &
    \rightarrow
    \big(iq_0=0\big)
    -C_{(c)}\,\delta^{}_{W,0}\,\beta\big[\nu(1-\nu)\big]^{2}\,\delta^{}_{l,f}\,\delta^{}_{l',f}\,
    \frac{\delta^{}_{m,f}\,\big(1-\delta^{}_{m',f}\big)\,\delta^{}_{n',f}\,\big(1-\delta^{}_{n,f}\big)
    +\delta^{}_{m',f}\,\big(1-\delta^{}_{m,f}\big)\,\delta^{}_{n,f}\,\big(1-\delta^{}_{n',f}\big)
    }{\big(\epsilon_{m,\vec{p}}-\epsilon_{m',\vec{p}}\big)\big(\epsilon_{n',\vec{p}+\vec{k}'}-\epsilon_{n,\vec{p}+\vec{k}'}\big)} \, ,\\
    &
    \frac{C_{(d)}}{\beta^{2}}\sum_{ik_0,ik_0'}
    S_{F}^{(6)}(\epsilon_{m,\vec{p}},\epsilon_{m',\vec{p}}-iq_0,\epsilon_{n',\vec{p}+\vec{k}}-ik_0-iq_0,\epsilon_{n,\vec{p}+\vec{k}}-ik_0,\epsilon_{l',\vec{p}+\vec{k}+\vec{k}'}-ik_0-ik_0',\epsilon_{l,\vec{p}+\vec{k}'}-ik_0')
    \Big)
    \nonumber\\
    &=\frac{C_{(d)}}{\beta}\sum_{ik_0}
    S_{F}^{(3)}(\epsilon_{m',\vec{p}}-iq_0,\epsilon_{m,\vec{p}},\epsilon_{l,\vec{p}+\vec{k}'}-ik_0)\,
    S_{F}^{(3)}(\epsilon_{n',\vec{p}+\vec{k}}-iq_0,\epsilon_{n,\vec{p}+\vec{k}},\epsilon_{l',\vec{p}+\vec{k}+\vec{k}'}-ik_0
    )
    \Big)
    \nonumber\\
    &
    \rightarrow
    \big(iq_0=0\big)
    -C_{(d)}\,\delta^{}_{W,0}\,\beta\big[\nu(1-\nu)\big]^{2}\,\delta^{}_{l,f}\,\delta^{}_{l',f}\,
    \frac{\delta^{}_{m,f}\,\big(1-\delta^{}_{m',f}\big)\,\delta^{}_{n',f}\,\big(1-\delta^{}_{n,f}\big)
    +\delta^{}_{m',f}\,\big(1-\delta^{}_{m,f}\big)\,\delta^{}_{n,f}\,\big(1-\delta^{}_{n',f}\big)
    }{\big(\epsilon_{m,\vec{p}}-\epsilon_{m',\vec{p}}\big)\big(\epsilon_{n',\vec{p}+\vec{k}}-\epsilon_{n,\vec{p}+\vec{k}}\big)} \, , \\
    &
    \frac{C_{(e)}}{\beta}\sum_{ik_0}
    S_{F}^{(4)}\big(\epsilon_{m,\vec{p}},\epsilon_{m',\vec{p}}-iq_0,\epsilon_{n',\vec{p}+\vec{k}}-ik_0-iq_0,\epsilon_{n,\vec{p}+\vec{k}}-ik_0\big)\,
    S_{F}^{(2)}\big(\epsilon_{l',\vec{p}'+\vec{k}}-ik_0,\epsilon_{l,\vec{p}'}\big)
    \nonumber\\
    &=
    \frac{C_{(e)}}{\beta}\sum_{ik_0}
    S_{F}^{(3)}\big(\epsilon_{m',\vec{p}}-iq_0,\epsilon_{m,\vec{p}},\epsilon_{l,\vec{p}'}-ik_0\big)\,
    S_{F}^{(3)}\big(\epsilon_{n',\vec{p}+\vec{k}}-iq_0,\epsilon_{n,\vec{p}+\vec{k}},\epsilon_{l',\vec{p}'+\vec{k}}-ik_0\big)
    \nonumber\\
    &
    \rightarrow
    \big(iq_0=0\big)
    -C_{(e)}\,\delta^{}_{W,0}\,\beta\big[\nu(1-\nu)\big]^{2}\,\delta^{}_{l,f}\,\delta^{}_{l',f}\,
    \frac{\delta^{}_{m,f}\,\big(1-\delta^{}_{m',f}\big)\,\delta^{}_{n',f}\,\big(1-\delta^{}_{n,f}\big)
    +\delta^{}_{m',f}\,\big(1-\delta^{}_{m,f}\big)\,\delta^{}_{n,f}\,\big(1-\delta^{}_{n',f}\big)
    }{\big(\epsilon_{m,\vec{p}}-\epsilon_{m',\vec{p}}\big)\big(\epsilon_{n',\vec{p}+\vec{k}}-\epsilon_{n,\vec{p}+\vec{k}}\big)} \, , \\
    &
    \frac{C_{(f)}}{\beta}\sum_{ik_0}
    S_{F}^{(3)}\big(\epsilon_{m,\vec{p}},\epsilon_{m',\vec{p}}-iq_0,\epsilon_{l',\vec{p}+\vec{k}}-ik_0\big)\,
    S_{F}^{(3)}\big(\epsilon_{n',\vec{p}'}-iq_0,\epsilon_{n,\vec{p}'},\epsilon_{l,\vec{p}'+\vec{k}}-ik_0\big)
    \nonumber\\
    &
    \rightarrow
    \big(iq_0=0\big)
    -C_{(f)}\,\delta^{}_{W,0}\,\beta\big[\nu(1-\nu)\big]^{2}\,\delta^{}_{l,f}\,\delta^{}_{l',f}\,
    \frac{\delta^{}_{m,f}\,\big(1-\delta^{}_{m',f}\big)\,\delta^{}_{n',f}\,\big(1-\delta^{}_{n,f}\big)
    +\delta^{}_{m',f}\,\big(1-\delta^{}_{m,f})\,\delta^{}_{n,f}\,\big(1-\delta^{}_{n',f}\big)
    }{\big(\epsilon_{m,\vec{p}}-\epsilon_{m',\vec{p}}\big)\big(\epsilon_{n',\vec{p}'}-\epsilon_{n,\vec{p}'}\big)} \, , \\
    &
    \frac{C_{(g)}}{\beta}\sum_{ik_0}
    S_{F}^{(3)}\big(\epsilon_{m,\vec{p}},\epsilon_{m',\vec{p}}-iq_0,\epsilon_{l',\vec{p}+\vec{k}}-ik_0\big)\,
    S_{F}^{(3)}\big(\epsilon_{n',\vec{p}'+\vec{k}}-iq_0-ik_0,\epsilon_{n,\vec{p}'+\vec{k}}-ik_0,\epsilon_{l,\vec{p}'}-iq_0\big)
    \nonumber\\
    &
    \rightarrow
    \big(iq_0=0\big)
    -C_{(g)}\,\delta^{}_{W,0}\,\beta\big[\nu(1-\nu)\big]^{2}\,\delta^{}_{l,f}\,\delta^{}_{l',f}\,
    \frac{\delta^{}_{m,f}\,\big(1-\delta^{}_{m',f}\big)\,\delta^{}_{n,f}\,\big(1-\delta^{}_{n',f}\big)
    +\delta^{}_{m',f}\,\big(1-\delta^{}_{m,f}\big)\,\delta^{}_{n',f}\,\big(1-\delta^{}_{n,f}\big)
    }{\big(\epsilon_{m,\vec{p}}-\epsilon_{m',\vec{p}}\big)\big(\epsilon_{n,\vec{p}'+\vec{k}}-\epsilon_{n',\vec{p}'+\vec{k}}\big)} \, .
\end{align}
}
\!\!Finally, we find that diagrams $(h)$, $(i)$, and $(j)$ do not contribute to the Drude weight at the leading order. Here, the part of the diagram that involves two Green's functions, one coupling to the electromagnetic gauge potential, and one electron density-density interaction, vanishes for interband transitions $m\neq m'$ and $n\neq n'$. Potential contributions due to a finite bandwidth are subleading within the semiquantum regime. To conclude, we obtained seven diagrams that contribute to the Drude weight in the semiquantum regime. Diagram $(a)$ yields contributions for narrow bands whereas the other six diagrams $(b)$ to $(g)$ only contribute for perfectly flat bands indicated by $\delta_{W,0}$. Note that each Matsubara summation naturally leads to $1/\Delta^2$ where $\Delta$ is the band gap to remote bands. In the next step, we will show that these contributions are essential for the occurrence of a minimal set of relevant quantum states after band summation, that is, flat-band projectors and projectors onto occupied states when combined with the coupling to the external electromagnetic gauge potential.

\subsubsection{Recombination with quantum state geometric part}
\label{sec:recombination}

So far, we have not evaluated the derivative of the Bloch Hamiltonian and the summation over all band indices in Eq.~\eqref{eq:diagrams}. For this step, we first derive three important identities that occur repeatedly. Decomposing the Bloch Hamiltonian into its band representation we have the convenient form 
\begin{align}
     \bH_{\vec{p}}=\sum_{m}\epsilon_{m,\vec{p}} \bP_{m,\vec{p}}
\end{align}
involving the band dispersions (measured to the chemical potential) $\epsilon_{m,\vec{p}}$ and the band projectors $ \bP_{m,\vec{p}}=|u_{m,\vec{p}}\rangle\langle u_{m,\vec{p}}|$. The projectors are orthogonal and satisfy $ \bP_{m,\vec{p}} \bP_{n,\vec{p}}=\delta_{nm} \bP_{n,\vec{p}}$. Using the decomposition of the Bloch Hamiltonian and this identity for distinct band indices we obtain 
\begin{align}
    m\neq m' \implies 
    \frac{ \bP_{m,\vec{p}}(\partial_{a} \bH_{\vec{p}}) \bP_{m',\vec{p}}}{\epsilon_{m,\vec{p}}-\epsilon_{m',\vec{p}}}
    &= \bP_{m,\vec{p}}(\partial_{a} \bP_{m,\vec{p}}) \bP_{m',\vec{p}}
    =- \bP_{m,\vec{p}}(\partial_{a} \bP_{m',\vec{p}}) \bP_{m',\vec{p}} \, .
\end{align}
Note that the momentum derivative of the projector is straight-forwardly evaluated since it is gauge-invariant under the $U(1)$-gauge ambiguity of the Bloch eigenstates $|u_{m,\vec{p}}\rangle\rightarrow e^{i\phi_{m,\vec{p}}}|u_{m,\vec{p}}\rangle$. Combined with the completeness of the basis $\sum_m  \bP_{m,\vec{p}}=1$ we obtain 
\begin{align}
    \sum_{m,m'}\delta_{m,f}(1-\delta_{m',f})\frac{ \bP_{m,\vec{p}}(\partial_{a} \bH_{\vec{p}}) \bP_{m',\vec{p}}}{\epsilon_{m,\vec{p}}-\epsilon_{m',\vec{p}}} & =\sum_{m'\neq f} \frac{ \bP_{f,\vec{p}}(\partial_a  \bH_\vec{p}) \bP_{m',\vec{p}}}{\epsilon_{f,\vec{p}}-\epsilon_{m',\vec{p}}} =  \bP_{f,\vec{p}}\,(\partial_{a} \bP_{f,\vec{p}}) \, ,
    \\
    \sum_{m,m'}\delta_{m',f}(1-\delta_{m,f})\frac{ \bP_{m,\vec{p}}(\partial_{a} \bH_{\vec{p}}) \bP_{m',\vec{p}}}{\epsilon_{m',\vec{p}}-\epsilon_{m,\vec{p}}} & =\sum_{m\neq f} \,\frac{ \bP_{m,\vec{p}}(\partial_a  \bH_\vec{p}) \bP_{f,\vec{p}}}{\epsilon_{f,\vec{p}}-\epsilon_{m,\vec{p}}} \,= (\partial_{a} \bP_{f,\vec{p}})\, \bP_{f,\vec{p}} \, ,
\end{align}
which involves only the projector onto the flat band. Note that the projector and its derivative generically do not commute. As a third identity, we calculate
\begin{align}
    &\sum_{\underset{m\neq m'}{m,m'}}[n_{F}(\epsilon_{m,\vec{p}})-n_{F}(\epsilon_{m',\vec{p}})]\frac{ \bP_{m,\vec{p}}(\partial_{a} \bH_{\vec{p}}) \bP_{m',\vec{p}}}{\epsilon_{m,\vec{p}}-\epsilon_{m',\vec{p}}}\\ 
    &= \sum_{m}n_{F}(\epsilon_{m,\vec{p}}) \bP_{m,\vec{p}}(\partial_{a} \bP_{m,\vec{p}})+\sum_{m'}n_{F}(\epsilon_{m',\vec{p}})(\partial_{a} \bP_{m',\vec{p}}) \bP_{m',\vec{p}}
    \\[2mm]
    &=
    \sum_{m}n_{F}(\epsilon_{m,\vec{p}})(\partial_{a} \bP_{m,\vec{p}})
    \\[2mm]
    &= 
    \partial_{a} \bP^{[\nu]}_{\text{occ},\vec{p}} +\mathcal{O}(\beta W) + \mathcal{O}(\exp{(-\beta\Delta)})\, .
\end{align}
In the third step, we used the product rule of the momentum derivative to combine the two terms into a single derivative of the band projector. Defining 
\begin{align}
     \bP^{[\nu]}_{\text{occ},\vec{p}}\equiv\sum_{n<f} \bP_{n,\vec{p}}+\nu\,  \bP_{f,\vec{p}} \, ,
\end{align}
where $n<f$ denotes the summation over all bands below the flat band, we can approximately combine the band summation into a single projector-like object. Note that $ \bP^{[\nu]}_{\text{occ},\vec{p}}$ is a projector only if $\nu=0$ or $\nu=1$. The approximation in the last step arises from the momentum derivative of the $n_F(\epsilon_{m,\vec{p}})$ of the narrow band and the remote bands, which are both subleading in the semiquantum regime, see Eq.~\eqref{eq:semiquantum_limit} in the main text. Combining all the results for the Drude weight originating from the diagrams shown in Fig.~\ref{fig:second_order_vertex_corrections} we obtain
\begin{align}
    D^{ab}_{\text{no-SE}}&=\pi\lim_{\omega+i\eta\rightarrow 0}\big(\Pi^{ab}_{\text{no-SE}}(iq_0)-(iq_0=0)\big)\big|_{iq_0\rightarrow\omega+i\eta}
    \\
    &= -\frac{\pi e^2}{N_c^{3}V_c}
    \sum_{\vec{p},\vec{k},\vec{k}'}
    U(\vec{k})U(\vec{k}')\nonumber\\
    &\hspace{20mm}\times
    \Big(
    C_{(a)}\,\beta\,\nu(1-\nu)\,
    \text{tr}\big[(\partial_{a} \bP^{[\nu]}_{\text{occ},\vec{p}}) \bP_{\vec{p}+\vec{k}}(\partial_{b} \bP^{[\nu]}_{\text{occ},\vec{p}+\vec{k}+\vec{k}'}) \bP_{\vec{p}+\vec{k}}\big]
    \nonumber \\
    &\hspace{26.5mm}
    -
    C_{(b)}\,\delta_{W,0}\,\beta\,[\nu(1-\nu)]^{2}\,
    \text{tr}\big[ \bP_{\vec{p}}(\partial_{a} \bP_{\vec{p}}) \bP_{\vec{p}+\vec{k}}(\partial_{b} \bP_{\vec{p}+\vec{k}+\vec{k}'}) \bP_{\vec{p}+\vec{k}+\vec{k}'} \bP_{\vec{p}+\vec{k}'}\big]
    \nonumber \\
    &\hspace{26.5mm}
    -
    C_{(b)}\,\delta_{W,0}\,\beta\,[\nu(1-\nu)]^{2}\,
    \text{tr}\big[(\partial_{a} \bP_{\vec{p}}) \bP_{\vec{p}} \bP_{\vec{p}+\vec{k}} \bP_{\vec{p}+\vec{k}+\vec{k}'}(\partial_{b} \bP_{\vec{p}+\vec{k}+\vec{k}'}) \bP_{\vec{p}+\vec{k}'}\big]
    \nonumber \\
    &\hspace{26.5mm}
    -
    C_{(c)}\,\delta_{W,0}\,\beta\,[\nu(1-\nu)]^{2}\,
    \text{tr}\big[ \bP_{\vec{p}}(\partial_{a} \bP_{\vec{p}}) \bP_{\vec{p}+\vec{k}} \bP_{\vec{p}+\vec{k}+\vec{k}'} \bP_{\vec{p}+\vec{k}'}(\partial_{b} \bP_{\vec{p}+\vec{k}'})\big]
    \nonumber \\
    &\hspace{26.5mm}
    -
    C_{(c)}\,\delta_{W,0}\,\beta\,[\nu(1-\nu)]^{2}\,
    \text{tr}\big[(\partial_{a} \bP_{\vec{p}}) \bP_{\vec{p}} \bP_{\vec{p}+\vec{k}} \bP_{\vec{p}+\vec{k}+\vec{k}'}(\partial_{b} \bP_{\vec{p}+\vec{k}'}) \bP_{\vec{p}+\vec{k}'}\big]
    \nonumber \\
    &\hspace{26.5mm}
    -
    C_{(d)}\,\delta_{W,0}\,\beta\,[\nu(1-\nu)]^{2}\,
    \text{tr}\big[ \bP_{\vec{p}}(\partial_{a} \bP_{\vec{p}}) \bP_{\vec{p}+\vec{k}}(\partial_{b} \bP_{\vec{p}+\vec{k}}) \bP_{\vec{p}+\vec{k}+\vec{k}'} \bP_{\vec{p}+\vec{k'}}\big]
    \nonumber \\
    &\hspace{26.5mm}
    -
    C_{(d)}\,\delta_{W,0}\,\beta\,[\nu(1-\nu)]^{2}\,
    \text{tr}\big[(\partial_{a} \bP_{\vec{p}}) \bP_{\vec{p}}(\partial_{b} \bP_{\vec{p}+\vec{k}}) \bP_{\vec{p}+\vec{k}} \bP_{\vec{p}+\vec{k}+\vec{k}'} \bP_{\vec{p}+\vec{k'}}\big]
    \Big)
    \nonumber \\
    &-\frac{\pi e^2}{N_c^{3}V_c}\sum_{\vec{p},\vec{p}',\vec{k}}
    U(\vec{k})U(\vec{k})\nonumber\\
    &\hspace{20mm}\times
    \Big(
    -
    C_{(e)}\,\delta_{W,0}\,\beta\,[\nu(1-\nu)]^{2}\,
    \text{tr}\big[ \bP_{\vec{p}}(\partial_{a} \bP_{\vec{p}}) \bP_{\vec{p}+\vec{k}}(\partial_{b} \bP_{\vec{p}+\vec{k}})\big]
    \text{tr}\big[ \bP_{\vec{p}'+\vec{k}} \bP_{\vec{p}'}\big]
    \nonumber \\
    &\hspace{26.5mm}
    -
    C_{(e)}\,\delta_{W,0}\,\beta\,[\nu(1-\nu)]^{2}\,
    \text{tr}\big[(\partial_{a} \bP_{\vec{p}}) \bP_{\vec{p}}(\partial_{b} \bP_{\vec{p}+\vec{k}}) \bP_{\vec{p}+\vec{k}}\big]
    \text{tr}\big[ \bP_{\vec{p}'+\vec{k}} \bP_{\vec{p}'}\big]
    \nonumber \\
    &\hspace{26.5mm}
    -
    C_{(f)}\,\delta_{W,0}\,\beta\,[\nu(1-\nu)]^{2}\,
    \text{tr}\big[ \bP_{\vec{p}}(\partial_{a} \bP_{\vec{p}}) \bP_{\vec{p}+\vec{k}}\big]\,
    \text{tr}\big[ \bP_{\vec{p}'}(\partial_{b} \bP_{\vec{p}'}) \bP_{\vec{p}'+\vec{k}}\big]
    \nonumber \\
    &\hspace{26.5mm}
    -
    C_{(f)}\,\delta_{W,0}\,\beta\,[\nu(1-\nu)]^{2}\,
    \text{tr}\big[(\partial_{a} \bP_{\vec{p}}) \bP_{\vec{p}} \bP_{\vec{p}+\vec{k}}\big]\,
    \text{tr}\big[(\partial_{b} \bP_{\vec{p}'}) \bP_{\vec{p}'} \bP_{\vec{p}'+\vec{k}}\big]
    \nonumber \\
    &\hspace{26.5mm}
    -
    C_{(g)}\,\delta_{W,0}\,\beta\,[\nu(1-\nu)]^{2}\,
    \text{tr}\big[ \bP_{\vec{p}}(\partial_{a} \bP_{\vec{p}}) \bP_{\vec{p}+\vec{k}}\big]\,
    \text{tr}\big[(\partial_{b} \bP_{\vec{p}'+\vec{k}}) \bP_{\vec{p}'+\vec{k}} \bP_{\vec{p}'}\big]
    \nonumber \\
    &\hspace{26.5mm}
    -
    C_{(g)}\,\delta_{W,0}\,\beta\,[\nu(1-\nu)]^{2}\,
    \text{tr}\big[(\partial_{a} \bP_{\vec{p}}) \bP_{\vec{p}} \bP_{\vec{p}+\vec{k}}\big]\,
    \text{tr}\big[ \bP_{\vec{p}'+\vec{k}}(\partial_{b} \bP_{\vec{p}'+\vec{k}}) \bP_{\vec{p}'}\big]
    \Big) \, .
\end{align}
Note the distinct structure of diagram $(a)$ as discussed in Sec.~\ref{sec:matsubara_summation} and the different momentum dependencies on the electron-electron interaction discussed in Sec.~\ref{sec:mathematical_formulation}. We combine the expression by including the fermionc signs $C_{(a)}=C_{(b)}=C_{(c)}=C_{(d)}=-1$ and $C_{(e)}=C_{(f)}=C_{(g)}=+1$. Using the hermiticity of projectors and the cyclic property of the trace we identify the complex-conjugated (c.c.) partners of each term so that the final expression is entirely real. We obtain
{\allowdisplaybreaks
\begin{align}\label{eq:D_no_se}
    D^{ab}_{\text{no-SE}}
    = \frac{\pi e^2 \beta}{V_c}\frac{1}{N_c^{2}}\sum_{\vec{k},\vec{k}'}
    &\,\,U(\vec{k})\,U(\vec{k}')\nonumber\\
    \times&\bigg[
    \nu(1-\nu)
    \frac{1}{N_{c}}\sum_{\vec{p}}\text{tr}\Big[\big(\partial_{a} \bP^{[\nu]}_{\text{occ},\vec{p}}\big) \bP_{\vec{p}+\vec{k}}\big(\partial_{b} \bP^{[\nu]}_{\text{occ},\vec{p}+\vec{k}+\vec{k}'}\big) \bP_{\vec{p}+\vec{k}}\Big]
    \nonumber \\
    &
    -
    \delta_{W,0}\,[\nu(1-\nu)]^{2}
    \frac{1}{N_{c}}\sum_{\vec{p}}
    \Big(
    \text{tr}\big[ \bP_{\vec{p}}(\partial_{a} \bP_{\vec{p}}) \bP_{\vec{p}+\vec{k}}(\partial_{b} \bP_{\vec{p}+\vec{k}+\vec{k}'}) \bP_{\vec{p}+\vec{k}+\vec{k}'} \bP_{\vec{p}+\vec{k}'}\big]
    +\text{c.c.}
    \nonumber \\ &
    \hspace{36mm}+
    \text{tr}\big[ \bP_{\vec{p}}(\partial_{a} \bP_{\vec{p}}) \bP_{\vec{p}+\vec{k}} \bP_{\vec{p}+\vec{k}+\vec{k}'} \bP_{\vec{p}+\vec{k}'}(\partial_{b} \bP_{\vec{p}+\vec{k}'})\big]
    +\text{c.c.}
    \nonumber \\[2mm] &
    \hspace{36mm}+
    \text{tr}\big[ \bP_{\vec{p}}(\partial_{a} \bP_{\vec{p}}) \bP_{\vec{p}+\vec{k}}(\partial_{b} \bP_{\vec{p}+\vec{k}}) \bP_{\vec{p}+\vec{k}+\vec{k}'} \bP_{\vec{p}+\vec{k'}}\big]
    +\text{c.c.}
    \Big)
    \nonumber \\[1mm]
    &
    +
    \delta_{W,0}\,[\nu(1-\nu)]^{2}\,N_c\,
    \delta_{\vec{k},\vec{k}'}
    \frac{1}{N^2_{c}}\sum_{\vec{p},\vec{p}'}
    \Big(
    \text{tr}\big[ \bP_{\vec{p}}(\partial_{a} \bP_{\vec{p}}) \bP_{\vec{p}+\vec{k}}(\partial_{b} \bP_{\vec{p}+\vec{k}})\big]
    \text{tr}\big[ \bP_{\vec{p}'+\vec{k}} \bP_{\vec{p}'}\big]
    +\text{c.c.}
    \nonumber \\ &
    \hspace{48mm}+
    \text{tr}\big[ \bP_{\vec{p}}(\partial_{a} \bP_{\vec{p}}) \bP_{\vec{p}+\vec{k}}\big]
    \text{tr}\big[ \bP_{\vec{p}'}(\partial_{b} \bP_{\vec{p}'}) \bP_{\vec{p}'+\vec{k}}\big]
    +\text{c.c.}
    \nonumber \\[1mm] &
    \hspace{48mm}+
    \text{tr}\big[ \bP_{\vec{p}}(\partial_{a} \bP_{\vec{p}}) \bP_{\vec{p}+\vec{k}}\big]
    \text{tr}\big[(\partial_{b} \bP_{\vec{p}'+\vec{k}}) \bP_{\vec{p}'+\vec{k}} \bP_{\vec{p}'}\big]
    +\text{c.c.}
    \Big)\!
    \bigg] \, .
\end{align}
}
\!\!We introduced $\delta_{\vec{k},\vec{k'}}$ to combine the expression into a single momentum summation. This expression can be further simplified, but we turn our attention to the remaining second-order terms due to the self-energy contributions summarized in Fig.~\ref{fig:second_order_self_energy_corrections} beforehand.
 
\subsection{Second-order diagrams in interaction involving a self-energy diagram}

The five diagrams, relevant to the second-order expansion of the current-current correlation function, that involve a self-energy contribution are depicted and labeled in Fig.~\ref{fig:second_order_self_energy_corrections}. Note that the second-order of the self-energy $\hat\Sigma$ will be relevant for diagram (SE1). In contrast, only the first order of $\hat\Sigma$ contributes to the other four diagrams for an overall second-order expansion. However, we will keep the discussion as general as possible when calculating their contributions to the Drude weight in the semiquantum limit. 

\subsubsection{The fermionic self-energies to first and second order in interaction}

Before we proceed with the diagrammatic expansion of the current-current correlator, it will be useful to separately evaluate the self-energy diagram to first and second order in interaction. We will use the superscript to distinguish the order of expansion, i.e. $\hat\Sigma_{\vec{p}}(ip_0)=\hat\Sigma_{\vec{p}}^{(1)}(ip_0)+\hat\Sigma_{\vec{p}}^{(2)}(ip_0)+\mathcal{O}(U^3)$. The self-energy to first order in the interaction reads
\begin{align}
    \hat\Sigma_{\vec{p}}^{(1)}(ip_0) &= \frac{1}{N_c}\sum_{\vec{k}}\sum_{l}\bigg[U(0)\frac{1}{\beta}\sum_{ik_0}\frac{\text{tr}[\bP_{l,\vec{k}}]}{ip_0+ik_0-\epsilon_{l,\vec{k}}}-
    U(\vec{k})\frac{1}{\beta}\sum_{ik_0}\frac{\bP_{l,\vec{p}+\vec{k}}}{ip_0+ik_0-\epsilon_{l,\vec{p}+\vec{k}}}
    \bigg]
    \\&
    =\frac{1}{N_c}\sum_{\vec{k}}\sum_{l}\bigg[U(0)\big(n_{F}(\epsilon_{l,\vec{k}})-1\big)
    -U(\vec{k})\bP_{l,\vec{p}+\vec{k}}\big(n_{F}(\epsilon_{l,\vec{p}+\vec{k}})-1\big)
    \bigg]
    \\&\label{eq:self_energy_corr1}
    =-\frac{1}{N_c}\sum_{\vec{k}}U(\vec{k})\bP_{\text{occ},\vec{p}+\vec{k}}^{[\nu]}+\frac{1}{N_c}\sum_{\vec{k}}\bigg[U(\vec{k})+U(0)\sum_{l}\big(n_{F}(\epsilon_{l,\vec{k}})-1\big)
    \bigg] \, ,
\end{align}
where the Matsubara summations are calculated via the weighting function $h(z)=-\beta(n_{F}(z)-1)$ to assure convergence. We used the completeness of the band projectors as well as the definition $ \bP^{[\nu]}_\text{occ}$, which allowed us to split the self-energy into a term proportional to the projection onto the fraction of filled bands and a term proportional to the identity within the band basis. Note that the self-energy to first order in interaction is independent of the fermionic Matsubara frequency $ip_0$. We will use this property when inserting $\hat\Sigma^{(1)}_{\vec{p}}$ into the current-current correlation function. The self-energy to second order in interaction reads
\begin{align}
    \hat\Sigma_{\vec{p}}^{(2)}(ip_0) &= \frac{1}{N_c}\sum_{\vec{k}}\sum_{l,l'}\bigg[U(0)\frac{1}{\beta}\sum_{ik_0}\frac{\text{tr}[\bP_{l,\vec{k}}\,\hat\Sigma_{\vec{k}}^{(1)}\,\bP_{l',\vec{k}}]}{(ip_0+ik_0-\epsilon_{l,\vec{k}})(ip_0+ik_0-\epsilon_{l',\vec{k}})}\nonumber \\
    &\hspace{24mm}-
    U(\vec{k})\frac{1}{\beta}\sum_{ik_0}\frac{\bP_{l,\vec{p}+\vec{k}}\,\hat\Sigma_{\vec{p}+\vec{k}}^{(1)}\,\bP_{l',\vec{p}+\vec{k}}}{(ip_0+ik_0-\epsilon_{l,\vec{p}+\vec{k}})(ip_0+ik_0-\epsilon_{l',\vec{p}+\vec{k}})}
    \bigg]
    \nonumber\\
    &-\frac{1}{N_c^2}\sum_{\vec{k},\vec{k}'}\sum_{l,l',l''}\bigg[U(\vec{k})U(\vec{k}')\frac{1}{\beta^2}\!\!\sum_{ik_0,ik_0'}\frac{\bP_{l,\vec{p}+\vec{k}}\,\bP_{l',\vec{p}+\vec{k}+\vec{k}'}\,\bP_{l'',\vec{p}+\vec{k}'}}{(ip_0+ik_0-\epsilon_{l,\vec{p}+\vec{k}})(ip_0+ik_0+ik_0'-\epsilon_{l',\vec{p}+\vec{k}+\vec{k}'})(ip_0+ik_0'-\epsilon_{l'',\vec{p}+\vec{k}'})}
    \bigg]
    \nonumber\\
    &+\frac{1}{N_c^2}\sum_{\vec{k},\vec{k}'}\sum_{l,l',l''}\bigg[U(\vec{k})U(\vec{k})\frac{1}{\beta^2}\!\!\sum_{ik_0,ik_0'}\frac{\bP_{l,\vec{p}+\vec{k}}\,\text{tr}[\bP_{l',\vec{k}'}\bP_{l'',\vec{k}+\vec{k}'}]}{(ip_0+ik_0-\epsilon_{l,\vec{p}+\vec{k}})(ip_0+ik_0'-\epsilon_{l',\vec{k}'})(ip_0+ik_0+ik_0'-\epsilon_{l'',\vec{k}+\vec{k}'})}
    \bigg]\,.
\end{align}
In addition to the $S^{(2)}_F$ contributions in the first terms, we identify $L_{3}(ip_0;\epsilon_1;\epsilon_2;\epsilon_3)$ as evaluated in Sec.~\ref{sec:L3} in the last two lines by converting the bosonic summations over $ik_0$ and $ik_0'$ into fermionic summations. Using the result gives
\begin{align}\label{eq:self_energy_corr2}
    \hat\Sigma_{\vec{p}}^{(2)}(ip_0)&= \frac{1}{N_c}\sum_{\vec{k}}\sum_{l,l'}\bigg[U(0)\,\text{tr}[\bP_{l,\vec{k}}\hat\Sigma_{\vec{k}}^{(1)}]\,S_{F}^{(2)}(\epsilon_{l,\vec{k}},\epsilon_{l',\vec{k}})\delta_{l,l'}
    -
    U(\vec{k})\bP_{l,\vec{p}+\vec{k}}\hat\Sigma_{\vec{p}+\vec{k}}^{(1)}\bP_{l',\vec{p}+\vec{k}}S_{F}^{(2)}(\epsilon_{l,\vec{p}+\vec{k}},\epsilon_{l',\vec{p}+\vec{k}})
    \bigg]
    \nonumber\\
    &-\frac{1}{N_c^2}\sum_{\vec{k},\vec{k}'}\sum_{l,l',l''}\bigg[U(\vec{k})U(\vec{k}')\bP_{l,\vec{p}+\vec{k}}\bP_{l',\vec{p}+\vec{k}+\vec{k}'}\bP_{l'',\vec{p}+\vec{k}'}
    \nonumber\\
    &\hspace{25mm}\times
    \frac{
    -n_{F}(\epsilon_{l',\vec{p}+\vec{k}+\vec{k}'})\big(1-[n_{F}(\epsilon_{l,\vec{p}+\vec{k}})+n_{F}(\epsilon_{l'',\vec{p}+\vec{k}'})]\big)-n_{F}(\epsilon_{l,\vec{p}+\vec{k}})n_{F}(\epsilon_{l'',\vec{p}+\vec{k}'})
    }{ip_0-(\epsilon_{l,\vec{p}+\vec{k}}+\epsilon_{l'',\vec{p}+\vec{k}'}-\epsilon_{l',\vec{p}+\vec{k}+\vec{k}'})}
    \bigg]
    \nonumber\\
    &+\frac{1}{N_c^2}\sum_{\vec{k},\vec{k}'}\sum_{l,l',l''}\bigg[U(\vec{k})U(\vec{k})\bP_{l,\vec{p}+\vec{k}}\text{tr}[\bP_{l',\vec{k}'}\bP_{l'',\vec{k}+\vec{k}'}]
    \nonumber\\
    &\hspace{25mm}\times
    \frac{-n_{F}(\epsilon_{l'',\vec{k}+\vec{k}'})\big(1-[n_{F}(\epsilon_{l,\vec{p}+\vec{k}})+n_{F}(\epsilon_{l',\vec{k}'})]\big)-n_{F}(\epsilon_{l,\vec{p}+\vec{k}})n_{F}(\epsilon_{l',\vec{k}'})}{ip_0-(\epsilon_{l,\vec{p}+\vec{k}}+\epsilon_{l',\vec{k}'}-\epsilon_{l'',\vec{k}+\vec{k}'})}
    \bigg]\,.
\end{align}
In contrast to the first-order self-energy, this expression depends on $ip_0$, and introduces poles into any external Matsubara summation. In our calculations, these poles will be accompanied by a negligible weight unless all bands are flat, i.e. $l=l'=l''=f$, which significantly simplifies the following evaluation.

\subsubsection{The contributing diagrams (SE1)}

The diagram (SE1) reads
\begin{align}
    &\Pi^{ab}_{\text{SE1}}(iq_0) 
    =-\frac{e^{2}}{N_c V_c}
    C_{(\text{SE1})}
    \sum_{\vec{p}}
    \sum_{m,m'}\sum_{n,n'}
    \frac{1}{\beta}\sum_{ip_0}
    \Bigg(
    \frac{\text{tr}\big[\bP_{m,\vec{p}}(\partial_{a} \bH_{\vec{p}})\bP_{m',\vec{p}}\,\hat\Sigma_{\vec{p}}^{(2)}(ip_0+iq_0)\bP_{n',\vec{p}}(\partial_{b} \bH_{\vec{p}})\bP_{n,\vec{p}}\big]}{(ip_0+iq_0-\epsilon_{m',\vec{p}})(ip_0+iq_0-\epsilon_{n',\vec{p}})(ip_0-\epsilon_{m,\vec{p}})}
    \nonumber\\
    &\hspace{70mm}+
    \frac{\text{tr}\big[\bP_{m,\vec{p}}(\partial_{a} \bH_{\vec{p}})\bP_{m',\vec{p}}\,\bP_{n',\vec{p}}(\partial_{b} \bH_{\vec{p}})\bP_{n,\vec{p}}\hat\Sigma_{\vec{p}}^{(2)}(ip_0)\big]}{(ip_0+iq_0-\epsilon_{m',\vec{p}})(ip_0-\epsilon_{n,\vec{p}})(ip_0-\epsilon_{m,\vec{p}})}\Bigg) \, .
\end{align}
We note that the self-energy to second order in interaction has poles, which impact the Matsubara summation. We split the discussion into two cases. First, terms in $\hat\Sigma_{\vec{p}}^{(2)}(ip_0)$ that are independent of $ip_0$ lead to Matsubara sums of the form $\delta_{m,n}\,S_{F}^{(3)}(\epsilon_{m',\vec{p}}-iq_0,\epsilon_{n',\vec{p}}-iq_0,\epsilon_{m,\vec{p}})$ in the expression for $\Pi^{ab}_{\text{SE1}}(iq_0)$. Considering only interband current vertices, i.e. $m\neq m'$ and $n\neq n'$, one finds these do not contribute to leading order in the semiquantum limit. Secondly, we consider the terms in $\hat\Sigma_{\vec{p}}^{(2)}(ip_0)$ that involve a simple pole $\epsilon_0=\epsilon_{l,\vec{p}+\vec{k}}+\epsilon_{l'',\vec{p}+\vec{k}'}-\epsilon_{l',\vec{p}+\vec{k}+\vec{k}'}$, such that $\hat\Sigma_{\vec{p}}^{(2)}(ip_0)\propto\frac{1}{ip_0-\epsilon_0}$. The first term in $\Pi^{ab}_{\text{SE1}}(iq_0)$ is proportional to $\delta_{m,n}\,S_{F}^{(4)}(\epsilon_{m',\vec{p}}-iq_0,\epsilon_{n',\vec{p}}-iq_0,\epsilon_{m,\vec{p}},\epsilon_0-iq_0)$ and the second one is proportional to $\delta_{m',n'}\,S_{F}^{(4)}(\epsilon_{m',\vec{p}}-iq_0,\epsilon_{n,\vec{p}},\epsilon_{m,\vec{p}},\epsilon_0)$. Considering only interband current vertices, one finds that a contribution to the Drude weight occurs only if the pole $\epsilon_0$ of the self-energy matches the energy on the fermion propagator that directly connects the two current vertices. Since $\epsilon_0$ depends on three different momenta, the only way for this term to contribute in the thermodynamic limits is if the band is perfectly flat ($W=0$) and all band indices $l,l',l''$ are in the flat band. The resulting coefficient $n_{F}'(\epsilon_0)$ in the Drude weight implies that this flat band must be the one near the chemical potential, i.e. band $f$. In conclusion, we rewrite the effective form contributing to leading order in the semiquantum limit to the Drude weight as
\begin{align}
    \hat\Sigma_{\vec{p}}^{(2)}(ip_0)&\approx 
    \delta_{W,0}\,\nu(1-\nu)\frac{1}{N_c^2}\sum_{\vec{k},\vec{k}'}
    \Big[
    U(\vec{k})U(\vec{k}')\bP_{\vec{p}+\vec{k}}\bP_{\vec{p}+\vec{k}+\vec{k}'}\bP_{\vec{p}+\vec{k}'}
    -U(\vec{k})U(\vec{k})\bP_{\vec{p}+\vec{k}}\,\text{tr}[\bP_{\vec{k}'}\bP_{\vec{k}+\vec{k}'}]
    \Big]\,
    \frac{1}{ip_0-\epsilon_f} \, .
\end{align}
Inserting this into the current-current correlation function we obtain 
\begin{align}
    &\Pi^{ab}_{\text{SE1}}(iq_0) 
    \approx-\frac{e^{2}}{N_c^{3} V_c}
    C_{(\text{SE1})}
    \sum_{\vec{p},\vec{k},\vec{k}'}
    \sum_{m,m'}\sum_{n,n'} \nonumber
    \\
    \Big(&U(\vec{k})U(\vec{k}')\,
    \delta_{W,0}\,\nu(1-\nu)\,
    \text{tr}\big[\bP_{m,\vec{p}}(\partial_{a} \bH_{\vec{p}})\bP_{m',\vec{p}}\bP_{\vec{p}+\vec{k}}\bP_{\vec{p}+\vec{k}+\vec{k}'}\bP_{\vec{p}+\vec{k}'}\bP_{n',\vec{p}}(\partial_{b} \bH_{\vec{p}})\bP_{n,\vec{p}}\big]\,
    S_{F}^{(4)}(\epsilon_{m',\vec{p}},\epsilon_{m,\vec{p}}+iq_0,\epsilon_{n',\vec{p}},\epsilon_f)
    \nonumber\\[2mm]
    +\,&U(\vec{k})U(\vec{k}')\,
    \delta_{W,0}\,\nu(1-\nu)\,
    \text{tr}\big[\bP_{m,\vec{p}}(\partial_{a} \bH_{\vec{p}})\bP_{m',\vec{p}}\,\bP_{n',\vec{p}}(\partial_{b} \bH_{\vec{p}})\bP_{n,\vec{p}}\bP_{\vec{p}+\vec{k}}\bP_{\vec{p}+\vec{k}+\vec{k}'}\bP_{\vec{p}+\vec{k}'}\big]\,
    S_{F}^{(4)}(\epsilon_{m',\vec{p}}-iq_0,\epsilon_{m,\vec{p}},\epsilon_{n,\vec{p}},\epsilon_f)
    \nonumber\\[2mm]
    -\,&U(\vec{k})U(\vec{k})\,
    \delta_{W,0}\nu(1-\nu)\,
    \text{tr}\big[\bP_{m,\vec{p}}(\partial_{a} \bH_{\vec{p}})\bP_{m',\vec{p}}\bP_{\vec{p}+\vec{k}}\bP_{n',\vec{p}}(\partial_{b} \bH_{\vec{p}})\bP_{n,\vec{p}}\big]\,
    \text{tr}[\bP_{\vec{k}'}\bP_{\vec{k}+\vec{k}'}]\,
    S_{F}^{(4)}(\epsilon_{m',\vec{p}},\epsilon_{m,\vec{p}}+iq_0,\epsilon_{n',\vec{p}},\epsilon_f)
    \nonumber\\[2mm]
    -\,&U(\vec{k})U(\vec{k})\,
    \delta_{W,0}\,\nu(1-\nu)\,
    \text{tr}\big[\bP_{m,\vec{p}}(\partial_{a} \bH_{\vec{p}})\bP_{m',\vec{p}}\,\bP_{n',\vec{p}}(\partial_{b} \bH_{\vec{p}})\bP_{n,\vec{p}}\bP_{\vec{p}+\vec{k}}\big]\,
    \text{tr}[\bP_{\vec{k}'}\bP_{\vec{k}+\vec{k}'}]\,
    S_{F}^{(4)}(\epsilon_{m',\vec{p}}-iq_0,\epsilon_{m,\vec{p}},\epsilon_{n,\vec{p}},\epsilon_f)
    \Big) \, .
\end{align}
Using the identities in Sec.~\ref{sec:limits}, we obtain the Drude weight contribution of diagram (SE1)
\begin{align}
    D^{ab}_{\text{SE1}} &= \pi \lim_{\omega+i\eta\rightarrow 0}\big[\Pi^{ab}_{\text{SE1}}(iq_0)-\Pi^{ab}_{\text{SE1}}(0)\big] 
    \\
    &\approx-\frac{\pi e^{2}}{N_c^{3} V_c}\,
    C_{(\text{SE1})}\,
    \beta\,
    \delta_{W,0}[\nu(1-\nu)]^{2}\,
    \sum_{\vec{p},\vec{k},\vec{k}'}
    \sum_{m,m'}\sum_{n,n'}
    \nonumber\\
    &\hspace{8mm}
    \Bigg[U(\vec{k})U(\vec{k}')\,
    \text{tr}\Big[\bP_{m,\vec{p}}(\partial_{a} \bH_{\vec{p}})\bP_{m',\vec{p}}\bP_{\vec{p}+\vec{k}}\bP_{\vec{p}+\vec{k}+\vec{k}'}\bP_{\vec{p}+\vec{k}'}\bP_{n',\vec{p}}(\partial_{b} \bH_{\vec{p}})\bP_{n,\vec{p}}\Big]\,\delta_{m,n}\,
    \frac{\delta_{m,f}(1-\delta_{m',f})(1-\delta_{n',f})}{(\epsilon_{f}-\epsilon_{m',\vec{p}})(\epsilon_{f}-\epsilon_{n',\vec{p}})}
    \nonumber\\[2mm]
    &\hspace{8mm}
    +U(\vec{k})U(\vec{k}')v
    \text{tr}\Big[\bP_{m,\vec{p}}(\partial_{a} \bH_{\vec{p}})\bP_{m',\vec{p}}\,\bP_{n',\vec{p}}(\partial_{b} \bH_{\vec{p}})\bP_{n,\vec{p}}\bP_{\vec{p}+\vec{k}}\bP_{\vec{p}+\vec{k}+\vec{k}'}\bP_{\vec{p}+\vec{k}'}\Big]\,
    \delta_{m',n'}\,
    \frac{\delta_{m',f}(1-\delta_{m,f})(1-\delta_{n,f})}{(\epsilon_{f}-\epsilon_{m,\vec{p}})(\epsilon_{f}-\epsilon_{n,\vec{p}})}
    \nonumber\\[2mm]
    &\hspace{8mm}
    -U(\vec{k})U(\vec{k})\,
    \text{tr}\Big[\bP_{m,\vec{p}}(\partial_{a} \bH_{\vec{p}})\bP_{m',\vec{p}}\bP_{\vec{p}+\vec{k}}\bP_{n',\vec{p}}(\partial_{b} \bH_{\vec{p}})\bP_{n,\vec{p}}\Big]\,
    \text{tr}[\bP_{\vec{k}'}\bP_{\vec{k}+\vec{k}'}]\,\delta_{m,n}\,
    \frac{\delta_{m,f}(1-\delta_{m',f})(1-\delta_{n',f})}{(\epsilon_{f}-\epsilon_{m',\vec{p}})(\epsilon_{f}-\epsilon_{n',\vec{p}})}
    \nonumber\\[2mm]
    &\hspace{8mm}
    -U(\vec{k})U(\vec{k})\,
    \text{tr}\Big[\bP_{m,\vec{p}}(\partial_{a} \bH_{\vec{p}})\bP_{m',\vec{p}}\,\bP_{n',\vec{p}}(\partial_{b} \bH_{\vec{p}})\bP_{n,\vec{p}}\bP_{\vec{p}+\vec{k}}\Big]\,
    \text{tr}[\bP_{\vec{k}'}\bP_{\vec{k}+\vec{k}'}]\,
    \delta_{m',n'}\,
    \frac{\delta_{m',f}(1-\delta_{m,f})(1-\delta_{n,f})}{(\epsilon_{f}-\epsilon_{m,\vec{p}})(\epsilon_{f}-\epsilon_{n,\vec{p}})}
    \Bigg]
    \\\label{eq:D_se1}
    &=-\frac{\pi e^{2}}{N_c^{3} V_c}\,
    \beta\,
    \delta_{W,0}\,[\nu(1-\nu)]^{2}\,
    \sum_{\vec{p},\vec{k},\vec{k}'}
    \nonumber\\
    &\hspace{10mm}
    \bigg[U(\vec{k})U(\vec{k}')
    \Big[
    \text{tr}\big[\bP_{\vec{p}}(\partial_{a}\bP_{\vec{p}})\bP_{\vec{p}+\vec{k}}\bP_{\vec{p}+\vec{k}+\vec{k}'}\bP_{\vec{p}+\vec{k}'}(\partial_{b}\bP_{\vec{p}})\bP_{\vec{p}}\big]
    +\text{tr}\big[(\partial_{a}\bP_{\vec{p}})\bP_{\vec{p}}\,\bP_{\vec{p}}(\partial_{b}\bP_{\vec{p}})\bP_{\vec{p}+\vec{k}}\bP_{\vec{p}+\vec{k}+\vec{k}'}\bP_{\vec{p}+\vec{k}'}\big]
    \Big]
    \nonumber\\
    &\hspace{10mm}
    -U(\vec{k})U(\vec{k})
    \Big[
    \text{tr}\big[\bP_{\vec{p}}(\partial_{a}\bP_{\vec{p}})\bP_{\vec{p}+\vec{k}}(\partial_{b}\bP_{\vec{p}})\bP_{\vec{p}}\big]\,
    \text{tr}[\bP_{\vec{k}'}\bP_{\vec{k}+\vec{k}'}]
    +
    \text{tr}\big[(\partial_{a}\bP_{\vec{p}})\bP_{\vec{p}}\,\bP_{\vec{p}}(\partial_{b}\bP_{\vec{p}})\bP_{\vec{p}+\vec{k}}\big]\,
    \text{tr}[\bP_{\vec{k}'}\bP_{\vec{k}+\vec{k}'}]\,
    \Big]
    \bigg] \, .
\end{align}
We used that $C_{(\text{SE1})}=1$ since the diagram encounters a single loop with an odd number of Green functions (excluding those inside the self-energy).

\subsubsection{The contributing diagrams (SE2)}

The diagram (SE2) reads
\begin{align}
    &\Pi^{ab}_{\text{SE2}}(iq_0) 
    =-\frac{e^{2}}{N_c V_c}
    C_{(\text{SE2})}
    \sum_{\vec{p}}
    \sum_{m,m'}\sum_{n,n'}
    \text{tr}\Big[\bP_{m,\vec{p}}(\partial_{a} \bH_{\vec{p}})\bP_{m',\vec{p}}\,\hat\Sigma_{\vec{p}}^{(1)}\bP_{n',\vec{p}}(\partial_{b} \bH_{\vec{p}})\bP_{n,\vec{p}}\,\hat\Sigma_{\vec{p}}^{(1)}\Big]\nonumber \\
    &\hspace{80mm}\times S_{F}^{(4)}(\epsilon_{m',\vec{p}}-iq_0,\epsilon_{m,\vec{p}},\epsilon_{n',\vec{p}}-iq_0,\epsilon_{n,\vec{p}}) \, ,
\end{align}
where the superscript on $\hat\Sigma_{\vec{p}}^{(1)}$ signifies that we only consider the self-energy up to the first order in interaction. As before, we focus on the leading-order contribution in the semiquantum limit and consider only interband contributions arising from the coupling to the electromagnetic gauge potentials,  $m\neq m'$ and $n\neq n'$. By using the identities for the Matsubara summation given in Sec.~\ref{sec:matsubara_summations}, we find the leading order contribution to the Drude weight arising from the diagram (SE2) to be
\begin{align}\label{eq:drude_se2_with_se}
    D^{ab}_{\text{SE2}} &= \pi \lim_{\omega+i\eta\rightarrow 0}[\Pi^{ab}_{\text{SE2}}(iq_0)-\Pi^{ab}_{\text{SE2}}(0)] 
    \\
    &\approx-\frac{\pi e^{2}}{N_c V_c}
    C_{(\text{SE2})}
    \sum_{\vec{p}}
    \sum_{m,m'}\sum_{n,n'}
    \text{tr}\Big[\bP_{m,\vec{p}}(\partial_{a} \bH_{\vec{p}})\bP_{m',\vec{p}}\,\hat\Sigma_{\vec{p}}^{(1)}\bP_{n',\vec{p}}(\partial_{b} \bH_{\vec{p}})\bP_{n,\vec{p}}\,\hat\Sigma_{\vec{p}}^{(1)}\Big]
    \nonumber\\
    &\hspace{35mm}\times
    [-n_{F}'(\epsilon_{f})]
    \frac{\delta_{m',f}(1-\delta_{f,m})\delta_{n,f}(1-\delta_{f,n'})+\delta_{m,f}(1-\delta_{f,m'})\delta_{n',f}(1-\delta_{f,n})}{(\epsilon_{n,\vec{p}}-\epsilon_{n',\vec{p}})(\epsilon_{m',\vec{p}}-\epsilon_{m,\vec{p}})}
    \\
    &\approx-\frac{\pi e^{2}}{N_c V_c}\beta\,\nu(1-\nu)
    \,C_{(\text{SE2})}
    \sum_{\vec{p}}
    \Big(
    \text{tr}\Big[\bP_{\vec{p}}(\partial_{a}\bP_{\vec{p}})\,\hat\Sigma_{\vec{p}}^{(1)}\bP_{\vec{p}}(\partial_{b}\bP_{\vec{p}})\,\hat\Sigma_{\vec{p}}^{(1)}\Big]
    +
    \text{tr}\Big[(\partial_{a}\bP_{\vec{p}})\bP_{\vec{p}}\,\hat\Sigma_{\vec{p}}^{(1)}(\partial_{b}\bP_{\vec{p}})\bP_{\vec{p}}\,\hat\Sigma_{\vec{p}}^{(1)}\Big]
    \Big) \, ,
\end{align}
where we used the identities for the summation over the band indices given in Sec.~\ref{sec:recombination}. We insert the self-energy to first-order in interaction, given in Eq.~\eqref{eq:self_energy_corr1}, and note that any scalar term in $\hat\Sigma_{\vec{p}}^{(1)}$ drops due to the projector identity $\bP_{\vec{p}}(\partial_{a}\bP_{\vec{p}})\bP_{\vec{p}}=0$. Thus, using $C_{\text{SE2}}=-1$ and inserting the non-vanishing first term of the self-energy into Eq.~\eqref{eq:drude_se2_with_se} leads to
\begin{align}\label{eq:D_se2}
    D^{ab}_{\text{SE2}}
    &=\frac{\pi e^{2}\beta}{N_c^{3} V_c}\nu(1-\nu)
    \sum_{\vec{p},\vec{k},\vec{k}'}U(\vec{k})U(\vec{k}')
    \Big(
    \text{tr}\big[\bP_{\vec{p}}(\partial_{a}\bP_{\vec{p}}) \bP^{[\nu]}_{\text{occ},\vec{p}+\vec{k}}\bP_{\vec{p}}(\partial_{b}\bP_{\vec{p}}) \bP^{[\nu]}_{\text{occ},\vec{p}+\vec{k}'}\big]
    +
    \text{c.c.}
    \Big) \, .
\end{align}
Note that the contribution is entirely real due to the addition of its complex conjugate (c.c.) partner. 

\subsubsection{The contributing diagrams (SE3)}

We continue our derivation by evaluating the diagram (SE3) in Fig.~\ref{fig:second_order_self_energy_corrections}, where we show only a single example of the topologically distinct contributions. The four topologically equivalent (SE3) diagrams combined yield the mathematical form
\begin{align}
    &\Pi^{ab}_{\text{SE3}}(iq_0) 
    =-\frac{e^{2}}{N_c^{2} V_c}
    C_{(\text{SE3})}
    \sum_{\vec{p},\vec{k}}
    \sum_{m,m'}\sum_{n,n'}\sum_{l'}
    U(\vec{p}-\vec{k})
    \nonumber\\
    &\hspace{8mm}\times
    \bigg(
    \text{tr}\Big[\bP_{m,\vec{p}}(\partial_{a} \bH_{\vec{p}})\bP_{m',\vec{p}}\,\bP_{l',\vec{k}}\,\hat\Sigma_{\vec{k}}^{(1)}\bP_{n',\vec{k}}(\partial_{b} \bH_{\vec{k}})\bP_{n,\vec{k}}\Big]
    S_{F}^{(2)}(\epsilon_{m',\vec{p}}-iq_0,\epsilon_{m,\vec{p}})\,
    S_{F}^{(3)}(\epsilon_{n',\vec{k}}-iq_0,\epsilon_{n,\vec{k}},\epsilon_{l',\vec{k}}-iq_0)
    \nonumber\\&\hspace{11mm}+
    \text{tr}\Big[\bP_{m,\vec{p}}(\partial_{a} \bH_{\vec{p}})\bP_{m',\vec{p}}\,\bP_{n',\vec{k}}(\partial_{b} \bH_{\vec{k}})\bP_{n,\vec{k}}\,\hat\Sigma_{\vec{k}}^{(1)}\bP_{l',\vec{k}}\Big]
    S_{F}^{(2)}(\epsilon_{m',\vec{p}}-iq_0,\epsilon_{m,\vec{p}})\,
    S_{F}^{(3)}(\epsilon_{n',\vec{k}}-iq_0,\epsilon_{n,\vec{k}},\epsilon_{l',\vec{k}})
    \nonumber\\&\hspace{11mm}+
    \text{tr}\Big[\bP_{l',\vec{p}}\,\hat\Sigma_{\vec{p}}^{(1)}\bP_{m,\vec{p}}(\partial_{a} \bH_{\vec{p}})\bP_{m',\vec{p}}\,\bP_{n',\vec{k}}(\partial_{b} \bH_{\vec{k}})\bP_{n,\vec{k}}\Big]
    S_{F}^{(3)}(\epsilon_{m',\vec{p}}-iq_0,\epsilon_{m,\vec{p}},\epsilon_{l',\vec{p}})\,
    S_{F}^{(2)}(\epsilon_{n',\vec{k}}-iq_0,\epsilon_{n,\vec{k}})
    \nonumber\\&\hspace{11mm}+
    \text{tr}\Big[\bP_{m,\vec{p}}(\partial_{a} \bH_{\vec{p}})\bP_{m',\vec{p}}\,\hat\Sigma_{\vec{p}}^{(1)}\bP_{l',\vec{p}}\,\bP_{n',\vec{k}}(\partial_{b} \bH_{\vec{k}})\bP_{n,\vec{k}}\Big]
    S_{F}^{(3)}(\epsilon_{m',\vec{p}}-iq_0,\epsilon_{m,\vec{p}},\epsilon_{l',\vec{p}}-iq_0)\,
    S_{F}^{(2)}(\epsilon_{n',\vec{k}}-iq_0,\epsilon_{n,\vec{k}})
    \bigg) \, ,
\end{align}
which reduces to the Drude weight in leading order within the semiquantum limit
\begin{align}
    D^{ab}_{\text{SE3}} &= \pi \lim_{\omega+i\eta\rightarrow 0}[\Pi^{ab}_{\text{SE3}}(iq_0)-\Pi^{ab}_{\text{SE3}}(0)] 
    \\
    &=-\frac{\pi e^{2}}{N_c^{2} V_c}\,[-n_{F}'(\epsilon_{f})]\,
    C_{(\text{SE3})}\,
    \sum_{\vec{p},\vec{k}}
    U(\vec{p}-\vec{k})
    \nonumber\\
    &\hspace{30mm}\times
    \bigg(\,\,
    \text{tr}\Big[\big(\partial_{a} \bP^{[\nu]}_{\text{occ},\vec{p}}\big)\bP_{\vec{k}}\,\hat\Sigma_{\vec{k}}^{(1)}\big(\partial_{b}\bP_{\vec{k}}\big)\bP_{\vec{k}}\Big]
    +
    \text{tr}\Big[\big(\partial_{a} \bP^{[\nu]}_{\text{occ},\vec{p}}\big)\bP_{\vec{k}}\big(\partial_{b}\bP_{\vec{k}}\big)\hat\Sigma_{\vec{k}}^{(1)}\bP_{\vec{k}}\Big]
    \nonumber\\&\hspace{33mm}\,\,+
    \text{tr}\Big[\bP_{\vec{p}}\,\hat\Sigma_{\vec{p}}^{(1)}\big(\partial_{a}\bP_{\vec{p}}\big)\bP_{\vec{p}}\big(\partial_{b} \bP^{[\nu]}_{\text{occ},\vec{k}}\big)\Big]
    +
    \text{tr}\Big[\bP_{\vec{p}}\big(\partial_{a}\bP_{\vec{p}}\big)\hat\Sigma_{\vec{p}}^{(1)}\bP_{\vec{p}}\big(\partial_{b} \bP^{[\nu]}_{\text{occ},\vec{k}}\big)\Big]
    \bigg)
    \\
    &=
    -\frac{\pi e^{2}\beta}{N_c^{2} V_c}\,\nu(1-\nu)\,
    C_{(\text{SE3})}\,
    \sum_{\vec{p},\vec{k}}
    U(\vec{k})
    \Big[
    \Big(
    \text{tr}\big[\bP_{\vec{p}}(\partial_{a}\bP_{\vec{p}})\hat\Sigma_{\vec{p}}^{(1)}\bP_{\vec{p}}(\partial_{b} \bP^{[\nu]}_{\text{occ},\vec{p}+\vec{k}})\big]
    +c.c.
    \Big)
    +(a\leftrightarrow b)
    \Big] \, .
\end{align}
We denote the addition of the same terms with exchange indices by $(a\leftrightarrow b)$. The diagrams involve a single fermion loop and an odd number of Green's function, which leads to $C_{(\text{SE3})}=+1$. Thus, we end up with 
\begin{align}\label{eq:D_se3}
    D^{ab}_{\text{SE3}}
    &=
    \frac{\pi e^{2}\beta}{N_c^{3} V_c}\nu(1-\nu)
    \sum_{\vec{p},\vec{k},\vec{k}'}
    U(\vec{k})\,U(\vec{k}')
    \bigg[
    \Big(
    \text{tr}\Big[\bP_{\vec{p}}\big(\partial_{a}\bP_{\vec{p}}\big) \bP^{[\nu]}_{\text{occ},\vec{p}+\vec{k}'}\bP_{\vec{p}}\big(\partial_{b} \bP^{[\nu]}_{\text{occ},\vec{p}+\vec{k}}\big)\Big]
    +\text{c.c.}
    \Big)
    +(a\leftrightarrow b)
    \bigg] \, .
\end{align}

\subsubsection{The non-contributing diagram (SE4)}

The diagram (SE4) does not contribute to the leading order in the semiquantum limit \eqref{eq:semiquantum_limit} in the main text for the same reasons as the diagrams $(h)$, $(i)$, and $(j)$ in Fig.~\ref{fig:second_order_vertex_corrections}.

\subsubsection{The contributing diagrams (SE5)}

The diagram (SE5) has two self-energies on one side and none on the other and is representative of two topologically equivalent diagrams. They read
\begin{align}
    \Pi^{ab}_{\text{SE5}}(iq_0) 
    =&-\frac{e^{2}}{N_c V_c}
    C_{(\text{SE5})}
    \sum_{\vec{p}}
    \sum_{m,m'}\sum_{n,n'}\sum_{l}\nonumber \\
    &\Big(
    \text{tr}\big[\bP_{m,\vec{p}}(\partial_{a} \bH_{\vec{p}})\bP_{m',\vec{p}}\,\hat\Sigma_{\vec{p}}^{(1)}\bP_{l,\vec{p}}\hat\Sigma_{\vec{p}}^{(1)}\bP_{n',\vec{p}}(\partial_{b} \bH_{\vec{p}})\bP_{n,\vec{p}}\big]\,
    S_{F}^{(4)}(\epsilon_{m',\vec{p}}-iq_0,\epsilon_{l,\vec{p}}-iq_0,\epsilon_{n',\vec{p}}-iq_0,\epsilon_{m,\vec{p}})
    \nonumber\\
    &+
    \text{tr}\big[\bP_{m,\vec{p}}(\partial_{a} \bH_{\vec{p}})\bP_{m',\vec{p}}\bP_{n',\vec{p}}(\partial_{b} \bH_{\vec{p}})\bP_{n,\vec{p}}\,\hat\Sigma_{\vec{p}}^{(1)}\bP_{l,\vec{p}}\hat\Sigma_{\vec{p}}^{(1)}\big]\,
    S_{F}^{(4)}(\epsilon_{m',\vec{p}}-iq_0,\epsilon_{n,\vec{p}},\epsilon_{l,\vec{p}},\epsilon_{m,\vec{p}})\Big) \, ,
\end{align}
where we used the fact that the self-energy does not depend on Matsubara frequency to first order in the interaction.
Using the identities in Sec.~\ref{sec:limits}, we obtain the Drude weight contribution of diagram (SE5) 
\begin{align}
    D^{ab}_{\text{SE5}} &= \pi \lim_{\omega+i\eta\rightarrow 0}[\Pi^{ab}_{\text{SE5}}(iq_0)-\Pi^{ab}_{\text{SE5}}(0)] 
    \\
    &\approx-\frac{\pi e^{2}}{N_c V_c}\,
    C_{(\text{SE5})}\,
    \beta\,
    [\nu(1-\nu)]\,
    \sum_{\vec{p}}
    \sum_{m,m'}\sum_{n,n'}\sum_{l}
    \nonumber\\
    &\hspace{8mm}\times\bigg(
    \text{tr}\Big[\bP_{m,\vec{p}}(\partial_{a} \bH_{\vec{p}})\bP_{m',\vec{p}}\,\hat\Sigma_{\vec{p}}^{(1)}\bP_{l,\vec{p}}\hat\Sigma_{\vec{p}}^{(1)}\bP_{n',\vec{p}}(\partial_{b} \bH_{\vec{p}})\bP_{n,\vec{p}}\Big]\delta_{m,n}\delta_{l,f}
    \frac{\delta_{m,f}(1-\delta_{m',f})(1-\delta_{n',f})}{(\epsilon_{f}-\epsilon_{m',\vec{p}})(\epsilon_{f}-\epsilon_{n',\vec{p}})}
    \nonumber\\
    &\hspace{10mm}
    +\text{tr}\Big[\bP_{m,\vec{p}}(\partial_{a} \bH_{\vec{p}})\bP_{m',\vec{p}}\bP_{n',\vec{p}}(\partial_{b} \bH_{\vec{p}})\bP_{n,\vec{p}}\,\hat\Sigma_{\vec{p}}^{(1)}\bP_{l,\vec{p}}\hat\Sigma_{\vec{p}}^{(1)}\Big]
    \delta_{m',n'}\delta_{l,f}
    \frac{\delta_{m',f}(1-\delta_{m,f})(1-\delta_{n,f})}{(\epsilon_{f}-\epsilon_{m,\vec{p}})(\epsilon_{f}-\epsilon_{n,\vec{p}})}\bigg)
    \\
    &=\frac{\pi e^{2}}{N_c V_c}
    \beta
    [\nu(1-\nu)]
    \sum_{\vec{p}}\bigg(
    \text{tr}\Big[\bP_{\vec{p}}(\partial_{a}\bP_{\vec{p}})\,\hat\Sigma_{\vec{p}}^{(1)}\bP_{\vec{p}}\hat\Sigma_{\vec{p}}^{(1)}(\partial_{b}\bP_{\vec{p}})\bP_{\vec{p}}\Big]
    +\text{tr}\Big[(\partial_{a}\bP_{\vec{p}})\bP_{\vec{p}}(\partial_{b}\bP_{\vec{p}})\,\hat\Sigma_{\vec{p}}^{(1)}\bP_{\vec{p}}\hat\Sigma_{\vec{p}}^{(1)}\Big]\bigg)
    \\\label{eq:D_se5}
    &=\frac{\pi e^{2}}{N_c^{3} V_c}
    \beta
    [\nu(1-\nu)]
    \sum_{\vec{p},\vec{k},\vec{k}'}U(\vec{k})U(\vec{k}')\bigg(
    \text{tr}\Big[\bP_{\vec{p}}(\partial_{a}\bP_{\vec{p}})\bP_{\text{occ},\vec{p}+\vec{k}}^{[\nu]}\bP_{\vec{p}}\bP_{\text{occ},\vec{p}+\vec{k}'}^{[\nu]}(\partial_{b}\bP_{\vec{p}})\bP_{\vec{p}}\Big]
    \nonumber\\
    &\hspace{70mm}
    +\text{tr}\Big[(\partial_{a}\bP_{\vec{p}})\bP_{\vec{p}}(\partial_{b}\bP_{\vec{p}})\bP_{\text{occ},\vec{p}+\vec{k}}^{[\nu]}\bP_{\vec{p}}\bP_{\text{occ},\vec{p}+\vec{k}'}^{[\nu]}\Big]\bigg) \, .
\end{align}

\subsection{The Drude weight of a flatband metal in the semiquantum limit}

We conclude by combining the various terms that we obtained throughout the SM. The Drude weight including all diagrams to second order in interaction and leading order in the semiquantum limit as specified in Eq.~\eqref{eq:semiquantum_limit} in the main text takes the form 
\begin{align}
    D^{ab}&=D^{ab}_{\text{no-SE}}+D^{ab}_{\text{SE1}}+D^{ab}_{\text{SE2}}+D^{ab}_{\text{SE3}}+D^{ab}_{\text{SE5}}
    \\
    &= \frac{\pi e^2 \beta}{N_c^{2}V_c}\sum_{\vec{k},\vec{k}'}
    U(\vec{k})U(\vec{k}')
    \bigg(
    \nu(1-\nu)\,
    \mathfrak{g}^{ab}_{1}(\vec{k},\vec{k}';\nu)
    +
    \delta_{W,0}\,[\nu(1-\nu)]^{2}\Big[\mathfrak{g}^{ab}_{2}(\vec{k}) \,N_{c}\,\delta_{\vec{k},\vec{k}'}-\mathfrak{g}^{ab}_{3}(\vec{k},\vec{k}')\Big]
    \bigg) \, ,
\end{align}
which we obtained by combining our intermediate results given in Eqs.~\eqref{eq:D_no_se},~\eqref{eq:D_se1},~\eqref{eq:D_se2},~\eqref{eq:D_se3}, and~\eqref{eq:D_se5}. We restructured the different contributions into three geometric tensors defined as
\begin{align}
    &\mathfrak{g}^{ab}_{1}(\vec{k},\vec{k}';\nu) = \frac{1}{2}\text{Re}\bigg[\frac{1}{N_{c}}\sum_{\vec{p}}\text{tr}\Big[\bP_{\vec{p}}\partial_{a}(\bP_{\vec{p}} \bP^{[\nu]}_{\text{occ},\vec{p}+\vec{k}}\bP_{\vec{p}})\bP_{\vec{p}}\partial_{b}(\bP_{\vec{p}} \bP^{[\nu]}_{\text{occ},\vec{p}+\vec{k}'}\bP_{\vec{p}})\Big]\bigg]+(a\leftrightarrow b)\, ,
    \\[2mm]
    &\mathfrak{g}_{2}^{{a}{b}}(\vec{k})
    =
    \text{Re}\bigg[\frac{1}{N_{c}}\sum_{\vec{p}}
    \text{tr}\Big[ \bP^{}_{\vec{p}}\partial^{}_{{a}}( \bP^{}_{\vec{p}} \bP^{}_{\vec{p}+\vec{k}}) \bP^{}_{\vec{p}+\vec{k}}\Big]\,\frac{1}{N_{c}}
    \sum_{\vec{p}'}
    \text{tr}\Big[ \bP^{}_{\vec{p}'}\partial^{}_{{b}}( \bP^{}_{\vec{p}'} \bP^{}_{\vec{p}'+\vec{k}}) \bP^{}_{\vec{p}'+\vec{k}}\Big]
    \nonumber
    \\&\hspace{12mm}
    +
    \frac{1}{N_{c}}\sum_{\vec{p}}
    \text{tr}\Big[ \bP^{}_{\vec{p}}\partial_{{a}}( \bP^{}_{\vec{p}} \bP^{}_{\vec{p}+\vec{k}}) \bP^{}_{\vec{p}+\vec{k}}\partial^{}_{{b}}( \bP^{}_{\vec{p}+\vec{k}} \bP^{}_{\vec{p}})\Big]\,
    \frac{1}{N_{c}}\sum_{\vec{p}'}\text{tr}\big[ \bP^{}_{\vec{p}'+\vec{k}} \bP^{}_{\vec{p}'}\big]\bigg]\,,
    \\[2mm]
    &\mathfrak{g}^{ab}_{3}(\vec{k},\vec{k}') = 
    \text{Re}\bigg[
    \frac{1}{N_{c}}\sum_{\vec{p}}
    \text{tr}\Big[ \bP^{}_{\vec{p}}\partial^{}_{{a}}( \bP^{}_{\vec{p}} \bP^{}_{\vec{p}+\vec{k}}) \bP^{}_{\vec{p}+\vec{k}} \bP^{}_{\vec{p}+\vec{k}+\vec{k}'} \bP^{}_{\vec{p}+\vec{k}'}\partial^{}_{{b}}( \bP^{}_{\vec{p}+\vec{k}'} \bP^{}_{\vec{p}})\Big]
    \bigg]
    +(a\leftrightarrow b)\,.
\end{align}
Note that each contribution is real, symmetric in $a\leftrightarrow b$ (and in $\vec{k}\leftrightarrow\vec{k}'$ where applicable), and invariant under the $U(1)$-gauge ambiguity of Bloch states as required. We take the thermodynamic limit and present the final result in Eqs.~\eqref{eq:drude_result}, \eqref{eq:g1_tensor}, \eqref{eq:g2_tensor}, and \eqref{eq:g3_tensor} in the main text.

\newpage
\end{widetext}

\end{document}